\documentclass{jaa}
\usepackage{natbib}
\usepackage{makecell}
\usepackage{url}

\DeclareFontFamily{OT1}{pzc}{}
\DeclareFontShape{OT1}{pzc}{m}{it}{<-> s * [1.10] pzcmi7t}{}
\DeclareMathAlphabet{\mathpzc}{OT1}{pzc}{m}{it}

\newcommand{\MSUN}{{\rm M}_{\odot}}

\usepackage{graphicx}

\begin{document}\sloppy

\title{Probing the Epoch of Reionization using synergies of line intensity\\ mapping}



\author{
Chandra Shekhar Murmu\textsuperscript{1,*},
Raghunath Ghara\textsuperscript{2},
Suman Majumdar\textsuperscript{1,3}, \and
Kanan K. Datta\textsuperscript{4}
}

\affilOne{\textsuperscript{1}
Department of Astronomy, Astrophysics and Space Engineering, Indian Institute of Technology Indore, Khandwa Rd., Simrol 453552, India.\\}
\affilTwo{\textsuperscript{2}
Astrophysics Research Center (ARCO), Department of Natural Sciences, The Open University of Israel, 1 University Rd., PO Box 808, Ra'anana 4353701, Israel.\\}
\affilThree{\textsuperscript{3}
Department of Physics, Blackett Laboratory, Imperial College, London SW7 2AZ, U.K.\\}
\affilFive{\textsuperscript{4}
Department of Physics, Jadavpur University, Kolkata 700032, India.}


\twocolumn[{

\maketitle

\corres{chandra0murmu@gmail.com}

\msinfo{11 Mar 2022; revised 15 Jun 2022}{13 Jul 2022}

\begin{abstract}
The Epoch of Reionization (EoR) remains a poorly understood cosmic era for the most part. Yet, efforts are still going on to probe and understand this epoch. We present a review of the latest developments in the techniques (especially line-intensity mapping) to study the EoR and try to highlight the contribution of the Indian community in this field. Line-emissions like [H {\scriptsize I}]$_{\rm 21cm}$, Lyman-$\alpha$, [C {\scriptsize II}]$_{\text{158}\mu\text{m}}$ and their role as tracers in probing the EoR are discussed. While the [H {\scriptsize I}]$_{\rm 21cm}$ is an excellent probe of the early IGM, the others are mainly targeted to do an unresolved and large-scale survey of the reionizing sources. Techniques to model these signals include simulations and machine learning approaches, along with the challenge to tackle foregrounds or interlopers. We also discuss synergy opportunities among the various tracers that we mention. Synergy addresses different aspects of the problem, which otherwise is difficult or impossible to tackle. They include statistics like cross-power spectrum, cross-bispectrum, and other techniques such as follow-up studies. We present updates on the relevant experiments; these include the upper limits on the [H {\scriptsize I}]$_{\rm 21cm}$ power spectrum, along with some highlights on high-redshift galaxy surveys. Finally, we highlight what can be improved further within the community: applying machine learning and simulations based on hydrodynamic and radiative-transfer techniques. Next-generation experiments also need to be conceived to address issues currently beyond our reach.
\end{abstract}

}]


\doinum{12.3456/s78910-011-012-3}
\artcitid{\#\#\#\#}
\volnum{000}
\year{0000}
\pgrange{1--}
\setcounter{page}{1}
\lp{1}

\section{Introduction}
The strive to understand the history of our Universe represents a major goal of modern cosmology. Precise measurements of the Cosmic Microwave Background Radiation (CMB) and accurate maps of the galaxies in the nearby universe have revolutionized our understanding of this history. Together they provide a somewhat detailed picture of the very early phase of the Universe and of its present state. However, our knowledge of how the Universe has evolved between these two extreme ends has actually been very limited so far.

One of the most important missing links to this history is the phase that includes the Cosmic Dawn (CD) and the Epoch of Reionization (EoR), the period during which the very first sources of light were formed. The ultraviolet (UV) and X-ray radiation emitted by these and by the subsequent population of sources gradually heated and ``re''-ionized the cold and neutral inter-galactic medium (IGM), consisting of mostly neutral hydrogen (H{\scriptsize I}) (see e.g. \citealt{fan+2006,furlanetto+2006,choudhury+2009,pritchard+2012} etc. for reviews). Our current understanding of this epoch is mainly guided by different indirect observations such as the Thompson scattering optical depth  of the CMB \citep{komatsu+2011,planck+2020}, the absorption spectra of high redshift quasars (see e.g. \citealt{becker+2001,fan+2003,barnett+2017} etc) and the luminosity function and clustering properties of Lyman-$\alpha$ emitters (e.g. \citealt{ouchi+2010,jensen+2013,choudhury+2015,bouwens+2016a,Zheng+2017} etc). These observations suggest that the reionization may have been an extended process, spanning over the redshift range $6 \leq z \leq 15$ (see e.g. \citealt{alvarez+2006,mitra+2013,mitra+2015,choudhury+2015, bouwens+2015, robertson+2015} etc.). However, these observations are unable to resolve the many of the pressing questions regarding the CD-EoR in a definitive manner: When did it start? How did it develop over time? What were the major sources of ionizing photons during this period? 

There are fundamentally two different direct approaches to observe this era and answer these questions: 
\begin{itemize}
    \item One is by observing the sources of light present in this period through photometry or spectroscopy with optical and near and far infrared telescopes.
    \item The other approach is by observing the time evolving neutral hydrogen (H{\scriptsize I}) distribution in the IGM with radio telescopes. These observations have the potential to probe the state of the IGM gas as it evolves with time due to the heating and ionization by the first light sources.
\end{itemize}

The present time is particularly exciting in the context of observing the CD as a large number of next-generation telescopes will become functional within a few years, which will be able to directly observe this era for the first time with unprecedented resolution and sensitivity. One of them is the recently launched James Webb Space Telescope (JWST), operating in optical and near infrared. The JWST will be able to show us the glimpses of the individual stars and galaxies from deep into the CD-EoR. 

The next-generation telescopes in far infrared and radio will be observing this era with a radically different approach: with the so-called Line Intensity Mapping (LIM) technique (see e.g. \citealt{Bharadwaj+2001,bharadwaj+2001b,Wyithe+2009,carilli+2011,Gong_2011,Gong+2012,silva13,Mashian+2015,Silva+2015,Sun+2018,Kovetz+2019,Silva+2021} etc). While this method relies on observations of specific spectral lines, it does not try to detect line emission from individual sources. Instead, to increase the detectability of the signal, all photons associated with a specific redshifted line frequency within a coarse resolution element (several times coarser than the angular size of the individual source) of the telescope are added up. Thus, it allows the observer to map out the distribution of sources in coarser resolution but in a larger field-of-view (FoV).

The LIM comes in two flavours:
\begin{itemize}
    \item One can observe the sources of light which produced the ionizing photons and reionized the IGM.
    \item One can also observe the H {\scriptsize I} in the IGM.
\end{itemize}

The 21cm signal emerging from the neutral H {\scriptsize I} atoms due to hyperfine spin-flip transition can be used to observe the early IGM. The first galaxies can be traced by multiple line emissions such as the [C {\scriptsize II}]$_{\text{158}\mu\text{m}}$, CO, [O {\scriptsize III}]$_{\text{88}\mu\text{m}}$, Lyman-$\alpha$ etc. These line emissions will be observed as redshifted lines due to the cosmological expansion of the Universe. We can combine all these tracers to develop a tomographic map of the early Universe.

The Square Kilometre Array (SKA) - the largest radio interferometric array ever built and one of the seven international Mega Science Projects in which India is a major partner - will become operational in $\sim$2027 and is expected to be capable of imaging the neutral hydrogen distribution during the CD using the redshifted [H {\scriptsize I}]$_{\text{21cm}}$ signal \citep{koopmans+2015,mellema+2015}. This signal will probe the physical processes in the IGM driving the reionization and the evolving topology of the H {\scriptsize I} distribution during the CD-EoR.

LIM experiments targeting galaxies using CO, [C {\scriptsize II}]$_{\text{158}\mu\text{m}}$ are coming online, and more will be operational before SKA is set up. LIM using [C {\scriptsize II}]$_{\text{158}\mu\text{m}}$ and CO lines~\citep{Lidz+2011,Gong+2012,Pullen+2013,Mashian+2015,Silva+2015,Yue+2015,Li+2016,Lidz_and_Taylor+2016,Serra+2016,Breysse17,Padmanabhan+2018,Sun+2018,Bernal+2019a,Bernal+2019b,Breyesse+2019,Dumitru+2019,Ihle+2019,Moradinezhad_Dizgah+2019,Padmanabhan+2019a,Sun+2019,Moradinezhad_Dizgah+2022a,Moradinezhad_Dizgah+2022b,Karoumpis+2022,Yang+2021,Yang+2022} are expected to be powerful probes of the galaxies responsible for reionization. Experiments like CONCERTO~\citep{Dumitru+2019,Catalano+2022}, TIME~\citep{crites+2014,Sun+2021}, and FYST~\citep{Dumitru+2019,Karoumpis+2022}, will be targeting the [C {\scriptsize II}]$_{\text{158}\mu\text{m}}$ line.

Experiments like COPSS~\citep{Keating+2015,Keating+2016} and mmIME~\citep{Breysse+2022} have reported the detection of the CO LIM signal from the early Universe.

To optimize the utilization of these next-generation experiments to understand the CD-EoR, the astronomers around the world on one hand  are developing strategies for synergistic multi-wavelength observations of this era and on the other hand building comprehensive interpretation pipelines for extracting maximum amount of information from these observations. In this article we review the recent advancements on such multi-wavelength observation strategies for the CD-EoR, specifically focusing on the LIM mapping approach and also the development in the front of novel statistics and other interpretation techniques for such future LIM studies. 

\section{Expected signals from the EoR:\newline [H {\scriptsize I}]$_{\text{21cm}}$ and others}

One of the ways to probe the EoR, is to try to detect the [H {\scriptsize I}]$_{\text{21cm}}$ signal emitted from the neutral hydrogen of the early IGM. This emission arises due to the hyperfine transition in the H{\scriptsize I} atoms, with excitation from CMB and Lyman-$\alpha$ photons, as well as collisional excitation from H {\scriptsize I} atoms. Therefore, by mapping this emission, we can effectively map the evolving state of the ionized IGM with cosmic time and uncover the astrophysics responsible. But building instruments that can detect this signal by effectively taking care of the foreground noise, which is orders of magnitudes higher than the expected signal, is a significant challenge.

The other ways of probing EoR are to map the reionizing galaxies themselves. Spectroscopic detection of galaxies are expensive but can provide the most detailed information. Ongoing experiments (e.g. ALMA) and upcoming ones (e.g. JWST) will play significant roles. These methods are not suitable for large-volume surveys. To overcome this, a newly emerging tool known as LIM has gained substantial popularity, with more new instruments coming online. This approach maps the integrated flux from numerous sources within a given voxel (2D pixel + frequency resolution). It can therefore do large-volume surveys within sustainable observational times. In the following sections, we discuss the physics of some of the line emissions that are a potential candidate to probe the Universe from the EoR, using LIM.

\subsection{[H {\scriptsize I}]$_{\text{21cm}}$}

The [H {\scriptsize I}]$_{\text{21cm}}$ signal arises from the neutral hydrogen due to the hyperfine spin-flip transition, resulting in the emission of photons with a rest-frame wavelength of approximately 21cm. This signal is detectable against the Cosmic Microwave Background Radiation (CMB) as a change in the brightness temperature in the relevant redshifted wavelengths. The strength of this signal depends on the population ratio of the hyperfine excited spin states, which one can parametrize with a quantity called the spin temperature ($T_s$), given as
\begin{equation}
    \frac{n_e}{n_g} = \frac{g_e}{g_g} \exp{-\frac{T_*}{T_s}}
    \label{eq: pop_state}
\end{equation}
\citep{Field+1958}. $n_e$ and $n_g$ are the number density of H {\scriptsize I} atoms in hyperfine excited and ground states, respectively, with $g_e$ and $g_g$ being the corresponding degeneracies. $T_* = h_p c / (k_B \lambda_e) = 68 \text{mK}$, with $h_p$, $c$ and $k_B$ being the Planck's constant, speed of light, and Boltzmann's constant respectively, and $\lambda_e$ = 21cm. The excitation of the H{\scriptsize I} atoms is caused by CMB photons, Lyman-$\alpha$ coupling (Wouthuysen-Field effect, \citep{Wouthuysen+1952,Field+1958}) and collisional coupling from H {\scriptsize I} atoms. These processes, in principle, determine the spin temperature, following the relation
\begin{equation}
    1 - \frac{T_\gamma}{T_s} = \frac{x_\alpha+x_c}{1+x_\alpha+x_c}\Bigg(1 - \frac{T_\gamma}{T_k}\Bigg)
    \label{eq: spin_temp}
\end{equation}
\citep{Field+1958,Madau+1997,Barkana_&_Loeb+2005}, where $x_{\alpha} = 4P_{\alpha}T_*/(27A_{10}T_{\gamma}$) is the Lyman-$\alpha$ coupling strength with $P_{\alpha}$ being the the Lyman-$\alpha$ scattering rate. $x_c = 4\kappa_{1-0}(T_k)n_HT_*/(3A_{10}T_{\gamma})$ is the collisional coupling rate of the spin-temperature to the gas temperature. $T_\gamma$ = $2.725(1+z)$K is the CMB temperature, $T_{\alpha}$ is the colour temperature of the Lyman-$\alpha$ radiation, defined as
\begin{equation}
    T_\alpha = -\frac{1}{k_B}\:\Bigg(\frac{\partial\log \mathcal{N}_{\nu}}{\partial (h_p\nu)}\Bigg)^{-1}
    \label{eq: lyman_color_temp}
\end{equation}
\citep{Madau+1997}, with $\mathcal{N}_{\nu}$ being the photon-occupation number. $T_k$ is the kinetic temperature of the gas distribution.

The brightness temperature $T_b$ of the [H {\scriptsize I}]$_{\text{21cm}}$ signal against the CMB is given by
\begin{equation}
    \begin{split}
    T_b = 4 \text{mK}\:x_{\text{HI}} (1+\delta_{\text{H}})\bigg(\frac{\Omega_\text{b} h^2}{0.02}\bigg)\bigg(\frac{0.7}{h}\bigg)\sqrt{\frac{1+z}{\Omega_\text{m}}} \\
    \times\:\Bigg(1-\frac{T_{\gamma}}{T_s}\Bigg)\Bigg[1 - \frac{1+z}{H(z)}\frac{\partial v}{\partial r}\Bigg]
    \label{eq: brightness_temp}
\end{split}
\end{equation}
\citep{Bharadwaj_&_Ali+2005}. In the above expression, $x_{\text{HI}} = \rho_{\text{HI}}/\rho_{\text{H}}$ is the mass-averaged neutral fraction, $\delta_{\text{H}}$ is the hydrogen over-density, and $\partial v/\partial r$ is the rate of change of peculiar velocity along the line of sight (LoS) with co-moving distance. Therefore, quantitatively, $T_b$ is a probe of the neutral hydrogen distribution in the Universe.

However, the problem of foreground contamination poses a significant challenge in the extraction of this [H {\scriptsize I}]$_{\text{21cm}}$ signal. This foreground signal is contributed by: Galactic synchrotron radiation and extra-galactic radio sources. These contamination signals can fall within the wavelength band of the redshifted, cosmological [H {\scriptsize I}]$_{\text{21cm}}$ signal, which typically dominates it by orders-of-magnitude~\citep{Di_Matteo+2002,Santos+2005}. In the subsequent sections, we will discuss how to mitigate this foreground challenge using clever techniques.

\subsection{Lyman-$\alpha$}

There are several avenues for studying the EoR using Lyman-$\alpha$ emitters (LAEs). One potential approach is to look into the impact of increasing neutral IGM on the statistical properties of the population of LAEs at redshift $z \gtrsim 6$. One obvious effect is the decrease in the fraction of UV-selected galaxies that also show Lyman-$\alpha$ emission as the H {\scriptsize I} density increases at higher redshifts. There are some observations which indicate that the LAE fraction starts decreasing at higher redshifts~\citep{ouchi+2010,Kashikawa+2011,Konno+2014,Matthee+2015,Bagley+2017,Ota+2017,Sadoun+2017,Shibuya+2019}. This can be used to probe ionization state of the IGM during the later stages of reionization \citep{choudhury15, weinberger18}. Another useful probe is the two-point correlation function of LAEs, which quantifies the clustering of these objects. It is expected that the clustering of galaxies will increase due to the patchy reionization process and, thus, will enhance the two-point correlation function of LAEs compared to UV-selected samples with the same number density \citep{jensen+2013}. Apart from probes based on observations of a handful of individual galaxies, there are proposals to study the EoR through Lyman-$\alpha$ intensity mapping \citep{silva13, heneka21}. The large scale power spectrum of Lyman-$\alpha$ intensity fluctuations carries signatures of the total Lyman-$\alpha$ luminosity from galaxies as well as the large scale matter power spectrum. The clustering power spectrum can be written as \citep{silva13},
\begin{equation}
P_{\rm GAL}^{\rm clus}(k, z)= b^2_{\rm Ly\alpha } {\bar I}^2_{\rm GAL} P_{\delta \delta}(z, k), 
\end{equation}
where $b_{\rm Ly\alpha }$ is the luminosity bias, ${\bar I}^2_{\rm GAL}$ is the total average intensity of Ly$\alpha$ emission and $P_{\delta \delta}$ is the matter power spectrum. There will also be a shot-noise power spectrum present, due to discrete galaxy distribution \citep{Gong_2011}. 
In addition to the approaches discussed above there are also a handful of studies which explore prospects of measuring cross-correlations of large scale [H {\scriptsize I}]$_{\text{21cm}}$ maps and Lyman-$\alpha$ emitters  using ongoing/upcoming radio interferometric experiments and Lyman-$\alpha$ surveys through the Subaru's Hyper Suprime-Cam (HSC)~\citep{furlanetto07,vrbanec16}.
Other than this, there are approaches to cross-correlate 21cm maps with Lyman-$\alpha$ emitters. It uses ongoing/upcoming radio interferometric experiments and Lyman-$\alpha$ surveys using Subaru's Hyper Suprime-Cam (HSC)~\citep{furlanetto07,vrbanec16} and SDSS/BOSS~\citep{Croft+2018}.
This cross-correlation signal can be measured using observations carried out by experiments like LOFAR/SKA and HSC and should be able to distinguish different EoR scenarios such the inside-out and outside-in \citep{sobacchi16,hutter17}. Impact of foreground contamination is likely to be much less severe in the cross-correlation signal and any detection of the cross-correlation signal will also confirm cosmological origin of the [H {\scriptsize I}]$_{\text{21cm}}$ signal \citep{feng17}.    

\subsection{[C {\scriptsize II}]$_{\text{158}\mu\text{m}}$}

The [C {\scriptsize II}]$_{\text{158}\mu\text{m}}$ line emission arises from the fine-structure transition \big($^2P_{3/2} \rightarrow {^2P_{1/2}}$\big) in a C$^+$ ion, resulting in a rest-frame wavelength of $\simeq 158\,\mu\rm m$. The population ratio of the ions can be written as $n_e/n_g = g_e/g_g \exp(-T_{*,\rm CII}/T_s)$, with $T_{*,\rm CII} = 91\,\rm K$. The dominant mechanism by which the C$^+$ ions are excited in the interstellar medium (ISM) of galaxies is collisions from electrons and atoms. This line emission is an excellent coolant for the ISM~\citep{Sun+2018} and arises from a variety of different environments: photo-dissociation regions, cold gas \citep{De_Looze+2011}, and CO dark clouds \citep{Olsen+2015}. Generally, it is also considered a good tracer of dust-enshrouded star-formation, with studies showing that it is well correlated with the star-formation rate (SFR) in a galaxy \citep{De_Looze+2011,De_Looze+2014}. It is a strong IR line emission, accessible within $4.5 \lesssim z \lesssim 8.5$ \citep{Kannan+2021}; all of these motivated upcoming experiments to map the EoR galaxies using [C {\scriptsize II}]$_{\text{158}\mu\text{m}}$. However, the CO line emissions from low-redshift galaxies pose a significant foreground interloper \citep{Gong+2012,Silva+2015}, making the signal detection more challenging. Later, we will discuss one of the many ways that might help in mitigating this problem.

\section{Modelling the signals and constraining astrophysics from the EoR}

There have been recent developments in the techniques to extract astrophysics of the EoR by analyzing the LIM signals. We summarize them in the following subsections.

\subsection{The [H {\scriptsize I}]$_{21\rm cm}$ signal}
\subsubsection{Modelling the [H {\scriptsize I}]$_{21\rm cm}$ signal from the EoR:}
A major component of developing any interpretation framework or pipeline for an observation is to build one or many forward models of the expected target signal or the target signal statistic. In case of the CD-EoR [H {\scriptsize I}]$_{\text{21cm}}$ signal the forward models come in three broad categories i.e. analytical, semi-numerical and numerical.

Most of the analytical models of the CD-EoR [H {\scriptsize I}]$_{\text{21cm}}$ signal are motivated by the fact that the signal fluctuations are determined by the size and distribution of the heated and ionized region in the IGM. The simplest approach to build such a model would be to consider the fluctuations in the signal via the distribution of non-overlapping ionized and heated spheres \citep{Bharadwaj_&_Ali+2005, Datta+2007}. These models have been further improved by accounting for the overlaps through excursion-set formalism proposed by \citet{Furlanetto+2004} and by taking photon conservation into account \citep{Paranjape+2012,Paranjape+2014,Paranjape+2016}. The analytical models provide the fastest avenue for estimating the large-scale fluctuations in the signal at the zeroth order approximation and have been extensively used for constraining the reionization histories while combining all available observations at the present time (see e.g. \citealt{Mitra+2011, mitra+2013, mitra+2015, Chatterjee+2021} etc).    
Though the improved analytical approaches discussed above can provide a reasonably good model for the signal, they are limited in their ability to include various complexities which have a major impact on the signal fluctuations. Some of these complexities are the non-spherical shape of the ionized regions, impact of the matter density fluctuations on ionization and recombination processes, inherent line of sight anisotropies e.g. redshfit space distortions and light cone effect etc.  These requirements have led to the development of a large number of semi-numerical models of the signal. In the semi-numerical models the focus is on simulating the important cosmological effects as accurately as possible within a large volume (comparable to the observational field of view) while approximating the radiative transfer process related to the reionization, thus optimizing the resources for computation. Being able to simulate the signal in large volumes with reasonable accuracy, these models allow us to explore the multi-dimensional CD-EoR parameter space at quicker pace and at the cost of moderate computing resources. 

Most of the popular semi-numerical models are based on the excursion set formalism proposed by \citep{Furlanetto+2004}. In these models the radiative transfer solutions are replaced by the comparison of smoothed fields of ionizing photon density with that of the neutral hydrogen density. The smoothing scale is then varied from an estimated mean free path of the photons to the resolution of the simulation, to check if at any scale the ionization condition is satisfied or not. An ionization map produced by this manner is then converted into the 21-cm brightness temperature map  via Equation \eqref{eq: brightness_temp}. To speed up the simulation process some of these models use Zeldovich approximation for generating the underlying dark matter density field \citep{Mesinger+2007,Mesinger+2011}. The other approach simulates the underlying dark matter field via an N-body simulation and also identifies the collapsed halos within the matter distribution as the potential hosts for the ionizing photon sources (see e.g. \citealt{Zahn+2007,Geil+2008, choudhury+2009, Majumdar+2012,Majumdar+2013,Majumdar+2014, Mondal+2015,Mondal+2016} etc). The second approach requires more computing resources but allows one to accurately implement various cosmological effects to the signal e.g. redshift space distortions and light cone effect which has a significant impact on the signal statistics. Several authors have compared the results from these semi-numerical techniques with that of the more accurate radiative transfer simulations \citep{choudhury+2009,Mesinger+2011,Majumdar+2014,Ghara+2018}. They have demonstrated that for most of the statistics of interest e.g. power spectrum at large scales and ionized bubble size distribution and their topology etc the results from  the semi-numerical approaches are within the sample variances limit of that of the radiative transfer simulations.                      
One of the major drawbacks of the excursion set based semi-numerical approach for simulating the [H {\scriptsize I}]$_{\text{21cm}}$ signal is the issue of photon non-conservation. This may lead to an amplitude change in the various signal statistics. This issue has been recently resolved by \citet{Paranjape+2016,Maity+2022} through effectively redistributing the ionizing photons of overlapping smoothing regions in its neighbourhood.

To include most of the physical processes of the IGM in the signal model one would need to use a radiative transfer simulation. There are many radiative transfer algorithms that have been developed over the years \citep{Gnedin+1997,Gnedin+2000a,Gnedin+2000b,Ciardi+2000,Mellema+2006,Iliev+2006,Iliev+2007,Iliev+2012,Iliev+2014,Kannan+2021,Kannan+2021a}. Among these different approaches of numerical radiative transfer simulations, \citet{Mellema+2006,Iliev+2006,Iliev+2007,Iliev+2012,Iliev+2014} are able to simulate the CD-EoR [H {\scriptsize I}]$_{\text{21cm}}$ signal in large enough cosmological volumes ($\sim [715\, \rm cMpc]^3$) which can mimic future observations of this era with the SKA. However, they also face the hindrance of not being able to rerun their simulations numerous times ($\geq 10^6$) to be able to explore the vast parameter space of the signal.  

An alternative to resolve this issue while keeping some of the details of the radiative transfer simulations, is to use a spherically symmetric one-dimensional radiative transfer
algorithm \citep{Thomas+2009}. It allows one to simulate the signal in very large-scales. This approach has been further improved and both hydrogen and helium reionization have been included in \citet{Ghara+2015a,Ghara+2015,ghara16,Ghara+2018}. It has also been demonstrated that this kind of one-dimensional radiative transfer algorithm is fast enough that it can be used for constraining CD-EoR parameters from the observed signal statistics \citep{2020MNRAS.493.4728G,2021MNRAS.503.4551G}.

\subsubsection{Reionization history:}
\label{sec:maps}
One of the crucial aspects that we are generally keen to learn about the EoR is how the reionization proceeded (history) and its duration. Usually, we use the neutral fraction, defined as $x_{\text{HI}} = \rho_{\text{HI}}/\rho_{\text{H}}$, which describes what fraction of hydrogen is still in a neutral state at a given cosmic time. The evolution of the globally averaged (averaged over all-sky) neutral fraction is one of the speed indicators of the reionization process and its history. Constraining this can help us understand many aspects, one of which is the role of source models in this process. One of the ways to constrain the reionization history, given the value of the $x_{\text{HI}}$ at a single redshift as an external input, is to employ the Multi-Frequency Angular Power Spectrum (MAPS) \citep{Datta+2007}. The brightness temperature can be decomposed in terms of spherical harmonics $(Y_l^m(\mathbf{\hat{n}}))$ as follows:
\begin{equation}
    T_b(\mathbf{\hat{n}},\nu) = \sum_{\ell,m} a_{\ell,m}(\nu) Y_{\ell}^m(\mathbf{\hat{n}}),
    \label{eq: sphr_harm}
\end{equation}
with $\mathbf{\hat{n}}$ and $\nu$ representing the direction and observed frequency of the signal respectively. Following \cite{Datta+2007}, the MAPS is defined as
\begin{equation}
   \mathcal{C}_{\ell}(\nu_1,\nu_2) =\Big \langle a_{\ell,m}(\nu_1) a^*_{\ell,m}(\nu_2)\Big \rangle.
    \label{eq: maps1}
\end{equation}
It can characterize the entire 2-point statistics of the signal in the presence of the light-cone effect since this statistic does not assume the signal to be statistically homogenous across the LoS \citep{Mondal+2018}. We can further decompose MAPS as
\begin{equation}
    \mathcal{C}_{\ell}(\nu_1, \nu_2) = \Bar{x}_{\text{HI}}(\nu_1)\Bar{x}_{\text{HI}}(\nu_2)\,\mathcal{C}_{\ell}^{E}(\Delta \nu)
    \label{eq: maps2}
\end{equation}
\citep{Mondal+2019}, assuming that the evolution of the mean neutral fraction ($\Bar{x}_{\text{HI}}$) across the LoS far surpasses that of the other quantities. With this, the evolutionary history of the neutral fraction could be extracted as
\begin{equation}
    x_{\text{HI}}(\nu) = A\sqrt{\mathcal{C}_{\ell}(\nu)/\overline{\mathcal{C}}_{\ell}},
    \label{eq: xhi_from_maps}
\end{equation}
with $\mathcal{C}_{\ell}(\nu)\equiv \mathcal{C}_{\ell}(\nu,\nu)$, and $\overline{\mathcal{C}}_{\ell}=B^{-1}\int_{-B/2}^{B/2}\mathcal{C}_{\ell}(\nu)d\nu$. $A$ is a normalization constant, which is determined from a single $x_{\text{HI}}$ value at a given redshift. The ratio $\mathcal{C}_{\ell}(\nu)/\overline{\mathcal{C}}_{\ell}$ is expected to show systematic variation with $\nu$, and thereby infer $x_{\rm HI}(\nu)$. However, for small $\ell$ bins this is not seen, and the analysis required restricting $\ell$ to $> 2571$.

Another promising way to extract $x_{\text{HI}}$ is to employ trained Convolutional Neural Networks (CNN) to analyze [H {\scriptsize I}]$_{\text{21cm}}$ images from future observations. \cite{Mangena+2020} explored the above, with simulated [H {\scriptsize I}]$_{\text{21cm}}$ maps from the Instantaneous version of S{\footnotesize IM}F{\footnotesize AST}21 \citep{Hassan+2016,Santos+2010} and generating mock images following SKA1-Low instrument design. The CNN is a set of different layers with specific tasks, such as extraction of features (convolutional layer) and up/down-sampling the output of the convolutional layer. Finally, the fully connected layer extracts features from a 1D input. Overall, this feature extraction is achieved by minimizing appropriate loss functions: the seperation between true values/labels and predicted values from the CNN.
When trained on the outputs of S{\footnotesize IM}F{\footnotesize AST}21, it achieves good accuracies of up to 99 per cent on the simulated data set and up to 98 per cent on the mock dataset, and it extracts $x_{\text{HI}}$ in a model-independent fashion.

\subsubsection{Bubble statistic and [H {\scriptsize I}]$_{\text{21cm}}$ morphology:}

We can characterize the reionization process with Minkowski Functionals (MFs) and Contour Minkowski Tensors (CMTs). \cite{Kapahtia+2018} discusses the prospects of using CMTs to constrain statistics of the EoR, such as the shape and mean size of the ionized bubble. If the boundary of the ionized bubble is represented with a curve, then we can correspondingly define a tensor
\begin{equation}
    \mathcal{W}_1 = \int_{\mathcal{C}} \hat{T}\otimes\hat{T} ds
    \label{eq: minkowski_tensor}
\end{equation}
$\hat{T}$ is the unit-tangent vector on the curve and $\otimes$ is the symmetric tensor product $\Big(\hat{T}\otimes\hat{T}\Big)_{ij} = \Big(\hat{T}_i\hat{T}_j + \hat{T}_j\hat{T}_i\Big)/2$. $\mathcal{W}_1$ transforms as a rank-2 tensor and has eigenvalues, $\lambda_1$ and $\lambda_2$. We can obtain two other quantities from this: $\beta = \lambda_1/\lambda_2$ and $r = (\lambda_1+\lambda_2)/2\pi$. Using further transformation on these quantities, we can precisely quantify the shape-anisotropy and mean size of the ionized bubbles and their redshift evolution \citep{Kapahtia+2018}. Furthermore, we can use this analysis to classify reionization morphology. Therefore, it can potentially constrain reionization models as well \citep{Kapahtia+2019}.

In a slightly different approach for analyzing the tomographic images one can try to follow the volume and morphology of the largest ionized region with cosmic time.  \citet{Bag+2018, Bag+2019} had used Largest Cluster Statistics (LCS) to follow the volume and morphology of the largest ionized region in 21-cm tomographic map. They have shown that  the LCS can reveal the stage of the reionization when percolation between the ionized regions takes place. Further, recently, \citet{Pathak+2022} have shown using a large suit of simulated reionization scenarios that the evolution of LCS can reveal the nature of the reionization i.e. whether the reionization is inside-out or outside-in. 

Another possible way of using the 21-cm images from the EoR obtained via SKA would be to conduct such imaging in the region of the sky where high redshift galaxies and quasars have been observed already via Euclid~\citep{Chary+2020}, Nancy Grace Roman~\citep{Chary+2020}, JWST~\citep{Steinhardt+2021} or ELT~\citep{Hammer+2021} through photometry or spectroscopy. The targeted 21-cm imaging of ionized via such synergistic observations, as discussed in \citet{Zackrisson+2020}, will be able to help us in constraining the contribution of bright galaxies and quasars in reionization.      


\subsubsection{[H {\scriptsize I}]$_{\text{21cm}}$ Fourier statistics:}
The main Fourier statistics that most of the present and next-generation radio interferometers are aiming for the detection of the [H {\scriptsize I}]$_{\text{21cm}}$ signal, is its power spectrum. The power spectrum measures the amplitude of fluctuations in the signal at different length scales. Therefore the shape and amplitude of the power spectrum and their evolution have the ability to constrain the CD-EoR parameters. This idea has been utilized to develop power spectrum based parameter estimation pipelines either under a Bayesian framework \citep{Greig+2015,Greig+2017,Greig+2018} or through the use of Artificial Neural Network (ANN) \citep{Choudhury+2021b} or a hybrid ANN aided Bayesian framework \citep{Schmit+2018,Binnie+2019,Tiwari+2021}. 

The power spectrum also gets affected by various line-of-sight anisotropies which are intrinsic to the signal e.g. the light-cone effect and the redshift space distortions. The light-cone effect arises due to the finite travel time of the cosmological signal from a distant source to the present day observer. The light cone effect can change the amplitude and the shape of the observed power spectrum depending on the reionization history and frequency bandwidth within which the power spectrum is being estimated \citep{Barkana+2006, Datta+2012,Datta+2014,Ghara+2015,Mondal+2018}.  
The other line-of-sight anisotropy that affects the signal power spectrum significantly is the redshift space distortions. The redshift space distortions, arising due to the gas peculiar velocities, will make any cosmological signal anisotropic along the line-of-sight of the observer. This will change both the shape and amplitude of the signal power spectrum \citep{Bharadwaj+2004,Bharadwaj+2005,Ali+2005, Majumdar+2013,Majumdar+2014,Majumdar+2016,Ghara+2015a}. Thus it is important to take this effect into account while using the signal power spectrum for the CD-EoR parameter estimation. It has also been demonstrated that the higher order non-zero multipole moments of the power spectrum, arising due to the redshift space distortions, has the potential to constrain the reionization history \citep{Majumdar+2016}.

However, the power spectrum is not the optimal statistics in terms of its information content. The CD-EoR [H {\scriptsize I}]$_{\text{21cm}}$ signal is expected to be highly non-Gaussian due to the non-uniform distribution of the heated and ionized regions in the IGM which dominates the signal fluctuations \citep{Mondal+2015,Mondal+2016,Shaw+2019}. As the power spectrum cannot capture this non-Gaussianity in the signal one has to estimate higher order statistics e.g. bispectrum \citep{Majumdar+2018,Majumdar+2020,Kamran+2021,Kamran+2021b,Saxena+2020, Tiwari+2021, Watkinson+2021,Mondal+2021}. It has also been demonstrated that the bispectrum have the potential to put tighter constraints on the CD-EoR parameters compared to the signal power spectrum \citep{Tiwari+2021}. 

\subsubsection{Applications of machine learning and neural networks: }

Machine learning (ML) have unfolded immense progress in studying the EoR. Also, over recent years, the applications of ML has developed rapidly. We have ML tools that can analyze data and extract the EoR astrophysics. \cite{Hassan+2019} report using convolutional neural networks that distinguish source models by analyzing [H {\scriptsize I}]$_{\text{21cm}}$ images. The accuracy can range from $92\text{--}100$ per cent. \cite{Choudhury+2020} study how the [H {\scriptsize I}]$_{\text{21cm}}$ signal can be extracted from the Cosmic Dawn using Artificial Neural Networks (ANNs). Their method can perform with 92 per cent accuracy and extract the foreground parameters. They also successfully constrained the EoR parameters from the global signal \citep{Choudhury+2021} and the [H {\scriptsize I}]$_{\text{21cm}}$ power spectrum \citep{Choudhury+2021b}. The work by \cite{Villanueva-Domingo+2021} has developed techniques to extract the underlying matter density field, given the [H {\scriptsize I}]$_{\text{21cm}}$ field in redshift space. The statistical properties match the true ones within error limits of a few per cent, down to scales of $\simeq 2\,\rm Mpc^{-1}$. Also, the astrophysical parameters can be constrained from the astrophysical information retained by the neural network. There is also the class of ML algorithms called emulators that, after being trained on input data, can re-create that input given some parameters; this happens to incur a much lesser computational cost. In that way, one can potentially use it for parameter estimation. \cite{Cohen+2020} use the 21{\footnotesize CM}GEM that emulates the global signal from the CD and EoR. It is trained on a dataset of $\sim 30,000$ simulated signals developed from varying seven parameters. The predictions achieve good accuracies with good runtime efficiencies. Also, emulators are applied for power spectrum \citep{Tiwari+2021,Sikder+2022} and bispectrum \citep{Tiwari+2021}, with bispectrum constraining parameters more tightly. The recent work by \cite{Zhou+2021} also points out the limitations of using only single semi-numeric models to train CNNs. They showed that when CNNs trained on one model tries to infer astrophysics from data produced with another model, they generally behave poorly, suggesting that CNNs need to be trained on diverse ranges of inputs that capture the full astrophysics of the EoR.

\subsubsection{Tackling foregrounds for [H {\scriptsize I}]$_{\text{21cm}}$ signal: }

We face foreground noise as the most significant challenge in detecting the [H {\scriptsize I}]$_{\text{21cm}}$ signal. It mainly arises from galactic synchrotron emission and extragalactic sources. Also, calibration errors and other instrumental systematics affect signal detection adversely. Therefore, appropriate tools and methods are necessary to deal with this problem. In this subsection, we discuss some of the recent developments to tackle the noise in the signal.

\cite{Kerrigan+2018} discusses a hybrid method of dealing with foreground noise. It involves subtracting foreground models in real space combined with the filtering of noise power-spectrum in Fourier space. When applied to PAPER and MWA data, these resulted in significant improvements compared to just filtering. For the MWA band centre, the improvement relative to just filtering is $\sim 10^{9}\,\text{mK}^2 (h^{-1}\,\text{Mpc})^3$. At the MWA band edges, one can expect improvements, in the order of $\sim 10^{12}\text{--}10^{13}\,\text{mK}^2 (h^{-1}\,\text{Mpc})^3$.

Even after handling the foreground noise, with avoidance and subtraction, errors may remain in calibration and other systematics. \cite{Mertens+2018} introduces the 'Gaussian Process Regression' (GPR) technique to statistically remove the contributions of the stochastic errors towards the [H {\scriptsize I}]$_{\text{21cm}}$ power spectrum. They demonstrated, using simulated LOFAR-EoR data, that this can recover the power spectrum on scales $k = 0.07\text{--}3\, h\,\text{Mpc}^{-1}$. This method is optimal when correlation in foregrounds is present on frequency scales $\gtrsim 3$ MHz and signal rms is $\sigma_{\text{21cm}} > 0.1\sigma_{\text{noise}}$. Compared to foreground avoidance, this improves the sensitivity by a factor of 3.

\subsection{[C {\scriptsize II}]$_{158\mu\text{m}}$}

Recent developments in using the [C {\scriptsize II}]$_{158\mu\text{m}}$ line for LIM include sophisticated line-emission modelling and instrumental forecasts. \cite{Daisy_Leung_2020} have used hydrodynamics based cosmological simulation, {\footnotesize SIMBA} \citep{Dave+2019}, to simulate the galaxy distribution. It models detailed physics such as radiative processes, the evolution of gas content from feedback mechanisms, and subgrid models to account for dust content. The simulation output is post-processed with {\footnotesize S\`{I}GAME}~\citep{Olsen+2015,Olsen+2016,Olsen+2017} for modelling line emissions. The results of such sophisticated modelling include a flatter [C {\scriptsize II}]$_{158\mu\text{m}}$ luminosity versus Star-Formation Rate ($L_{\text{CII}}$-SFR) relation, which can have implications for interpreting the [C {\scriptsize II}]$_{158\mu\text{m}}$ power spectrum from upcoming observations.
 
This relation has a significant scatter ranging from $0.3-0.6$ dex. Other sets of modelling include luminosity function and stellar-mass function. They have limited their study to $z=6$. Detailed line-emission modelling from early galaxies, including the [C {\scriptsize II}]$_{158\mu\text{m}}$ line, has also been addressed by \cite{Kannan+2021}, using the flagship simulation, the {\footnotesize THESAN} project. However, their predictions for the [C {\scriptsize II}]$_{158\mu\text{m}}$ line emission fall far below the \cite{Daisy_Leung_2020} prediction because it doesn't model the individual ISM phases (Photo-dissociation regions and CO dark molecular clouds) in detail responsible for this line emission, which they plan to improve in the future version of their simulations. Works such as \cite{Murmu+2021b} revisit the interpretation of [C {\scriptsize II}]$_{158\mu\text{m}}$ power spectrum by remapping the scatter in the [C {\scriptsize II}]$_{158\mu\text{m}}$ luminosities from \cite{Daisy_Leung_2020} in a large volume N-body simulation.

Further works such as \cite{Sun+2018} discuss foreground mitigation for the [C {\scriptsize II}]$_{158\mu\text{m}}$ line for a mock field of Tomographic Ionized-carbon Mapping Experiment (TIME). The idea is to use galaxy catalogues to identify voxels contaminated with at least one of the CO lines and discard it. It involves using the stellar mass of galaxies to determine the masking depth, which relates to a threshold CO flux to cut-off. Compared to blind masking of voxels based on brightness only, this technique uses the spectral information of galaxies. Therefore, it can reduce faint CO contamination better. They find that this criterion amounts to K-band magnitudes of $m_{\text{AB}} \lesssim 22$, and [C {\scriptsize II}]$_{158\mu\text{m}}$/CO power ratio of $\gtrsim 10$ might be achievable, at a cost of 8 percent loss in survey volume.

\begin{figure}[!ht]
    \centering
    \includegraphics[width=\columnwidth]{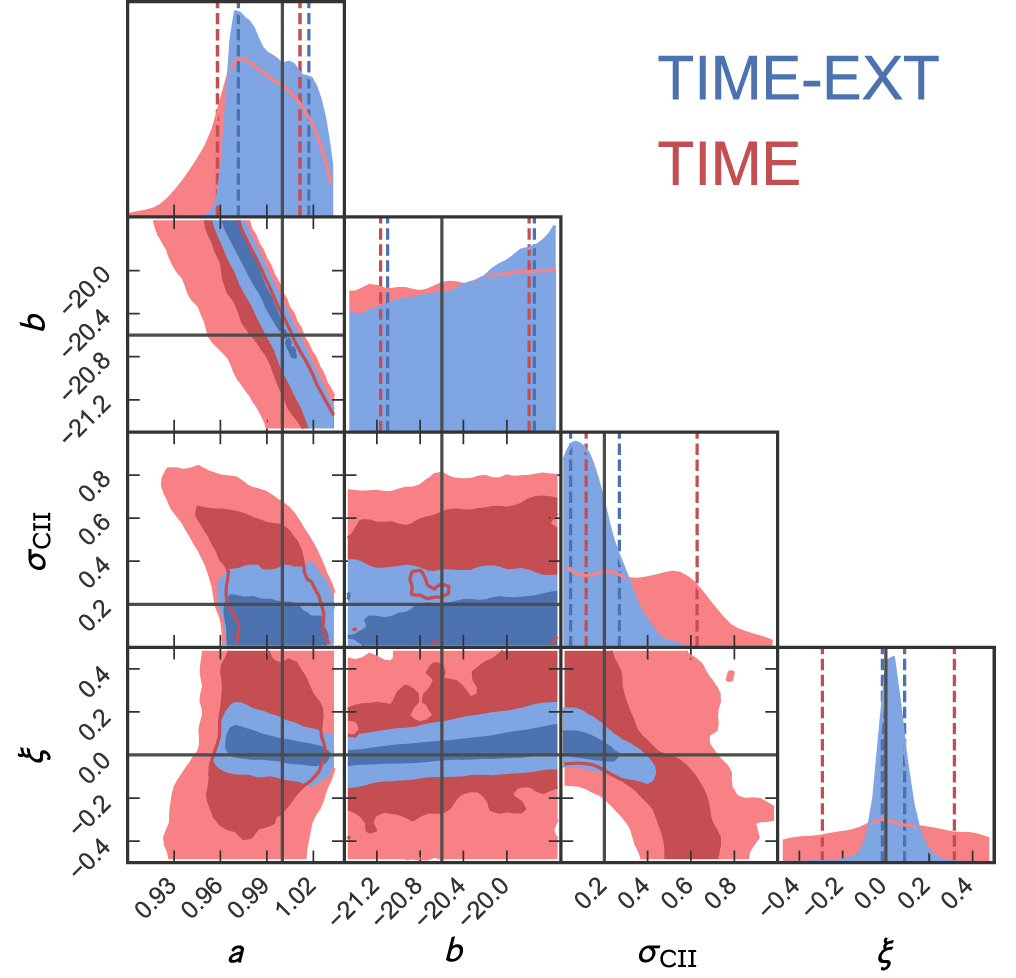}
    \caption{Constraints on the parameters relating to the [C {\scriptsize II}]$_{158\mu\text{m}}$ line emission, forecasted for TIME and TIME-EXT.~Source: \cite{Sun+2021}, \textit{Probing Cosmic Reionization and Molecular Gas Growth with TIME}, ApJ 915 33, Fig.~7, \copyright AAS. Reproduced with permission.}
    \label{fig: cii_constraint}
\end{figure}

The role of TIME in constraining the parameters relating to [C {\scriptsize II}]$_{158\mu\text{m}}$ line emission will be crucial. \cite{Sun+2021} had made forecasts for the TIME using MCMC analysis, on the kind of constraints it can provide. The [C {\scriptsize II}]$_{158\mu\text{m}}$ luminosity can be related to the UV continuum luminosity as
\begin{equation}
    \log \Bigg(\frac{L_{\text{CII}}}{L_{\odot}}\Bigg) = a \log \Bigg(\frac{L_{\rm UV}}{\rm erg\: \rm s^{-1}\: \rm Hz^{-1}}\Bigg) + b
    \label{eq: luv_to_lcii}
\end{equation}
$L_{\rm UV}$ is a proxy for the star-formation rate as $\Dot{M}_* = \mathcal{K} L_{\rm UV}$. One can also assume a scatter parameter, $\sigma_{\rm CII}$, to account for the luminosity scatter. The other parameter is $\xi$, which is used to model star formation efficiency as
\begin{equation}
    f_*(M) = \frac{f_{*,0}}{(M/M_p)^{\gamma_{\text{lo}}(M)}+(M/M_p)^{\gamma_{\rm hi}}},
\end{equation}
with $\gamma_{\text{lo}}(M) = -0.55\times10^{\xi/M/M_c}$. This model accounts for the deviation of the star-formation efficiency from a power law. The star-formation rate is, therefore, $\Dot{M}_* = f_* \Omega_b/\Omega_m \Dot{M}$, with $\Dot{M}$ being the growth rate of halo mass. Thus the potential parameters to be constrained are $a$, $b$, $\sigma_{\rm CII}$ and $\xi$. The analysis is done for TIME and TIME-EXT (which is an improvement over TIME in terms of survey parameters). In Fig.~\ref{fig: cii_constraint}, we can see that parameters a, $\sigma_{\rm CII}$ and $\xi$ are constrained well since they are related to the power spectrum shape, whereas b is prior dominated, which only controls the power spectrum normalization. We can expect TIME-EXT to provide a better constraint on the parameter $\xi$. These constraints can, in turn, infer the [C {\scriptsize II}]$_{158\mu\text{m}}$ luminosity function as well.

\section{Extracting the EoR-astrophysics with Synergy-studies}

In earlier sections, we discussed how we plan to use the individual tracers of the IGM and galaxies to probe the EoR. Studies find that there are several ways by which these line emissions can help understand the EoR-astrophysics and constrain the EoR-parameters. In this section, we talk about synergies: combining individual probes. Compared to single tracers, synergies can unfold an understanding that is otherwise not achievable. One of the important advantages of cross-correlation synergies is the mitigation of of common foreground and interloper contamination. Since these foreground and interlopers generally originate from a different redshift, with different source-distribution properties, these are expected to wash-out when two different tracers are cross-correlated. Therefore, in the prospect of studying the EoR, synergies between the different probes are crucial.

\subsection{Line-intensity mapping synergies:}

\subsubsection{The [C {\scriptsize II}]$_{\text{158}\mu\text{m}}\times$[H {\scriptsize I}]$_{\text{21cm}}$ cross-power spectra:}

We can use the cross-power spectrum to constrain the EoR parameters. Below, we discuss the [C {\scriptsize II}]$_{\text{158}\mu\text{m}}\times$[H {\scriptsize I}]$_{\text{21cm}}$ cross-power spectra in these scenarios. The work form \cite{Dumitru+2019} discusses the prospects of constraining EoR parameters like minimum-halo mass $(M_{\text{min}})$ for reionization. They analyze the detectability of this cross-power signal and conclude that it is detectable to within Signal-to-Noise Ratio (SNR) of 10 by Stage-II experiments like FYST~\citep{Karoumpis+2022}. Cross-power spectrum can constrain the parameters like $M_{\text{esc}}$ (minimum halo-mass capable of producing Lyman-C photons that can escape galaxies) better than 21cm measurements alone, by a factor of 3 and 10, when derived from 1000 hr and 5000 hr observational data of [C {\scriptsize II}]$_{\text{158}\mu\text{m}}$ experiments.

However, when extracting the cross-power spectrum from a large-volume survey, one must be careful about the impact of the light-cone effect. As shown in \cite{Murmu+2021}, the light-cone effect significantly affects the spherically-averaged [C {\scriptsize II}]$_{\text{158}\mu\text{m}}\times$[H {\scriptsize I}]$_{\text{21cm}}$ cross-power spectrum, up to 20 per cent on small $k$-modes ($k\sim0.1 \text{Mpc}^{-1}$).
\begin{figure*}[!ht]
    \centering
    \includegraphics[width=\textwidth]{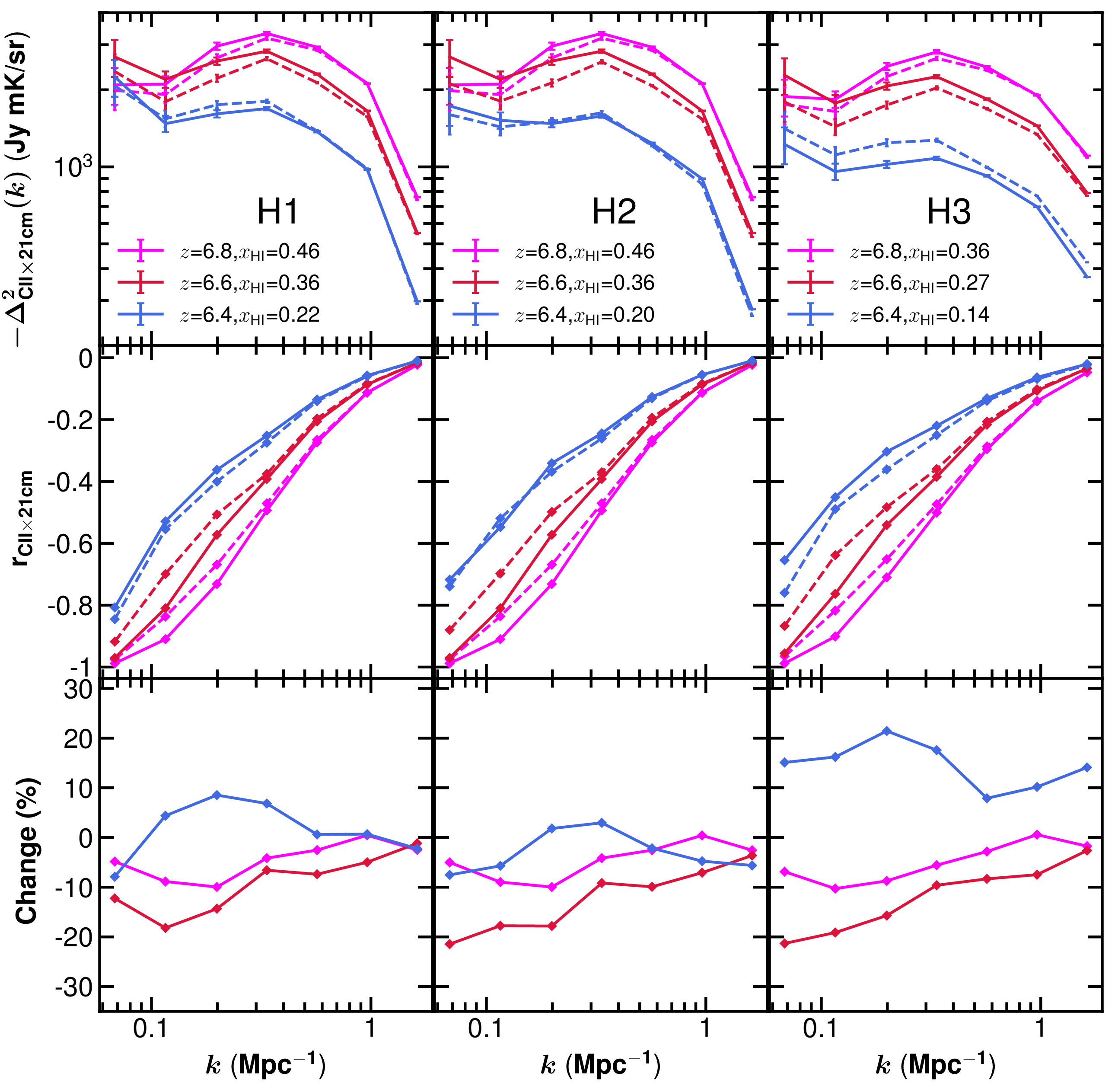}
    \caption{Light-cone impact on the [C {\scriptsize II}]$_{\text{158}\mu\text{m}}\times$[H {\scriptsize I}]$_{\text{21cm}}$ cross-power spectrum is shown in this figure \citep{Murmu+2021}. Top panel shows the cross-power with and without light-cone effect, middle panel shows the cross-correlation coefficient, and the bottom panel shows the impact of light-cone in percentage. H1, H2 and H3 represent different reionization histories.}
    \label{fig: c2hi_cps}
\end{figure*}
Moreover, the light-cone effect influences the [H {\scriptsize I}]$_{\text{21cm}}$ signal differently when the reionization history is different, and in Fig.~\ref{fig: c2hi_cps}, we can see that the cross-power spectrum is also affected accordingly. Since it is not straightforward to predict the impact of light-cone on the cross-power spectrum using the light-cone effect on the individual auto-power spectrum alone, we must model it carefully.

\subsubsection{Multi-tracer:}

\begin{figure}[ht]
    \centering
    \includegraphics{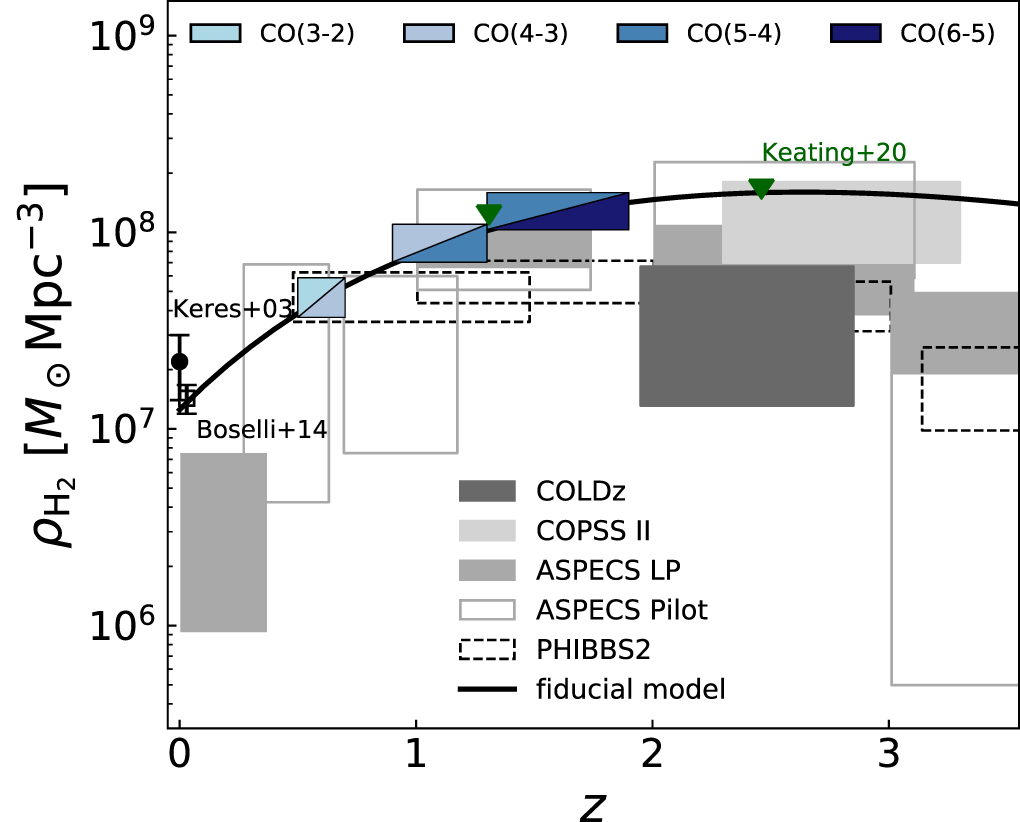}
    \caption{TIME will provide constraints on the molecular gas growth by cross-correlating adjacent pair of CO lines.~Source: \cite{Sun+2021}, \textit{Probing Cosmic Reionization and Molecular Gas Growth with TIME}, ApJ 915 33, Fig.~14, \copyright AAS. Reproduced with permission.}.
    \label{fig: rho_h2}
\end{figure}

TIME can constrain cosmic molecular-gas growth by cross-correlating pairs of adjacent CO rotational lines, e.g. CO(3-2)$\times$CO(4-3), CO(4-3)$\times$CO(5-4), CO(5-4)$\times$CO(6-5). It can determine the CO(1-0) luminosity, thereby constraining $\rho_{\text{H}_2}$ from the following relation
\begin{equation}
    \rho_{\text{H}_2} = \alpha_{\text{CO}}\int dM \frac{dn}{dM}L^{\prime}_{\text{CO}}(M, z),
\end{equation}
which relates $\rho_{\text{H}_2}$ to the the CO(1-0) luminosity. As shown in Fig.~\ref{fig: rho_h2}, TIME will provide this additional constraint besides other experiments, providing valuable cross-check.
\citep{Kannan+2021}\\

\subsubsection{Cross-bispectrum:}

Similar to the cross-power spectrum, we can devise the cross-bispectrum to extract astrophysics of the EoR and constrain parameters. The cross bispectrum can be defined as
\begin{equation}
\begin{split}
    \big\langle T_{21\text{cm}}(\mathbf{k_1})I_{\text{CII}}(\mathbf{k_2})I_{\text{CII}}(\mathbf{k_3}) \big\rangle = (2\pi^3)\delta_D(\mathbf{k_1+k_2+k_3})\\
    \times B_{21,\text{CII},\text{CII}}(\mathbf{k_1,k_2,k_3}),
\end{split}
    \label{eq: cross_bi}
\end{equation}
with the reduced cross-bispectrum defined as
\begin{equation}
    \begin{split}
        \hat{Q}_{21,\text{CII},\text{CII}}(\mathbf{k_1,k_2,k_3}) = \frac{B_{21,\text{CII},\text{CII}}(\mathbf{k_1,k_2,k_3})}{P_{21,\text{CII}}(k_1)P_{21, \text{CII}}(k_2)+2\, \text{perm}.}.
    \end{split}
    \label{eq: reduced_bi}
\end{equation}
\cite{Beane_&_Lidz+2018} show that we can write the reduced cross-bispectrum with an arbitrary field X, as
\begin{equation}
    \hat{Q}^{(0)}_{21,X,X} = \frac{Q_{\delta,\delta,\delta}}{\langle T_{21}\rangle b_{21}} + C_{21,X,X},
    \label{eq: reduced_cross_bi}
\end{equation}
with $C_{21,\delta,\delta} = b^{(2)}_{21}/\big(6\langle T_{21} \rangle b^2_{21}\big)$ and $C_{21,\text{CII},\text{CII}} = C_{21,\delta,\delta} + b^{(2)}_{\text{CII}}/\big(3\langle T_{21} \rangle b_{21}b_{\text{CII}}\big)$. Therefore, both $\hat{Q}^{(0)}_{21,\delta,\delta}$ and $\hat{Q}^{(0)}_{21,\text{CII},\text{CII}}$ follow $\hat{Q}_{\delta,\delta,\delta}$ within a constant offset. In this study, they demonstrate the effectiveness of $\hat{Q}^{(0)}_{21,\text{CII},\text{CII}}$ by using $\hat{Q}^{(0)}_{21,\delta,\delta}$ as a proxy to extract $\langle T_{21} \rangle b_{21}$. They find that it can be used to constrain $\langle T_{21} \rangle b_{21}$ within a 5 per cent error for $\langle x_i \rangle < 0.5$ and 10 per cent error for $\langle x_i \rangle > 0.5$, except for the lowest redshift ($z=6$) tested in their work.
\begin{figure}
    \centering
    \includegraphics[width=\columnwidth]{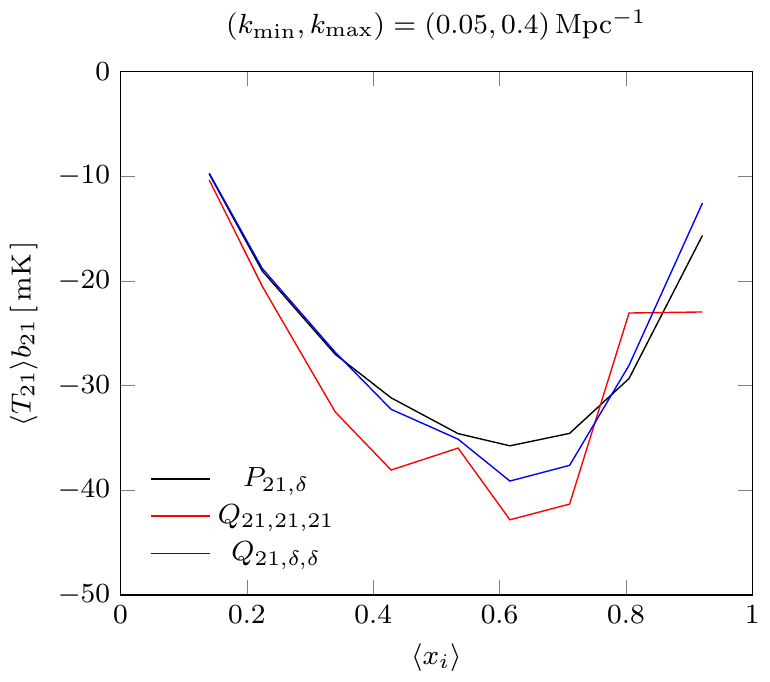}
    \caption{$\langle T_{21} \rangle b_{21}$ as inferred from $Q_{21,21,21}$ and $Q_{21,\delta,\delta}$, and compared against $P_{21,\delta}$.~Source: \cite{Beane_&_Lidz+2018}, \textit{Extracting Bias Using the Cross-bispectrum: An EoR and 21 cm–[C {\scriptsize II}]–[C {\scriptsize II}] Case Study}, ApJ 867 26, Fig.~5, \copyright AAS. Reproduced with permission.}
    \label{fig: cross_bi}
\end{figure}
\noindent
Fig.~\ref{fig: cross_bi} illustrates that the bias parameter inferred from $Q_{21,\delta,\delta}$ is in better agreement with the value from $P_{21,\delta}$. The inference from $Q_{21,21,21}$ is accurate to only 20 per cent. Therefore, $Q_{21,\delta,\delta}$ can be used as a consistency check on the extraction of the bias parameter when inferred from the cross-power spectrum. \cite{Beane_&_Lidz+2018} also analyze the detectability of the cross-power and find that it might be somewhat challenging to detect the [H {\scriptsize I}]$_{\text{21cm}}$-[C {\scriptsize II}]$_{\text{158}\mu\text{m}}$ cross bispectrum, even with 'Stage-II' [C {\scriptsize II}]$_{\text{158}\mu\text{m}}$ experiments~\citep{Silva+2015}. It might, therefore, require next-generation dedicated surveys for the [C {\scriptsize II}]$_{\text{158}\mu\text{m}}$ line. However, the proof-of-concept remains and can be applied to other line combinations such as the [H {\scriptsize I}]$_{\text{21cm}}$-Lyman-$\alpha$. It can even be feasible with SPHEREx \citep{Dore+2018} and HERA-350 \citep{DeBoer+2017}, given that SPHEREx has almost all-sky coverage and match the $\sim 1440 \deg^2$ of HERA-350 coverage.

\subsection{LIM and Galaxy-surveys}

\subsubsection{[C {\scriptsize II}]$_{\text{158}\mu\text{m}}$ and LAEs:}
\begin{figure}[ht]
    \centering
    \includegraphics{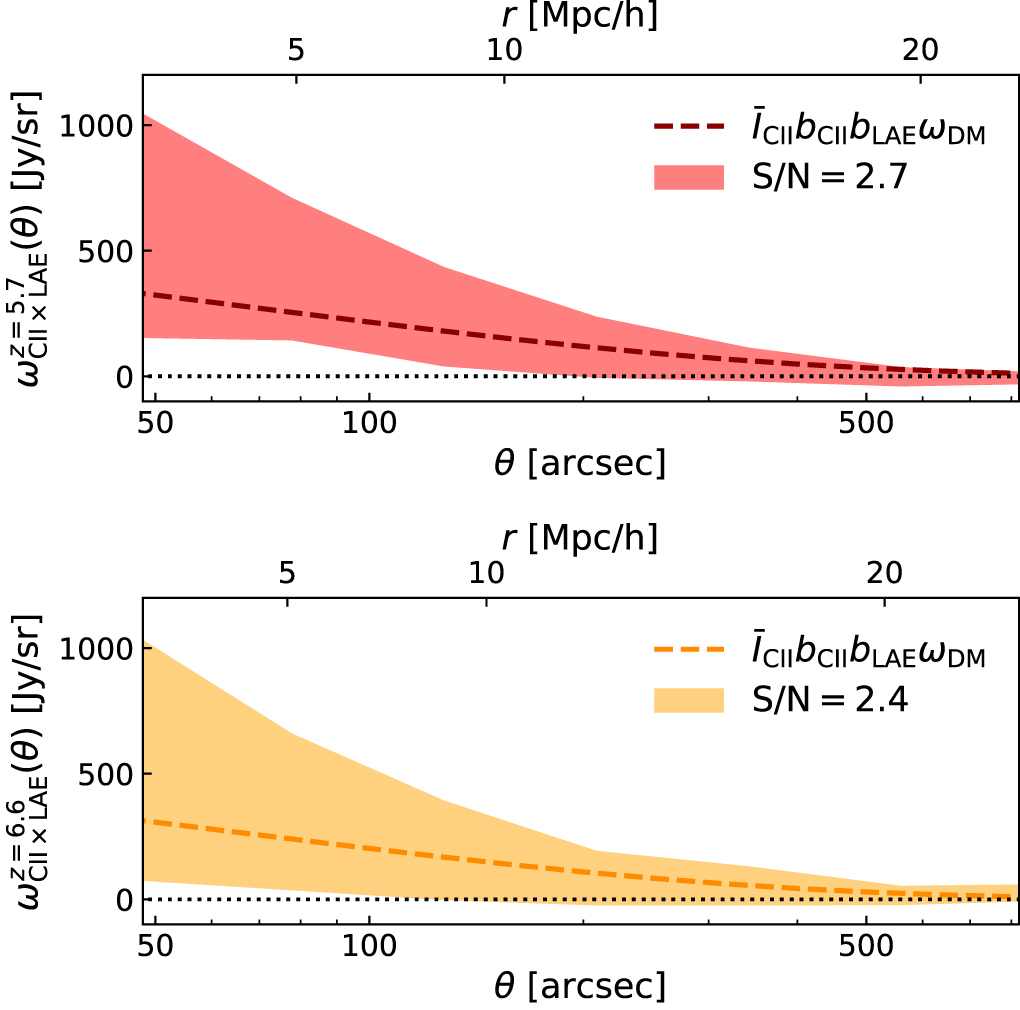}
    \caption{Sensitivity predictions for Subaru HSC for the angular cross-correlation (at $z=6.6\,\text{and}\,5.7$) are shown in this figure \citep{Sun+2021}. Shaded regions represent a 68\% confidence interval, and the dashed lines are the predictions for $\omega_{\text{CII}\times \text{LAE}}$.~Source: \cite{Sun+2021}, \textit{Probing Cosmic Reionization and Molecular Gas Growth with TIME}, ApJ 915 33, Fig.~12, \copyright AAS. Reproduced with permission.}
    \label{fig: cii_lae_cross}
\end{figure}

\cite{Sun+2021} explores the possibility of angular cross-correlation between [C {\scriptsize II}]$_{\text{158}\mu\text{m}}$ and LAEs, and analyze its performance in terms of constraining parameters and SNR. The angular cross-correlation is defined as
\begin{equation}
    \omega_{\text{CII}\times \text{LAE}}(\theta) \equiv \frac{\sum_i^{N(\theta)} (I^i_{\text{CII}(\theta)} -\bar{I}_{\text{CII}})}{N(\theta)} \approx b_{\text{LAE}}b_{\text{CII}}\bar{I}_{\text{CII}}\omega_{\text{DM}}(\theta),
    \label{eq: angular_cross}
\end{equation}
with 
\begin{equation}
    \omega(\theta, z) = \int dz^\prime \mathcal{N}(z^\prime)\int dz^{\prime\prime}\mathcal{N}(z^{\prime\prime})\, \xi(r(\theta, z^{\prime}, z^{\prime\prime}),\, z).
    \label{eq: angular_corr}
\end{equation}
Here, $\xi(r, \theta)$ is the spatial-correlation function. \cite{Sun+2021} demonstrate, from mock simulations, that the quantity $b_{\text{CII}}\bar{I}_{\text{CII}}$ can be constrained by using this cross-correlation. In Fig.~\ref{fig: cii_lae_cross}, the sensitivity of the cross-correlation measurement is shown when data from Subaru HSC and TIME are synergized.

\subsubsection{COMAP and HETDEX survey:}

Joint analysis of LIM voxels with a galaxy survey can output better constraints than that of the LIM survey alone. \cite{Silva+2021} explores the possibility of combining the CO Mapping Array Project (COMAP) data with the Lyman-$\alpha$ emitters (LAE) survey from the Hobby-Eberly Telescope Dark Energy Experiment using mock data from the IllustrisTNG300 galaxy-formation simulation. Table1 summarizes the large-scale structures contributing to the CO lines and Lyman-$\alpha$ emission.

\begin{table}[htb]
\tabularfont
\caption{Classifying the large-scale structures according to the combined CO and LAE signals \citep{Silva+2021}}\label{table1}
\begin{center}
\begin{tabular}{l|cr}
\topline
\textbf{Survey}&\textbf{LAE Bright}&\textbf{LAE Faint}\\
\midline
\textbf{CO Bright}&\makecell{Knots / Galaxy Cluster\\Mixed dust regions}&\makecell{Filaments\\High dust}\\
\hline
\textbf{CO Faint}&\makecell{Filaments\\Low dust}&\makecell{Void\\Low dust}\\
\hline
\end{tabular}
\end{center}
\end{table}
\noindent
Depending on the correlation between these lines, we can use the combined data from COMAP+HETDEX to appropriately to constrain a variety of quantities:
\begin{itemize}
    \item Total CO emission at $z\sim3$
    \item Bright CO voxels at $z=3$
    \item Equivalent width of CO
    \item Upper limit on total CO emission at $z=6$
    \item Estimate of the total CO emission at $z=6$
    \item Voxel intensity distribution (VID)
\end{itemize}

\subsubsection{[H {\scriptsize I}]$_{\text{21cm}}$ and galaxy-surveys:}

Probing the EoR with galaxy surveys also opens the opportunity to do synergy studies with the [H {\scriptsize I}]$_{\text{21cm}}$ survey. One such synergy would be the 21cm-LAE synergy. \cite{Hutter+2019_WhitePaper} explores the possibility of combining WFIRST with the SKA survey with the following ways to do synergy studies:
\begin{itemize}
    \item When reionization is at mid-stage, LAEs will be visible. These will be hosted in ionized regions and anti-correlated to the 21cm signal.
    \item It might constrain the reionization topology by measuring brightness temperature differences of regions with and without LAEs. An inside-out topology will have higher temperatures in under-dense parts of the IGM.
    \item We can also use the difference in brightness temperatures between regions with and without LAEs to infer the ionization state of the IGM as well.
\end{itemize}
A similar idea of synergy is also explored by \cite{Zackrisson+2020}. Since SKA-1 can locate ionized bubbles of size $\gtrsim 1000\, \text{cMpc}^3$ with sufficient resolution, there are high probabilities of detecting galaxies within these regions for the reionization process dominated by galaxies. We can accomplish this by observations with JWST, WFIRST, ELT etc. Towards the end of reionization, spectroscopy may reveal bubble galaxies for the smallest-detectable ionized region. A photometric survey might be done for higher redshifts $z=10$. In turn, these studies can constrain quantities, such as the minimum total stellar mass required to produce that ionized bubble and the photon number-weighted mean escape-fraction $\big\langle f_\text{esc} \big\rangle$.

\subsection{Other synergies}

Here, we discuss a unique synergy that explores the prospects of combining data from CMB, quasar spectra and the global 21cm signal. \cite{Chatterjee+2021} has developed a Markov Chain Monte Carlo (MCMC) based parameter-estimator, \texttt{CosmoReionMC}, that can constrain parameters of interest using these input data. We see that adding data from quasar spectra to the CMB-only data leads to tighter constraints on cosmological parameters \citep[][Fig.~1]{Chatterjee+2021}.

\texttt{CosmoReionMC} is based, on physically-motivated semi-analytical models of reionization \citep{Mitra+2011} and global [H {\scriptsize I}]$_{\text{21cm}}$ signal, including appropriate modifications to the CAMB code. It is also expected that the package can be extended to:
\begin{itemize}
    \item Input other kinds of data (e.g. BAO) and constrain other astrophysical parameters
    \item Deal with non-standard extensions to the lambda-CDM cosmological model
\end{itemize}

\section{Probing the signals from the EoR:\newline Instrumentations and Experiments}

Developments in the instrumentations to detect these signals are crucial in achieving our goal. Several ongoing and upcoming experiments will go hand-in-hand to uncover this mysterious epoch. In the field of detecting [H {\scriptsize I}]$_{\text{21cm}}$ line, we have observatories making good progress, such as the GMRT, LOFAR and many more. Other planned observatories like HERA and SKA are underway. Experiments for other line emissions include TIME~(CO and [C {\scriptsize II}]$_{\text{158}\mu\text{m}}$) and COMAP~(CO). In this section, we provide recent developments and updates on these experiments. We discuss what progress have been made in observing these cosmic signals, such as upper limits on the intensity power spectrum.

\subsection{Global [H {\scriptsize I}]$_{\text{21cm}}$ signal detection}

The global signal represents the brightness intensity averaged over all sky directions. From eq.~\ref{eq: brightness_temp}, we can write this as,
\begin{equation}
\begin{split}
    \big\langle T_b(z) \big\rangle = 4\text{mK}\,\big\langle x_{\text{HI}}(z) \big\rangle \Bigg( \frac{\Omega_{\rm b} h^2}{0.02} \Bigg) \Bigg( \frac{0.7}{h} \Bigg) \sqrt{\frac{1+z}{\Omega_{\rm m}}}\\ \times\, \Bigg( 1 - \frac{T_{\gamma}(z)}{T_s(z)} \Bigg).
\end{split}
\end{equation}
Here, we have neglected the $\partial v/\partial r$ term. Therefore, it is essentially a probe of how the average signal has evolved over a certain period of cosmic time. The first significant observation of this was made by the Experiment to Detect the Global EoR Signature \citep[EDGES, ][]{Bowman+2018}, which reported an absorption trough, which is unusually deep. Recently, we have had independent detections from the SARAS, and below, we discuss its updates on the signal detections and compare them to the EDGES result.

\subsubsection{Shaped Antenna measurement of the background RAdio Spectrum 3 (SARAS 3):}

SARAS 3 is an Indian effort to measure the sky averaged [H {\scriptsize I}]$_{\text{21cm}}$ signal as a function of redshift during the CD and EoR \citep{2021arXiv210401756N}. The single antenna based  spectral radiometer operates between  40-230 MHz frequency band. SARAS 3 is the third radiometer in the SARAS series. This version is based on a monocone antenna floating on lakes in  Southern India. The experiment uses an improved receiver, calibration technique and better systematic mitigation methods. 

The first radiometer in this series, SARAS 1 was a fat-dipole correlation spectrometer that was used to improve the absolute calibration of the previous 150-MHz all-sky map \citep{2013ExA....36..319P, 2015ApJ...801..138P}. The measurements using an electrically short spherical monopole antenna in SARAS 2 experiment, with a half-power beam width of $45^{\circ}$, started to rule out a set of CD and EoR scenarios that featured rapid reionization and weak X-ray heating~\citep{Singh+2018}. Recently, \citet{2021arXiv211206778S} analyze SARAS 3 observation data  in the 55--85 MHz band and rule out ( with $95.3\%$ confidence) the previously claimed strong [H {\scriptsize I}]$_{\text{21cm}}$ signal around redshift $\sim 17$ by \citet{Bowman+2018}. The non-detection in \citet{2021arXiv211206778S}  thus do not confirm the non-standard processes such as excess cooling \citep[see e.g.,][]{2018Natur.555...71B, 2018Natur.557..684M, PhysRevLett.121.011102} or excess radio background \citep[see e.g.,][]{2018ApJ...858L..17F, 2018ApJ...868...63E, 2021arXiv210813593G} at redshifts beyond $\sim 17$.

\subsection{LIM surveys with [H {\scriptsize I}]$_{\text{21cm}}$ and other line emissions}

The LIM experiments using tracers like CO and [C {\scriptsize II}]$_{\text{158}\mu\text{m}}$ will probe intensity fluctuations, $\delta I(\mathbf{x})\,= I(\mathbf{x}) - \overline{I}(\mathbf{x})$. Here, $I(\mathbf{x})$ is the intensity of the LIM signal probed directly across the sky, and $\overline{I}(\mathbf{x})$ is the average intensity. The [H {\scriptsize I}]$_{\text{21cm}}$ interferometers will measure the Fourier space fluctuations, $\Delta \Tilde{T}_b(\mathbf{k}, z) = \int d^3x ~\exp(-i\mathbf{k.x}) \Delta T_b(\mathbf{x}, z)$, of the [H {\scriptsize I}]$_{\text{21cm}}$ signal intensity, with $\Delta T_b(\mathbf{x},z) = T_b(\mathbf{x},z) - \langle T_b(z) \rangle$. We discuss below the LIM surveys that will study the skies by measuring the signal fluctuations.

\subsubsection{Giant Metrewave Radio Telescope (GMRT):}
The GMRT radio interferometer in western India aims to observe redshifted [H {\scriptsize I}]$_{\text{21cm}}$ signal within $50 - 1420$ MHz frequency range. The telescope consists of 30 dishes, each 45m in diameter, distributed in a Y-shaped array, over 25Km. The 325 and 610 MHz receivers are often used for the post-EoR studies, while the 150 MHz receiver is used for EoR studies. It has a FoV of $\sim 3.8^{\circ}$ at 150 MHz.

\citet{Bharadwaj+2001, Bharadwaj_&_Ali+2005, Ali+2008} started studying the angular modes of the [H {\scriptsize I}]$_{\text{21cm}}$ signal using GMRT observations and visibility correlations; later it was extended to MAPS \citep[][Section~\ref{sec:maps}]{Datta+2007}. Using a foreground removal technique that is based on a fourth-order polynomial and MAPS formalism, \citet{Ghosh+2011a} first attempted to measure the [H {\scriptsize I}]$_{\text{21cm}}$ fluctuations with GMRT observations (at $\nu_{\text{obs}}=601$ MHz and $z=1.32$). A further improvement of the foreground removal step was reported in \citet{Ghosh+2011b}, which reduced sidelobes and tapered the primary beam. \citet{Ghosh+2012} did the same analysis later for $z=8.1$ using GMRT observation at $\nu_{\text{obs}}=150$ MHz.

Later, \citet{Choudhuri+2014, Choudhuri+2016, Bharadwaj+2019} introduced and studied the optimized power spectrum estimators, based on visibility correlations, such as the Tapered Gridded Estimator (TGE) and Bare Estimator.

The first upper limit on the [H {\scriptsize I}]$_{\text{21cm}}$ power spectrum at $z=8.6$ was, however, estimated by \citet{Paciga2011MNRAS.413.1174P} using 50 hours of GMRT observation which reported a $2\sigma$ value of $ (70 ~\rm mK)^2$ at $k=0.65\,h~\rm Mpc^{-1}$. \citet{Paciga+2013} later accounted for the signal loss due to previously used piecewise-linear foreground subtraction method and updated the upper limit to 2$\sigma$ value of $(248)^2~\rm mK^2$ at $k\sim 0.5\,h~\rm Mpc^{-1}$. Recently, \citet{Chakraborty+2021} employ the foreground avoidance technique and estimate upper limits on the [H {\scriptsize I}]$_{\text{21cm}}$ power spectrum at $k\sim 1.0\,h~\rm Mpc^{-1}$ as $(58.87 ~\rm mK)^2, ~(61.49 ~\rm mK)^2, ~(60.89 ~\rm mK)^2$, and $(105.85 ~\rm mK)^2$ for $z = 1.96, 2.19, 2.62, \text{and}\, 3.58$ respectively.

\subsubsection{Low Frequency Array (LOFAR):}
LOFAR radio interferometer consists of 38 stations located in the Netherlands among which 24 are core stations located within 3 km and others up to $\sim 100$ km. LOFAR also has 14 international stations, but those are not used for EoR studies. The stations contain two types of receptors: (1) the High-Band Antennas (HBA), 120–240 MHz; (2) Low-Band Antennas (LBA), 30–90 MHz. Each station has a size of $\sim 30$ m of diameter and a FoV of $3^{\circ}$  at 150 MHz.

The LOFAR EoR Key Science Project team aims to model and subtract the foreground contaminants from the 21-cm observation. The team focussed on developing data analysis Methodology \citep[e.g.,][]{2011MNRAS.414.1656K, 2015MNRAS.449.4506Y},  foreground mitigation techniques \citep[e.g.,][]{2009MNRAS.397.1138H, 2013MNRAS.429..165C, Mertens+2018, 2021MNRAS.500.2264H}, investigate systematic effects \citep[e.g.,][]{2022MNRAS.509.3693M}, etc. 

The first LOFAR results on EoR H{\scriptsize I} fluctuations was published by \citet{2017ApJ...838...65P} which place upper limits on the [H {\scriptsize I}]$_{\text{21cm}}$ power spectrum at redshifts between 9.6-10.6 using 13 hours of LOFAR-HBA observations. Later, \citet{2019MNRAS.488.4271G} reported upper limits on the [H {\scriptsize I}]$_{\text{21cm}}$ power spectrum between redshift 20-25 using 14 hours of LOFAR-LBA observations. The best upper limit from LOFAR recently appear in \citet{2020MNRAS.493.1662M} which sets a 2$\sigma$ upper limit of $\Delta^2(k=0.075 ~h ~\rm Mpc^{-1})=(73)^2 \rm ~mK^2$ on the [H {\scriptsize I}]$_{\text{21cm}}$ power spectrum at $z\approx 9.1$ using 141 hours of LOFAR-HBA observation. While previous upper limits were unable to rule out standard EoR and CD scenarios, the upper limits as obtained in \citet{2020MNRAS.493.1662M} started ruling out extreme reionization scenarios and constraining the properties of the sources as well as the IGM at redshift 9.1 \citep{2020MNRAS.493.4728G, 2020MNRAS.498.4178M, 2020arXiv200802639G}.

\subsubsection{Murchison Widefield Array (MWA):}
The MWA array in Western Australia operates between 80-200 MHz and aims to measure [H {\scriptsize I}]$_{\text{21cm}}$ fluctuations. The interferometer consists of 256 stations among which 128 with long baselines belongs to  Extended Array and the rest 128 which includes two 36 stations redundant subarrays belongs to Compact Array with short baselines. The compact array is particularly used for EoR [H {\scriptsize I}]$_{\text{21cm}}$ observation.  Each station is $\sim  6$ times smaller than a LOFAR station. The smaller size of the station makes the FoV as large as 15-50 degrees at 200 MHz.  

The primary method used in MWA data analysis used a delay transform to produce a 2D power spectrum of the signal in $K_{\bot}, k_{\parallel}$ space. After calibration, the analysis uses the `EoR Window’ part which is expected to be without external contamination coming from inaccurate calibration, leakage of foregrounds, etc. Initial efforts of the team were on developing data calibration and source subtraction methodologies such as Real-Time System \citep{RTC4703504}, Fast Holographic Deconvolution \citep{2012ApJ...759...17S} and power spectra estimator pipeline such as CHIPS \citep{2016ApJ...818..139T} and $\varepsilon$ppsilon \citep{2019PASA...36...26B} for MWA EoR data analysis. 

\citet{2019ApJ...884....1B, 2019ApJ...887..141L} produced the initial upper limit results on the EoR [H {\scriptsize I}]$_{\text{21cm}}$ signal. \citet{2016MNRAS.460.4320E} published the first upper limits on the Cosmic Dawn H{\scriptsize I} signal at a higher redshift range 12–18.  Recently,  \citet{2020MNRAS.493.4711T} reported the so far best MWA  2$\sigma$ upper limit at redshift 6.5 of $\Delta_{21\rm cm}^2(k = 0.14 ~h ~\rm Mpc^{-1})\approx (43 ~\rm mK)^2$ using 110 hours of MWA high band observation on the EoR0 field. While the study considered 6 different redshifts between the redshift range 6.5 -- 8.8, the upper limits become weaker at the high redshift end. Further studies such as \citet{2021MNRAS.508.5954R} aimed to improve the upper limits by  understanding and mitigating the systematics arising from instrumental,  observational and analysis effects. However, as the data analysed in \citet{2021MNRAS.508.5954R} is only 14 hours, the achieved 2$\sigma$ upper limit at redshift 6.5 of $\Delta_{21\rm cm}^2(k = 0.13 ~h ~\rm Mpc^{-1})\approx (73.78 ~\rm mK)^2$ is still larger than the upper limits of  \citet{2020MNRAS.493.4711T}.

Considering the new MWA upper limits on the [H {\scriptsize I}]$_{\text{21cm}}$ signal power spectrum as reported in  \citet{2020MNRAS.493.4711T}, studies such as  \citet{2020arXiv200802639G, 2021MNRAS.503.4551G}  explored EoR scenarios that are disfavoured by the limits. These studies ruled out completely neutral and cold IGM at redshift 6.5 and suggest that the IGM must have undergone X-ray heating by that time.

\subsubsection{The Hydrogen Epoch of Reionization Array (HERA):}
HERA radio interferometer aims to observe  H{\scriptsize I} signal fluctuations from EoR and CD. This is currently under construction array located in the Karoo desert of South Africa. Phase I of the interferometer has $\sim 350$  fixed, zenith pointing dishes packed hexagonally within $\sim 300$m area where each is 14 meters in diameter. HERA Phase I uses the feeds and correlator from the previous H{\scriptsize I} experiment PAPER. The main goal of HERA phase I is to measure the [H {\scriptsize I}]$_{\text{21cm}}$ power spectrum in the redshift range 6 –12 with high significance. The new feeds for Phase II are under testing and will enable [H {\scriptsize I}]$_{\text{21cm}}$ observations in the high-redshift range 12–35. Hera-350 has a FoV of $9^{\circ}$.

Initial works of the HERA team include developing pipelines for calibration, power spectrum estimation and understanding systematic and error propagation, etc \citep{2020MNRAS.499.5840D, 2020ApJ...890..122K, 2020ApJ...888...70K}. HERA data analysis primarily aim to control the spectral systematics to keep the EoR window largely free from contamination. Recently,  \citet{2022ApJ...925..221A} analysed $\sim 36$ hours of observation with roughly 50 HERA antennas and published upper limits on the [H {\scriptsize I}]$_{\text{21cm}}$ power spectrum at redshifts 7.9 and 10.4. This HERA Phase I observation achieved the so far strongest $2\sigma$ limits of $(30.76 ~\rm mK)^2$ for $k$-scale of $ 0.192 ~h ~\rm Mpc^{-1}$  at $z = 7.9$ and $(95.74 ~\rm mK)^2$ for  $ 0.256 ~h ~\rm Mpc^{-1}$ at $z = 10.4$ . The interpretation of these recent results was done in \citet{2022ApJ...924...51A} which shows that the IGM temperature must be larger than the adiabatic cooling threshold by redshift 8.  The study also constrains the soft band X-ray luminosities per star formation rate to [$10^{40.2}, 10^{41.9}$] erg/s/($\MSUN$/yr) ($1\sigma$ level).

\subsubsection{The Square Kilometre Array (SKA):}
The planned SKA radio interferometer consists of two different types of arrays. (1) SKA-mid: the array is planned to build in the Karoo desert of South Africa, it covers 350 MHz to 15.3 GHz frequency range; (2) SKA-low:  to be built in Western Australia, this low-frequency array of 512 stations in phase I will cover 50 - 350 MHz frequency range ($3<z<27.4$) and thus relevant for probing [H {\scriptsize I}]$_{\text{21cm}}$ signal from the CD and EoR. Among these SKA-low stations, 212 stations will be built inside a compact core of about 600 m, while the remaining 300 stations will be placed on three `spiral’ arms outward up to about 65 km from the central core. In terms of sensitivity, the SKA-low is expected to achieve $\sim 10$ times higher sensitivity than the LOFAR.

\begin{figure}
    \centering
    \includegraphics[width=\columnwidth]{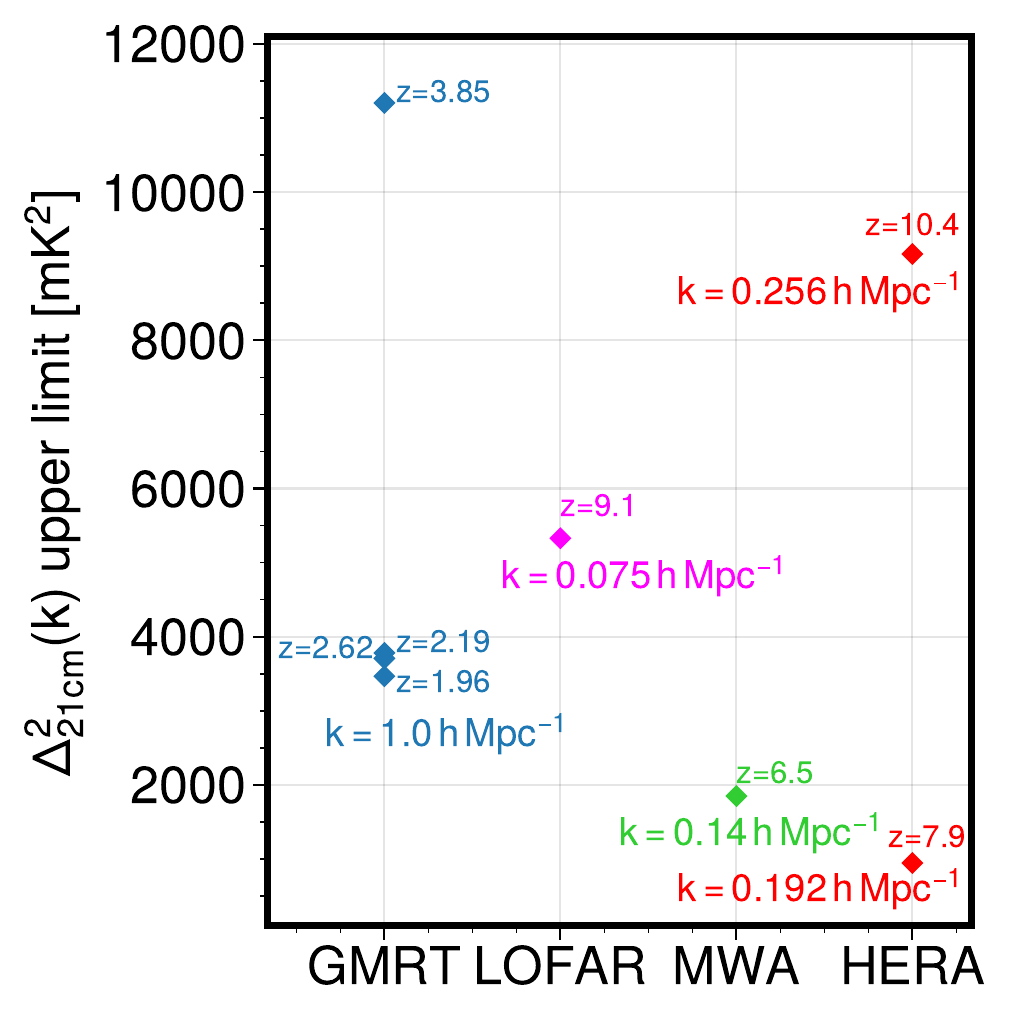}
    \caption{The various upper limits on the [H {\scriptsize I}]$_{\text{21cm}}$ power spectrum ($\Delta^2_{21cm}(k)$) as estimated by the instruments, are shown here. Blue, magenta, green, and red points correspond to the upper limits from GMRT, LOFAR, MWA and HERA, respectively.}
    \label{fig:upper_limit}
\end{figure}

Besides statistical measures of the [H {\scriptsize I}]$_{\text{21cm}}$ signal using quantities such as power spectrum, SKA-low’s sensitivity will allow making tomographic images \citep[e.g.,][]{ghara16}. While SKA-low is still in the construction state, several theoretical studies have started developing methodologies to extract information about the EoR as well as the CD from the SKA tomographic images. These methods include use of Minkowski functionals \citep[e.g.,][]{2021JCAP...05..026K}, Euler characteristic \citep[see e.g.,][]{2021MNRAS.505.1863G}, Bubble size distributions \citep[][]{2018MNRAS.479.5596G, 2020MNRAS.496..739G}, Fractal dimensions \citep[e.g.,][]{2017MNRAS.466.2302B}, Individual 2D maps of the [H {\scriptsize I}]$_{\text{21cm}}$ signal using convolutional neural network \citep[e.g.,][]{2019MNRAS.484..282G}, etc.  All these studies statistically characterise the features of the tomographic images. In Table~\ref{table2}, we summarize all the instrument parameters discussed here. Further, the various upper limits provided by the instruments so far are summarized in Fig.~\ref{fig:upper_limit}.
\begin{table*}[htb]
\tabularfont
\caption{Instrument parameters for GMRT, LOFAR, MWA, HERA, and SKA are summarized here with references.}\label{table2}
\begin{center}
\begin{tabular}{lcccr}
\topline
\textbf{Instrument} & \textbf{Dish size} & \textbf{FoV} & \textbf{Frequency (MHz)} & \makecell{\textbf{References}}\\
\midline
GMRT & 45m & 3.8 deg (at 150 MHz) & 50\text{--}1420 & \makecell{\cite{Mercier+2006},\\\cite{Ghosh+2012}}\\
LOFAR & $\sim 30$m & 3 deg (at 150 MHz)& 30\text{--}240 & \makecell{\cite{Fackle+2007},\\\cite{Haarlem+2013}}\\
MWA & $\sim 5$m & 15\text{--}50 deg (at 200 MHz) & 80\text{--}300 & \makecell{\cite{Lonsdale+2009},\\\cite{tingay+2012}}\\
HERA & 14m & 9 deg (at 150 MHz) & 50\text{--}250 & \makecell{\cite{DeBoer+2017}}\\
SKA1-Low & 35-40m (each station) & $\sim 5.45$ deg (at 110 MHz) & 50-350 & \makecell{\cite{braun+2019}}\\
SKA1-Mid & 15m & $\sim 1.8$ deg (at 770 MHz) & 350\text{--}15300 & \makecell{\cite{braun+2019}}\\
\hline
\end{tabular}
\end{center}
\end{table*}

\subsubsection{The CO Mapping Array Project (COMAP):}
The COMAP Pathfinder with a single 19-pixel spectrometer receiver mounted on a 10.4meter dish located at the Owens Valley Radio Observatory started observation in 2019.  It primarily operates in the frequency range 26-34 GHz while the receiver is sensitive to CO(1-0) line emission in the redshift range 2.4-3.4 (post-reionization) and  CO(2-1) emission in the redshift range  5.8-7.8 which corresponds to the last stage of the EoR.

A 5-year long survey of $\sim 12$ square degrees of the sky using this pathfinder is currently ongoing. This aims to detect the CO(1-0) signal from redshift $\sim 3$. This is also used  for validating the already developed technologies, developing new methodologies, understanding systematics and making new observational strategies \citep[][]{2021arXiv211105931C, 2021arXiv211105933B, 2021arXiv211105930I}. While the COMAP team forecast a detection of the CO power spectrum after 5 years of the survey with a signal-to-noise ratio of  9–17, \citet{2021arXiv211105927C} already estimated the first $2\sigma$ upper limits on the clustering component of CO(1–0) power spectrum using the first 13 months of observation. The reported upper limit is $P_{\rm CO}(k) = -2.7\pm 1.7 \times 10^4 ~\mu \rm K^2 ~\rm Mpc^3$ on $k$-scale of   $0.051-0.62 ~\rm Mpc^{-1}$. Future phases will tackle more challenging CO(2-1) line emission from redshift $\sim 6$.

\subsubsection{The Tomographic Ionized-carbon Mapping Experiment (TIME):}
TIME is an imaging spectrometer array with wide-bandwidth aiming for tomographic measurement of redshifted  [C {\scriptsize II}]$_{\text{158}\mu\text{m}}$ line intensity fluctuations from redshifts $6 \lesssim z \lesssim 9$ \citep{2018JLTP..193..893H, 2020ApJ...901..142C}. It also simultaneously detects the rotational CO line emission from galaxies at redshifts $z=0.5-2$. This instrument started operating recently in 2021. Initially, 1000 hrs of observing time is planned for TIME operating at the ALMA 12m Prototype Antenna at the Arizona Radio Observatory in Kitt Peak, Arizona. While the instrument is collecting data, \citet{Sun+2021} forecast a measurement of the [C {\scriptsize II}]$_{\text{158}\mu\text{m}}$ auto power spectrum during the EoR with an SNR $>5$.

\subsection{Galaxy surveys}

On the other hand, galaxy surveys are designed to detect individual galaxies using photometry or spectroscopy. These employ high-resolution detections as opposed to LIM surveys. In the process, they give up the capability to survey large volumes within sustainable observational times. Nonetheless, they can provide the most detailed information about the early galaxies responsible for reionizing the IGM. Below, we discuss the recent updates from the galaxy surveys based on the ALMA observatory.

ALMA has observed a variety of fields up until now to infer galaxy properties. One of the surveys conducted with ALMA is the ALMA Spectroscopic Survey (ASPECS) in the Hubble Deep Field. As reported in \cite{Decarli+2020}, this survey used the 1.2 and 3 mm, bands and consisted of two spatially overlapping mosaics. The CO(2-1) and CO(3-2) lines dominate the flux by 80 per cent in the 3 mm band at $z=1\text{--}3$. Also, at the 1.2 mm band, more than 50 per cent flux is observed from line emissions originating from intermediate CO transitions ($J_{\rm up}=3\text{--}6$) and 12 per cent from neutral carbon, while less than 1 per cent from [C {\scriptsize II}]$_{\text{158}\mu\text{m}}$. Therefore, it suggests that upcoming [C {\scriptsize II}]$_{\text{158}\mu\text{m}}$ experiments will face significant foreground challenges. Other results include the evolution of {CO} luminosity function, probed at 1.2 mm, and following the evolution of $\rho_{\rm H_2}$, from early cosmic times to $z=2\text{--}3$, which suggests that it is in qualitative agreement with the cosmic star-formation rate density. Also, the estimates of $\rho_{\rm H_2}$ at $z \gtrsim 0.5$ are not dominated by cosmic variance.

The other observed fields are the COSMOS and ECDFS, done by the ALMA Large Program to INvestigate C{\scriptsize II} at Early Times (ALPINE) survey. Literature reported predictions for different quantities, including line-luminosity functions (LF) and star-formation history. This survey detected as many as 118 sources, with various characteristics and upper limits. \cite{Yan+2020} find that 75 of the sources were significant [C {\scriptsize II}]$_{\text{158}\mu\text{m}}$ detections. The [C {\scriptsize II}]$_{\text{158}\mu\text{m}}$ LF were consistent with low redshift ($z \sim 0$) LF at $10^{8.25}\text{--}10^{9.75}\,L_{\odot}$. Combining estimates from other sources, they find that available model predictions underestimate the number density for [C {\scriptsize II}]$_{\text{158}\mu\text{m}}$ emitters at $z \sim 4\text{--}6$. Additionally, they set the constraint on $\rho_{\rm H_2}$ at $(2\text{--}7) \times 10^7\,\rm M_{\odot}\,Mpc^{-3}$ at $z \sim 4\text{--}6$, consistent broadly with existing studies. \cite{Loiacono+2021} further point out there could be an evolution of the [C {\scriptsize II}]$_{\text{158}\mu\text{m}}$ LF between $z \sim 0\text{--}5$. 
Also, the ALPINE survey is the first to probe the faint end of the infrared LF, suggesting little evolution between $z \approx 2.5\text{--}6$ \citep{Gruppioni+2020}. This survey has also helped constrain the star-formation rate density (SFRD) over the relevant redshift ranges. \cite{Gruppioni+2020} point out that SFRD derived from the infrared LF are significantly higher than that estimated from far-ultraviolet (FUV) observations, suggesting that the dust-obscured star formation plays a vital role at high redshifts ($z \approx 2\text{--}6$). At $z \sim 5.5$ dust-obscured fraction of SFRD is around 61 per cent of the total \citep{Khusanova+2021}.

\section{Discussion}

We summarize some of the latest developments in LIM surveys probing the EoR and their synergy opportunities. In particular, we try to highlight the Indian contribution, which comprises modelling the signal and its statistics (power spectrum, bispectrum) through analytic and numerical methods and observational analyses to the field of LIM. Given the active participation of many Indian researchers in developments related to the SKA (SKA-India Consortium), the Indian community has a vast contribution to the [H {\scriptsize I}]$_{\text{21cm}}$ field. Other than this, India is also a partner of the Thirty Metre Telescope (TMT) project. Below, we highlight what further avenues can be addressed and improved.

\subsection{Future scopes}

There are some avenues in which more effort can be invested by the community. Studies that require simulations are addressed with N-body simulations most of the time. Although given the limitations of computational resources and that we generally need predictions for large-scale statistics, N-body simulations might not suffice alone in the coming years. Similarly, for reionization, semi-numerical methods generally show good agreement with radiative-transfer approaches, only when compared in terms of 2-point statistics. 

Most of the signal statistics that have been worked on include the power spectrum, which is relatively easy to estimate and interpret. However, the bispectrum proves to be a more robust statistic for the [H {\scriptsize I}]$_{\text{21cm}}$ signal, given that we are indeed dealing with a highly non-Gaussian signal. Although its estimation and interpretation are not as straightforward as the power spectrum, we are likely to see more analysis on this. For accurate predictions on bispectrum, more sophisticated approaches like hydrodynamic and radiative-transfer simulations are more suitable. The caveats of this, like limited simulation volumes, can be addressed by more accelerating developments in ML algorithms and emulators. Although radiative-transfer codes have been developed within the community, a hydrodynamic simulator is still lacking. It is usually addressed by hydro-simulations already present outside the community. However, one can actively work on this avenue to develop independent codes that can be better understood and trusted within the community. Also, it provides a platform for independent cross-checks on the other simulations available. As mentioned earlier, the role of ML algorithms will be crucial here; after getting trained from hydro-simulation and radiative transfer outputs, it is more likely to predict reliable statistics (especially bispectrum) and infer accurate astrophysics from data. Also, it will find further engaging applications in the domain of image analysis techniques.

The community lacks sufficient participation in other LIM surveys except for the [H {\scriptsize I}]$_{\text{21cm}}$ and Lyman-$\alpha$ surveys. It has happened in both the observational analysis and modelling of the signal and its statistics. In the coming years, we might see more engagement in the fields to probe the EoR with line emissions like [C {\scriptsize II}]$_{158\mu\text{m}}$, CO, [O{\scriptsize III}]$_{88\mu\text{m}}$ and others. A multi-tracer approach to the problem provides necessary consistency checks, and more synergy opportunities, with a better understanding of astrophysics in terms of completeness.

\subsection{Next-Generation experiments}

We have had discussed the line emissions and the sciences that is made possible by a variety of experiments. The limits of current technology are being pushed with observatories like ALMA, JWST, and many more. The upcoming SKA will be a milestone in the history of radio astronomy in terms of observational capabilities. The diverse varieties of synergies with these experiments make it possible to address science goals that are not achievable or difficult to achieve in other ways. However, given that we have progressed enormously in observational capabilities, some science goals are beyond the current reach yet important. Below, we give a few examples of such and discuss what kind of future surveys we might require to address these science goals.

During reionization, the free electrons, liberated from hydrogen atoms that are being ionized, are expected to interact with both the CMB photons and the 21cm photons given off by the neutral hydrogen. The consequence of such interaction is to change the polarization of these radiations. To understand reionization, exploring this polarization scenario can be crucial. \cite{Ji+2021} has proposed doing cross-correlation of polarization in the context of testing two null hypotheses:
\begin{itemize}
    \item How well one can rule out that there is no reionization process?
    \item Given that reionization had indeed happened, how well we can rule out that there is no [H {\scriptsize I}]$_{\text{21cm}}$ polarization signal?
\end{itemize}
They find from their analysis that for a fiducial survey with $T_{\rm sys} = 40\rm\,K$, $\theta_{\rm B}(z=6.1) = 20\,\rm arcmin$ and $t_{\rm obs} = 2\,\rm yr$, the SNR would be 2.1 for the first hypothesis when cross-correlated with CMB, surveyed with thermal noise of $\Delta_T = 1\mu\rm K$-arcmin. However, for the second hypothesis, the SNR has an unacceptable value of 0.017. Therefore, although it is hard to reach a desired level of SNR, next-generation dedicated surveys can be conceived that can address these issues.

As discussed earlier, the [H {\scriptsize I}]$_{\text{21cm}}$-[C {\scriptsize II}]$_{\text{158}\mu\text{m}}$ cross bispectrum is tough to detect because a desirable SNR requires a demanding limit on the noise power spectrum for the [C {\scriptsize II}]$_{\text{158}\mu\text{m}}$ experiment. As explored in \cite{Beane_&_Lidz+2018}, the limit on the noise power spectrum for the [C {\scriptsize II}]$_{\text{158}\mu\text{m}}$ experiment needed, for a joint survey area of $50\,\rm deg^2$, would be $N_{\rm CII} \lesssim 1.6\times10^8\,\rm (Jy/sr)^2\,Mpc^3$. For a 'Stage II' experiment reported in \cite{Silva+2015}, it is $N_{\rm CII} \sim 2.5\times10^9\,\rm (Jy/sr)^2\,Mpc^3$; in future, this might be addressed by progress in detector technology. For a futuristic survey of $1000\,\rm deg^2$, the limit on noise power spectrum required would be $3.4\times10^9, 1.1\times10^9~\text{and}~1.4\times10^9\,\rm (Jy/sr)^2 \,Mpc^3$, at $z = 6.43, 7.37~\text{and}~9.41$, respectively. This would indeed help in characterizing the behaviour of the [H {\scriptsize I}]$_{\text{21cm}}$-[C {\scriptsize II}]$_{\text{158}\mu\text{m}}$ cross-bispectrum.

Similarly, we find from \cite{Sun+2021} that the performance of TIME can be improved with advancement in the detector technology, with more spectrometers. This next-generation TIME (TIME-NG) will have three times less noise-equivalent intensity (NEI) than the current specifications of TIME ($5\,\rm MJy\,sr^{-1}\,s^{1/2}$) and an order of magnitude improvement in survey power compared to TIME-EXT. More cross-correlations will thus become possible with significant SNR, like LAEs and Lyman-break galaxies (LGBs) from Nancy Grace Roman and Euclid telescopes and [H {\scriptsize I}]$_{\text{21cm}}$ from HERA and SKA.




\section*{Acknowledgements}

CSM is funded by the Council of Scientific and Industrial Research (CSIR), via a CSIR-SRF grant, File No. 09/1022(0080)/2019-EMR-I. RG furthermore acknowledge support by the Israel Science Foundation (grant no. 255/18). SM acknowledges financial support through the project titled ``Observing the Cosmic Dawn in Multicolour using Next Generation Telescopes'' funded by the Science and Engineering Research Board (SERB), Department of Science and Technology, Government of India through the Core Research Grant No. CRG/2021/004025. The authors acknowledge the use of NASA Astrophysics Data System Bibliographic Services (NASA ADS) and arXiv\footnote{\url{https://arxiv.org}} research-sharing platform.
\vspace{-1em}


\newpage
\bibliography{Refs}

\begin{thebibliography}{}
\expandafter\ifx\csname natexlab\endcsname\relax\def\natexlab#1{#1}\fi

\bibitem[{{Abdurashidova} {$et~al$.}(2022{\natexlab{a}}){Abdurashidova},
  {Aguirre}, {Alexander}, {Ali}, {Balfour}, {Beardsley}, {Bernardi},
  {Billings}, {Bowman}, {Bradley}, {Bull}, {Burba}, {Carey}, {Carilli},
  {Cheng}, {DeBoer}, {Dexter}, {de Lera Acedo}, {Dibblee-Barkman}, {Dillon},
  {Ely}, {Ewall-Wice}, {Fagnoni}, {Fritz}, {Furlanetto}, {Gale-Sides},
  {Glendenning}, {Gorthi}, {Greig}, {Grobbelaar}, {Halday}, {Hazelton},
  {Hewitt}, {Hickish}, {Jacobs}, {Julius}, {Kern}, {Kerrigan}, {Kittiwisit},
  {Kohn}, {Kolopanis}, {Lanman}, {La Plante}, {Lekalake}, {Lewis}, {Liu},
  {MacMahon}, {Malan}, {Malgas}, {Maree}, {Martinot}, {Matsetela}, {Mesinger},
  {Molewa}, {Morales}, {Mosiane}, {Murray}, {Neben}, {Nikolic}, {Nunhokee},
  {Parsons}, {Patra}, {Pascua}, {Pieterse}, {Pober}, {Razavi-Ghods},
  {Ringuette}, {Robnett}, {Rosie}, {Sims}, {Singh}, {Smith}, {Syce},
  {Thyagarajan}, {Williams}, {Zheng}, \& {HERA
  Collaboration}}]{2022ApJ...925..221A}
{Abdurashidova}, Z., {Aguirre}, J.~E., {Alexander}, P., {$et~al$.}
  2022{\natexlab{a}}, \apj, 925, 221

\bibitem[{{Abdurashidova} {$et~al$.}(2022{\natexlab{b}}){Abdurashidova},
  {Aguirre}, {Alexander}, {Ali}, {Balfour}, {Barkana}, {Beardsley}, {Bernardi},
  {Billings}, {Bowman}, {Bradley}, {Bull}, {Burba}, {Carey}, {Carilli},
  {Cheng}, {DeBoer}, {Dexter}, {de Lera Acedo}, {Dillon}, {Ely}, {Ewall-Wice},
  {Fagnoni}, {Fialkov}, {Fritz}, {Furlanetto}, {Gale-Sides}, {Glendenning},
  {Gorthi}, {Greig}, {Grobbelaar}, {Halday}, {Hazelton}, {Heimersheim},
  {Hewitt}, {Hickish}, {Jacobs}, {Julius}, {Kern}, {Kerrigan}, {Kittiwisit},
  {Kohn}, {Kolopanis}, {Lanman}, {La Plante}, {Lekalake}, {Lewis}, {Liu}, {Ma},
  {MacMahon}, {Malan}, {Malgas}, {Maree}, {Martinot}, {Matsetela}, {Mesinger},
  {Mirocha}, {Molewa}, {Morales}, {Mosiane}, {Mu{\~n}oz}, {Murray}, {Neben},
  {Nikolic}, {Nunhokee}, {Parsons}, {Patra}, {Pieterse}, {Pober}, {Qin},
  {Razavi-Ghods}, {Reis}, {Ringuette}, {Robnett}, {Rosie}, {Santos}, {Sikder},
  {Sims}, {Smith}, {Syce}, {Thyagarajan}, {Williams}, \&
  {Zheng}}]{2022ApJ...924...51A}
---. 2022{\natexlab{b}}, \apj, 924, 51

\bibitem[{{Ali} {$et~al$.}(2008){Ali}, {Bharadwaj}, \& {Chengalur}}]{Ali+2008}
{Ali}, S.~S., {Bharadwaj}, S., \& {Chengalur}, J.~N. 2008, \mnras, 385, 2166

\bibitem[{{Ali} {$et~al$.}(2005){Ali}, {Bharadwaj}, \& {Pandey}}]{Ali+2005}
{Ali}, S.~S., {Bharadwaj}, S., \& {Pandey}, B. 2005, \mnras, 363, 251

\bibitem[{{Alvarez} {$et~al$.}(2006){Alvarez}, {Komatsu}, {Dor{\'e}}, \&
  {Shapiro}}]{alvarez+2006}
{Alvarez}, M.~A., {Komatsu}, E., {Dor{\'e}}, O., \& {Shapiro}, P.~R. 2006,
  \apj, 647, 840

\bibitem[{{Bag} {$et~al$.}(2019){Bag}, {Mondal}, {Sarkar}, {Bharadwaj},
  {Choudhury}, \& {Sahni}}]{Bag+2019}
{Bag}, S., {Mondal}, R., {Sarkar}, P., {$et~al$.} 2019, \mnras, 485, 2235

\bibitem[{{Bag} {$et~al$.}(2018){Bag}, {Mondal}, {Sarkar}, {Bharadwaj}, \&
  {Sahni}}]{Bag+2018}
{Bag}, S., {Mondal}, R., {Sarkar}, P., {Bharadwaj}, S., \& {Sahni}, V. 2018,
  \mnras, 477, 1984

\bibitem[{{Bagley} {$et~al$.}(2017){Bagley}, {Scarlata}, {Henry}, {Rafelski},
  {Malkan}, {Teplitz}, {Dai}, {Baronchelli}, {Colbert}, {Rutkowski}, {Mehta},
  {Dressler}, {McCarthy}, {Bunker}, {Atek}, {Garel}, {Martin}, {Hathi}, \&
  {Siana}}]{Bagley+2017}
{Bagley}, M.~B., {Scarlata}, C., {Henry}, A., {$et~al$.} 2017, \apj, 837, 11

\bibitem[{{Bandyopadhyay} {$et~al$.}(2017){Bandyopadhyay}, {Choudhury}, \&
  {Seshadri}}]{2017MNRAS.466.2302B}
{Bandyopadhyay}, B., {Choudhury}, T.~R., \& {Seshadri}, T.~R. 2017, \mnras,
  466, 2302

\bibitem[{{Barkana}(2018)}]{2018Natur.555...71B}
{Barkana}, R. 2018, \nat, 555, 71

\bibitem[{{Barkana} \& {Loeb}(2005)}]{Barkana_&_Loeb+2005}
{Barkana}, R., \& {Loeb}, A. 2005, \apj, 626, 1

\bibitem[{{Barkana} \& {Loeb}(2006)}]{Barkana+2006}
---. 2006, \mnras, 372, L43

\bibitem[{{Barnett} {$et~al$.}(2017){Barnett}, {Warren}, {Becker}, {Mortlock},
  {Hewett}, {McMahon}, {Simpson}, \& {Venemans}}]{barnett+2017}
{Barnett}, R., {Warren}, S.~J., {Becker}, G.~D., {$et~al$.} 2017, \aap, 601,
  A16

\bibitem[{{Barry} {$et~al$.}(2019{\natexlab{a}}){Barry}, {Beardsley}, {Byrne},
  {Hazelton}, {Morales}, {Pober}, \& {Sullivan}}]{2019PASA...36...26B}
{Barry}, N., {Beardsley}, A.~P., {Byrne}, R., {$et~al$.} 2019{\natexlab{a}},
  \pasa, 36, e026

\bibitem[{{Barry} {$et~al$.}(2019{\natexlab{b}}){Barry}, {Wilensky}, {Trott},
  {Pindor}, {Beardsley}, {Hazelton}, {Sullivan}, {Morales}, {Pober}, {Line},
  {Greig}, {Byrne}, {Lanman}, {Li}, {Jordan}, {Joseph}, {McKinley}, {Rahimi},
  {Yoshiura}, {Bowman}, {Gaensler}, {Hewitt}, {Jacobs}, {Mitchell}, {Udaya
  Shankar}, {Sethi}, {Subrahmanyan}, {Tingay}, {Webster}, \&
  {Wyithe}}]{2019ApJ...884....1B}
{Barry}, N., {Wilensky}, M., {Trott}, C.~M., {$et~al$.} 2019{\natexlab{b}},
  \apj, 884, 1

\bibitem[{{Beane} \& {Lidz}(2018)}]{Beane_&_Lidz+2018}
{Beane}, A., \& {Lidz}, A. 2018, \apj, 867, 26

\bibitem[{{Becker} {$et~al$.}(2001){Becker}, {Fan}, {White}, {Strauss},
  {Narayanan}, {Lupton}, {Gunn}, {Annis}, {Bahcall}, {Brinkmann}, {Connolly},
  {Csabai}, {Czarapata}, {Doi}, {Heckman}, {Hennessy}, {Ivezi{\'c}}, {Knapp},
  {Lamb}, {McKay}, {Munn}, {Nash}, {Nichol}, {Pier}, {Richards}, {Schneider},
  {Stoughton}, {Szalay}, {Thakar}, \& {York}}]{becker+2001}
{Becker}, R.~H., {Fan}, X., {White}, R.~L., {$et~al$.} 2001, \aj, 122, 2850

\bibitem[{Berlin {$et~al$.}(2018)Berlin, Hooper, Krnjaic, \&
  McDermott}]{PhysRevLett.121.011102}
Berlin, A., Hooper, D., Krnjaic, G., \& McDermott, S.~D. 2018, Phys. Rev.
  Lett., 121, 011102

\bibitem[{{Bernal} {$et~al$.}(2019{\natexlab{a}}){Bernal}, {Breysse},
  {Gil-Mar{\'\i}n}, \& {Kovetz}}]{Bernal+2019a}
{Bernal}, J.~L., {Breysse}, P.~C., {Gil-Mar{\'\i}n}, H., \& {Kovetz}, E.~D.
  2019{\natexlab{a}}, \prd, 100, 123522

\bibitem[{{Bernal} {$et~al$.}(2019{\natexlab{b}}){Bernal}, {Breysse}, \&
  {Kovetz}}]{Bernal+2019b}
{Bernal}, J.~L., {Breysse}, P.~C., \& {Kovetz}, E.~D. 2019{\natexlab{b}}, \prl,
  123, 251301

\bibitem[{{Bharadwaj} \& {Ali}(2004)}]{Bharadwaj+2004}
{Bharadwaj}, S., \& {Ali}, S.~S. 2004, \mnras, 352, 142

\bibitem[{{Bharadwaj} \& {Ali}(2005{\natexlab{a}})}]{Bharadwaj_&_Ali+2005}
---. 2005{\natexlab{a}}, \mnras, 356, 1519

\bibitem[{{Bharadwaj} \& {Ali}(2005{\natexlab{b}})}]{Bharadwaj+2005}
---. 2005{\natexlab{b}}, \mnras, 356, 1519

\bibitem[{{Bharadwaj} {$et~al$.}(2001){Bharadwaj}, {Nath}, \&
  {Sethi}}]{bharadwaj+2001b}
{Bharadwaj}, S., {Nath}, B.~B., \& {Sethi}, S.~K. 2001, Journal of Astrophysics
  and Astronomy, 22, 21

\bibitem[{{Bharadwaj} {$et~al$.}(2019){Bharadwaj}, {Pal}, {Choudhuri}, \&
  {Dutta}}]{Bharadwaj+2019}
{Bharadwaj}, S., {Pal}, S., {Choudhuri}, S., \& {Dutta}, P. 2019, \mnras, 483,
  5694

\bibitem[{{Bharadwaj} \& {Sethi}(2001)}]{Bharadwaj+2001}
{Bharadwaj}, S., \& {Sethi}, S.~K. 2001, \jcap, 22, 293

\bibitem[{{Binnie} \& {Pritchard}(2019)}]{Binnie+2019}
{Binnie}, T., \& {Pritchard}, J.~R. 2019, \mnras, 487, 1160

\bibitem[{{Bouwens} {$et~al$.}(2015){Bouwens}, {Illingworth}, {Oesch},
  {Trenti}, {Labb{\'e}}, {Bradley}, {Carollo}, {van Dokkum}, {Gonzalez},
  {Holwerda}, {Franx}, {Spitler}, {Smit}, \& {Magee}}]{bouwens+2015}
{Bouwens}, R.~J., {Illingworth}, G.~D., {Oesch}, P.~A., {$et~al$.} 2015, \apj,
  803, 34

\bibitem[{{Bouwens} {$et~al$.}(2016){Bouwens}, {Aravena}, {Decarli}, {Walter},
  {da Cunha}, {Labb{\'e}}, {Bauer}, {Bertoldi}, {Carilli}, {Chapman}, {Daddi},
  {Hodge}, {Ivison}, {Karim}, {Le Fevre}, {Magnelli}, {Ota}, {Riechers},
  {Smail}, {van der Werf}, {Weiss}, {Cox}, {Elbaz}, {Gonzalez-Lopez},
  {Infante}, {Oesch}, {Wagg}, \& {Wilkins}}]{bouwens+2016a}
{Bouwens}, R.~J., {Aravena}, M., {Decarli}, R., {$et~al$.} 2016, \apj, 833, 72

\bibitem[{{Bowman} {$et~al$.}(2018){Bowman}, {Rogers}, {Monsalve}, {Mozdzen},
  \& {Mahesh}}]{Bowman+2018}
{Bowman}, J.~D., {Rogers}, A. E.~E., {Monsalve}, R.~A., {Mozdzen}, T.~J., \&
  {Mahesh}, N. 2018, \nat, 555, 67

\bibitem[{Braun {$et~al$.}(2019)Braun, Bonaldi, Bourke, Keane, \&
  Wagg}]{braun+2019}
Braun, R., Bonaldi, A., Bourke, T., Keane, E., \& Wagg, J. 2019, arXiv
  preprint, arXiv:1912.12699

\bibitem[{{Breysse} \& {Alexandroff}(2019)}]{Breyesse+2019}
{Breysse}, P.~C., \& {Alexandroff}, R.~M. 2019, \mnras, 490, 260

\bibitem[{{Breysse} {$et~al$.}(2017){Breysse}, {Kovetz}, {Behroozi}, {Dai}, \&
  {Kamionkowski}}]{Breysse17}
{Breysse}, P.~C., {Kovetz}, E.~D., {Behroozi}, P.~S., {Dai}, L., \&
  {Kamionkowski}, M. 2017, \mnras, 467, 2996

\bibitem[{Breysse {$et~al$.}(2022)Breysse, Yang, Somerville, Pullen, Popping,
  \& Maniyar}]{Breysse+2022}
Breysse, P.~C., Yang, S., Somerville, R.~S., {$et~al$.} 2022, \apj, 929, 30

\bibitem[{{Breysse} {$et~al$.}(2021){Breysse}, {Chung}, {Cleary}, {Ihle},
  {Padmanabhan}, {Silva}, {Bond}, {Borowska}, {Catha}, {Church}, {Dunne},
  {Eriksen}, {Foss}, {Gaier}, {Ott Gundersen}, {Harris}, {Hobbs}, {Keating},
  {Lamb}, {Lawrence}, {Lunde}, {Murray}, {Pearson}, {Philip}, {Rasmussen},
  {Readhead}, {Rennie}, {Stutzer}, {Viero}, {Watts}, {Wehus}, \&
  {Woody}}]{2021arXiv211105933B}
{Breysse}, P.~C., {Chung}, D.~T., {Cleary}, K.~A., {$et~al$.} 2021, arXiv
  e-prints, arXiv:2111.05933

\bibitem[{{Carilli}(2011)}]{carilli+2011}
{Carilli}, C.~L. 2011, \apjl, 730, L30

\bibitem[{{Catalano} {$et~al$.}(2022){Catalano}, {Ade}, {Aravena}, {Barria},
  {Beelen}, {Benoit}, {B{\'e}thermin}, {Bounmy}, {Bourrion}, {Bres}, {De
  Breuck}, {Calvo}, {D{\'e}sert}, {Dur{\'a}n}, {Duvauchelle}, {Eraud},
  {Fasano}, {Fenouillet}, {Garcia}, {Garde}, {Goupy}, {Groppi}, {Hoarau}, {Hu},
  {Lagache}, {Lambert}, {Leggeri}, {Levy-Bertrand}, {Mac{\'\i}as-P{\'e}rez},
  {Mani}, {Marpaud}, {Marton}, {Mauskopf}, {Monfardini}, {Pisano}, {Ponthieu},
  {Prieur}, {Raffin}, {Roni}, {Roudier}, {Tourres}, {Tucker}, \&
  {Vivargent}}]{Catalano+2022}
{Catalano}, A., {Ade}, P., {Aravena}, M., {$et~al$.} 2022, in European Physical
  Journal Web of Conferences, Vol. 257, European Physical Journal Web of
  Conferences, 00010

\bibitem[{{Chakraborty} {$et~al$.}(2021){Chakraborty}, {Datta}, {Roy},
  {Bharadwaj}, {Choudhury}, {Datta}, {Pal}, {Choudhury}, {Choudhuri}, {Dutta},
  \& {Sarkar}}]{Chakraborty+2021}
{Chakraborty}, A., {Datta}, A., {Roy}, N., {$et~al$.} 2021, \apjl, 907, L7

\bibitem[{{Chapman} {$et~al$.}(2013){Chapman}, {Abdalla}, {Bobin}, {Starck},
  {Harker}, {Jeli{\'c}}, {Labropoulos}, {Zaroubi}, {Brentjens}, {de Bruyn}, \&
  {Koopmans}}]{2013MNRAS.429..165C}
{Chapman}, E., {Abdalla}, F.~B., {Bobin}, J., {$et~al$.} 2013, \mnras, 429, 165

\bibitem[{{Chary} {$et~al$.}(2020){Chary}, {Helou}, {Brammer}, {Capak},
  {Faisst}, {Flynn}, {Groom}, {Ferguson}, {Grillmair}, {Hemmati}, {Koekemoer},
  {Lee}, {Malhotra}, {Miyatake}, {Melchior}, {Momcheva}, {Newman}, {Masiero},
  {Paladini}, {Prakash}, {Rusholme}, {Stickley}, {Smith}, {Wood-Vasey}, \&
  {Teplitz}}]{Chary+2020}
{Chary}, R., {Helou}, G., {Brammer}, G., {$et~al$.} 2020, arXiv e-prints,
  arXiv:2008.10663

\bibitem[{{Chatterjee} {$et~al$.}(2021){Chatterjee}, {Choudhury}, \&
  {Mitra}}]{Chatterjee+2021}
{Chatterjee}, A., {Choudhury}, T.~R., \& {Mitra}, S. 2021, \mnras, 507, 2405

\bibitem[{{Cheng} {$et~al$.}(2020){Cheng}, {Chang}, \&
  {Bock}}]{2020ApJ...901..142C}
{Cheng}, Y.-T., {Chang}, T.-C., \& {Bock}, J.~J. 2020, \apj, 901, 142

\bibitem[{{Choudhuri} {$et~al$.}(2016){Choudhuri}, {Bharadwaj}, {Chatterjee},
  {Ali}, {Roy}, \& {Ghosh}}]{Choudhuri+2016}
{Choudhuri}, S., {Bharadwaj}, S., {Chatterjee}, S., {$et~al$.} 2016, \mnras,
  463, 4093

\bibitem[{{Choudhuri} {$et~al$.}(2014){Choudhuri}, {Bharadwaj}, {Ghosh}, \&
  {Ali}}]{Choudhuri+2014}
{Choudhuri}, S., {Bharadwaj}, S., {Ghosh}, A., \& {Ali}, S.~S. 2014, \mnras,
  445, 4351

\bibitem[{{Choudhury} {$et~al$.}(2021{\natexlab{a}}){Choudhury}, {Chatterjee},
  {Datta}, \& {Choudhury}}]{Choudhury+2021}
{Choudhury}, M., {Chatterjee}, A., {Datta}, A., \& {Choudhury}, T.~R.
  2021{\natexlab{a}}, \mnras, 502, 2815

\bibitem[{{Choudhury} {$et~al$.}(2020){Choudhury}, {Datta}, \&
  {Chakraborty}}]{Choudhury+2020}
{Choudhury}, M., {Datta}, A., \& {Chakraborty}, A. 2020, \mnras, 491, 4031

\bibitem[{{Choudhury} {$et~al$.}(2021{\natexlab{b}}){Choudhury}, {Datta}, \&
  {Majumdar}}]{Choudhury+2021b}
{Choudhury}, M., {Datta}, A., \& {Majumdar}, S. 2021{\natexlab{b}}, arXiv
  e-prints, arXiv:2112.13866

\bibitem[{{Choudhury} {$et~al$.}(2009){Choudhury}, {Haehnelt}, \&
  {Regan}}]{choudhury+2009}
{Choudhury}, T.~R., {Haehnelt}, M.~G., \& {Regan}, J. 2009, \mnras, 394, 960

\bibitem[{{Choudhury} {$et~al$.}(2015){Choudhury}, {Puchwein}, {Haehnelt}, \&
  {Bolton}}]{choudhury+2015}
{Choudhury}, T.~R., {Puchwein}, E., {Haehnelt}, M.~G., \& {Bolton}, J.~S. 2015,
  \mnras, 452, 261

\bibitem[{Choudhury {$et~al$.}(2015)Choudhury, Puchwein, Haehnelt, \&
  Bolton}]{choudhury15}
Choudhury, T.~R., Puchwein, E., Haehnelt, M.~G., \& Bolton, J.~S. 2015, Monthly
  Notices of the Royal Astronomical Society, 452, 261

\bibitem[{{Chung} {$et~al$.}(2021){Chung}, {Breysse}, {Cleary}, {Ihle},
  {Padmanabhan}, {Silva}, {Bond}, {Borowska}, {Catha}, {Church}, {Dunne},
  {Eriksen}, {Foss}, {Gaier}, {Ott Gundersen}, {Harper}, {Harris}, {Hensley},
  {Hobbs}, {Keating}, {Kim}, {Lamb}, {Lawrence}, {Gahr Sturtzel Lunde},
  {Murray}, {Pearson}, {Philip}, {Rasmussen}, {Readhead}, {Rennie}, {Stutzer},
  {Uzgil}, {Viero}, {Watts}, {Wechsler}, {Kathrine Wehus}, \&
  {Woody}}]{2021arXiv211105931C}
{Chung}, D.~T., {Breysse}, P.~C., {Cleary}, K.~A., {$et~al$.} 2021, arXiv
  e-prints, arXiv:2111.05931

\bibitem[{{Ciardi} {$et~al$.}(2000){Ciardi}, {Ferrara}, {Governato}, \&
  {Jenkins}}]{Ciardi+2000}
{Ciardi}, B., {Ferrara}, A., {Governato}, F., \& {Jenkins}, A. 2000, \mnras,
  314, 611

\bibitem[{{Cleary} {$et~al$.}(2021){Cleary}, {Borowska}, {Breysse}, {Catha},
  {Chung}, {Church}, {Dickinson}, {Eriksen}, {Foss}, {Ott Gundersen}, {Harper},
  {Harris}, {Hobbs}, {H{\r{a}}vard}, {Ihle}, {Kim}, {Kocz}, {Lamb}, {Lunde},
  {Padmanabhan}, {Pearson}, {Philip}, {Powell}, {Rasmussen}, {Readhead},
  {Rennie}, {Silva}, {Stutzer}, {Uzgil}, {Watts}, {Kathrine Wehus}, {Woody},
  {Basoalto}, {Bond}, {Dunne}, {Gaier}, {Hensley}, {Keating}, {Lawrence},
  {Murray}, {Reeves}, {Viero}, \& {Wechsler}}]{2021arXiv211105927C}
{Cleary}, K.~A., {Borowska}, J., {Breysse}, P.~C., {$et~al$.} 2021, arXiv
  e-prints, arXiv:2111.05927

\bibitem[{{Cohen} {$et~al$.}(2020){Cohen}, {Fialkov}, {Barkana}, \&
  {Monsalve}}]{Cohen+2020}
{Cohen}, A., {Fialkov}, A., {Barkana}, R., \& {Monsalve}, R.~A. 2020, \mnras,
  495, 4845

\bibitem[{{Crites} {$et~al$.}(2014){Crites}, {Bock}, {Bradford}, {Chang},
  {Cooray}, {Duband}, {Gong}, {Hailey-Dunsheath}, {Hunacek}, {Koch}, {Li},
  {O'Brient}, {Prouve}, {Shirokoff}, {Silva}, {Staniszewski}, {Uzgil}, \&
  {Zemcov}}]{crites+2014}
{Crites}, A.~T., {Bock}, J.~J., {Bradford}, C.~M., {$et~al$.} 2014, in Society
  of Photo-Optical Instrumentation Engineers (SPIE) Conference Series, Vol.
  9153, Millimeter, Submillimeter, and Far-Infrared Detectors and
  Instrumentation for Astronomy VII, ed. W.~S. {Holland} \& J.~{Zmuidzinas},
  91531W

\bibitem[{{Croft} {$et~al$.}(2018){Croft}, {Miralda-Escud{\'e}}, {Zheng},
  {Blomqvist}, \& {Pieri}}]{Croft+2018}
{Croft}, R. A.~C., {Miralda-Escud{\'e}}, J., {Zheng}, Z., {Blomqvist}, M., \&
  {Pieri}, M. 2018, \mnras, 481, 1320

\bibitem[{{Datta} {$et~al$.}(2007){Datta}, {Choudhury}, \&
  {Bharadwaj}}]{Datta+2007}
{Datta}, K.~K., {Choudhury}, T.~R., \& {Bharadwaj}, S. 2007, \mnras, 378, 119

\bibitem[{{Datta} {$et~al$.}(2014){Datta}, {Jensen}, {Majumdar}, {Mellema},
  {Iliev}, {Mao}, {Shapiro}, \& {Ahn}}]{Datta+2014}
{Datta}, K.~K., {Jensen}, H., {Majumdar}, S., {$et~al$.} 2014, \mnras, 442,
  1491

\bibitem[{{Datta} {$et~al$.}(2012){Datta}, {Mellema}, {Mao}, {Iliev},
  {Shapiro}, \& {Ahn}}]{Datta+2012}
{Datta}, K.~K., {Mellema}, G., {Mao}, Y., {$et~al$.} 2012, \mnras, 424, 1877

\bibitem[{Davé {$et~al$.}(2019)Davé, Anglés-Alcázar, Narayanan, Li,
  Rafieferantsoa, \& Appleby}]{Dave+2019}
Davé, R., Anglés-Alcázar, D., Narayanan, D., {$et~al$.} 2019, Monthly
  Notices of the Royal Astronomical Society, 486, 2827

\bibitem[{De~Looze {$et~al$.}(2011)De~Looze, Baes, Bendo, Cortese, \&
  Fritz}]{De_Looze+2011}
De~Looze, I., Baes, M., Bendo, G.~J., Cortese, L., \& Fritz, J. 2011, Monthly
  Notices of the Royal Astronomical Society, 416, 2712

\bibitem[{{De Looze, Ilse} {$et~al$.}(2014){De Looze, Ilse}, {Cormier, Diane},
  {Lebouteiller, Vianney}, {Madden, Suzanne}, {Baes, Maarten}, {Bendo, George
  J.}, {Boquien, M\'ed\'eric}, {Boselli, Alessandro}, {Clements, David L.},
  {Cortese, Luca}, {Cooray, Asantha}, {Galametz, Maud}, {Galliano,
  Fr\'ed\'eric}, {Graci\'a-Carpio, Javier}, {Isaak, Kate}, {Karczewski, Oskar
  L.}, {Parkin, Tara J.}, {Pellegrini, Eric W.}, {R\'emy-Ruyer, Aur\'elie},
  {Spinoglio, Luigi}, {Smith, Matthew W. L.}, \& {Sturm,
  Eckhard}}]{De_Looze+2014}
{De Looze, Ilse}, {Cormier, Diane}, {Lebouteiller, Vianney}, {$et~al$.} 2014,
  A\&A, 568, A62

\bibitem[{DeBoer {$et~al$.}(2017)DeBoer, Parsons, Aguirre, Alexander, Ali,
  Beardsley, Bernardi, Bowman, Bradley, Carilli, Cheng, de~Lera~Acedo, Dillon,
  Ewall-Wice, Fadana, Fagnoni, Fritz, Furlanetto, Glendenning, Greig,
  Grobbelaar, Hazelton, Hewitt, Hickish, Jacobs, Julius, Kariseb, Kohn,
  Lekalake, Liu, Loots, MacMahon, Malan, Malgas, Maree, Martinot, Mathison,
  Matsetela, Mesinger, Morales, Neben, Patra, Pieterse, Pober, Razavi-Ghods,
  Ringuette, Robnett, Rosie, Sell, Smith, Syce, Tegmark, Thyagarajan, Williams,
  \& Zheng}]{DeBoer+2017}
DeBoer, D.~R., Parsons, A.~R., Aguirre, J.~E., {$et~al$.} 2017, \pasp, 129,
  045001

\bibitem[{{Decarli} {$et~al$.}(2020){Decarli}, {Aravena}, {Boogaard},
  {Carilli}, {Gonz{\'a}lez-L{\'o}pez}, {Walter}, {Cortes}, {Cox}, {da Cunha},
  {Daddi}, {D{\'\i}az-Santos}, {Hodge}, {Inami}, {Neeleman}, {Novak}, {Oesch},
  {Popping}, {Riechers}, {Smail}, {Uzgil}, {van der Werf}, {Wagg}, \&
  {Weiss}}]{Decarli+2020}
{Decarli}, R., {Aravena}, M., {Boogaard}, L., {$et~al$.} 2020, \apj, 902, 110

\bibitem[{{Di Matteo} {$et~al$.}(2002){Di Matteo}, {Perna}, {Abel}, \&
  {Rees}}]{Di_Matteo+2002}
{Di Matteo}, T., {Perna}, R., {Abel}, T., \& {Rees}, M.~J. 2002, \apj, 564, 576

\bibitem[{{Dillon} {$et~al$.}(2020){Dillon}, {Lee}, {Ali}, {Parsons}, {Orosz},
  {Nunhokee}, {La Plante}, {Beardsley}, {Kern}, {Abdurashidova}, {Aguirre},
  {Alexander}, {Balfour}, {Bernardi}, {Billings}, {Bowman}, {Bradley}, {Bull},
  {Burba}, {Carey}, {Carilli}, {Cheng}, {DeBoer}, {Dexter}, {de Lera Acedo},
  {Ely}, {Ewall-Wice}, {Fagnoni}, {Fritz}, {Furlanetto}, {Gale-Sides},
  {Glendenning}, {Gorthi}, {Greig}, {Grobbelaar}, {Halday}, {Hazelton},
  {Hewitt}, {Hickish}, {Jacobs}, {Julius}, {Kerrigan}, {Kittiwisit}, {Kohn},
  {Kolopanis}, {Lanman}, {Lekalake}, {Lewis}, {Liu}, {Ma}, {MacMahon}, {Malan},
  {Malgas}, {Maree}, {Martinot}, {Matsetela}, {Mesinger}, {Molewa}, {Morales},
  {Mosiane}, {Murray}, {Neben}, {Nikolic}, {Pascua}, {Patra}, {Pieterse},
  {Pober}, {Razavi-Ghods}, {Ringuette}, {Robnett}, {Rosie}, {Santos}, {Sims},
  {Smith}, {Syce}, {Tegmark}, {Thyagarajan}, {Williams}, \&
  {Zheng}}]{2020MNRAS.499.5840D}
{Dillon}, J.~S., {Lee}, M., {Ali}, Z.~S., {$et~al$.} 2020, \mnras, 499, 5840

\bibitem[{{Dor{\'e}} {$et~al$.}(2018){Dor{\'e}}, {Werner}, {Ashby}, {Bleem},
  {Bock}, {Burt}, {Capak}, {Chang}, {Chaves-Montero}, {Chen}, {Civano},
  {Cleeves}, {Cooray}, {Crill}, {Crossfield}, {Cushing}, {de la Torre},
  {DiMatteo}, {Dvory}, {Dvorkin}, {Espaillat}, {Ferraro}, {Finkbeiner},
  {Greene}, {Hewitt}, {Hogg}, {Huffenberger}, {Jun}, {Ilbert}, {Jeong},
  {Johnson}, {Kim}, {Kirkpatrick}, {Kowalski}, {Korngut}, {Li}, {Lisse},
  {MacGregor}, {Mamajek}, {Mauskopf}, {Melnick}, {M{\'e}nard}, {Neyrinck},
  {{\"O}berg}, {Pisani}, {Rocca}, {Salvato}, {Schaan}, {Scoville}, {Song},
  {Stevens}, {Tenneti}, {Teplitz}, {Tolls}, {Unwin}, {Urry}, {Wandelt},
  {Williams}, {Wilner}, {Windhorst}, {Wolk}, {Yorke}, \& {Zemcov}}]{Dore+2018}
{Dor{\'e}}, O., {Werner}, M.~W., {Ashby}, M. L.~N., {$et~al$.} 2018, arXiv
  e-prints, arXiv:1805.05489

\bibitem[{{Dumitru} {$et~al$.}(2019){Dumitru}, {Kulkarni}, {Lagache}, \&
  {Haehnelt}}]{Dumitru+2019}
{Dumitru}, S., {Kulkarni}, G., {Lagache}, G., \& {Haehnelt}, M.~G. 2019,
  \mnras, 485, 3486

\bibitem[{{Ewall-Wice} {$et~al$.}(2018){Ewall-Wice}, {Chang}, {Lazio},
  {Dor{\'e}}, {Seiffert}, \& {Monsalve}}]{2018ApJ...868...63E}
{Ewall-Wice}, A., {Chang}, T.~C., {Lazio}, J., {$et~al$.} 2018, \apj, 868, 63

\bibitem[{{Ewall-Wice} {$et~al$.}(2016){Ewall-Wice}, {Dillon}, {Hewitt},
  {Loeb}, {Mesinger}, {Neben}, {Offringa}, {Tegmark}, {Barry}, {Beardsley},
  {Bernardi}, {Bowman}, {Briggs}, {Cappallo}, {Carroll}, {Corey}, {de
  Oliveira-Costa}, {Emrich}, {Feng}, {Gaensler}, {Goeke}, {Greenhill},
  {Hazelton}, {Hurley-Walker}, {Johnston-Hollitt}, {Jacobs}, {Kaplan},
  {Kasper}, {Kim}, {Kratzenberg}, {Lenc}, {Line}, {Lonsdale}, {Lynch},
  {McKinley}, {McWhirter}, {Mitchell}, {Morales}, {Morgan}, {Thyagarajan},
  {Oberoi}, {Ord}, {Paul}, {Pindor}, {Pober}, {Prabu}, {Procopio}, {Riding},
  {Rogers}, {Roshi}, {Shankar}, {Sethi}, {Srivani}, {Subrahmanyan}, {Sullivan},
  {Tingay}, {Trott}, {Waterson}, {Wayth}, {Webster}, {Whitney}, {Williams},
  {Williams}, {Wu}, \& {Wyithe}}]{2016MNRAS.460.4320E}
{Ewall-Wice}, A., {Dillon}, J.~S., {Hewitt}, J.~N., {$et~al$.} 2016, \mnras,
  460, 4320

\bibitem[{{Falcke} {$et~al$.}(2007){Falcke}, {van Haarlem}, {de Bruyn},
  {Braun}, {R{\"o}ttgering}, {Stappers}, {Boland}, {Butcher}, {de Geus},
  {Koopmans}, {Fender}, {Kuijpers}, {Miley}, {Schilizzi}, {Vogt}, {Wijers},
  {Wise}, {Brouw}, {Hamaker}, {Noordam}, {Oosterloo}, {B{\"a}hren},
  {Brentjens}, {Wijnholds}, {Bregman}, {van Cappellen}, {Gunst}, {Kant},
  {Reitsma}, {van der Schaaf}, \& {de Vos}}]{Fackle+2007}
{Falcke}, H.~D., {van Haarlem}, M.~P., {de Bruyn}, A.~G., {$et~al$.} 2007,
  Highlights of Astronomy, 14, 386

\bibitem[{{Fan} {$et~al$.}(2006){Fan}, {Carilli}, \& {Keating}}]{fan+2006}
{Fan}, X., {Carilli}, C.~L., \& {Keating}, B. 2006, \araa, 44, 415

\bibitem[{{Fan} {$et~al$.}(2003){Fan}, {Strauss}, {Schneider}, {Becker},
  {White}, {Haiman}, {Gregg}, {Pentericci}, {Grebel}, {Narayanan}, {Loh},
  {Richards}, {Gunn}, {Lupton}, {Knapp}, {Ivezi{\'c}}, {Brandt}, {Collinge},
  {Hao}, {Harbeck}, {Prada}, {Schaye}, {Strateva}, {Zakamska}, {Anderson},
  {Brinkmann}, {Bahcall}, {Lamb}, {Okamura}, {Szalay}, \& {York}}]{fan+2003}
{Fan}, X., {Strauss}, M.~A., {Schneider}, D.~P., {$et~al$.} 2003, \aj, 125,
  1649

\bibitem[{{Feng} {$et~al$.}(2017){Feng}, {Cooray}, \& {Keating}}]{feng17}
{Feng}, C., {Cooray}, A., \& {Keating}, B. 2017, \apj, 846, 21

\bibitem[{{Feng} \& {Holder}(2018)}]{2018ApJ...858L..17F}
{Feng}, C., \& {Holder}, G. 2018, \apjl, 858, L17

\bibitem[{{Field}(1958)}]{Field+1958}
{Field}, G.~B. 1958, Proceedings of the IRE, 46, 240

\bibitem[{{Furlanetto} \& {Lidz}(2007)}]{furlanetto07}
{Furlanetto}, S.~R., \& {Lidz}, A. 2007, \apj, 660, 1030

\bibitem[{{Furlanetto} {$et~al$.}(2006){Furlanetto}, {Oh}, \&
  {Briggs}}]{furlanetto+2006}
{Furlanetto}, S.~R., {Oh}, S.~P., \& {Briggs}, F.~H. 2006, \physrep, 433, 181

\bibitem[{{Furlanetto} {$et~al$.}(2004){Furlanetto}, {Zaldarriaga}, \&
  {Hernquist}}]{Furlanetto+2004}
{Furlanetto}, S.~R., {Zaldarriaga}, M., \& {Hernquist}, L. 2004, \apj, 613, 1

\bibitem[{{Gehlot} {$et~al$.}(2019){Gehlot}, {Mertens}, {Koopmans},
  {Brentjens}, {Zaroubi}, {Ciardi}, {Ghosh}, {Hatef}, {Iliev}, {Jeli{\'c}}, {},
  {Kooistra}, {Krause}, {Mellema}, {Mevius}, {Mitra}, {Offringa}, {Pandey},
  {Sardarabadi}, {Schaye}, {Silva}, {Vedantham}, \&
  {Yatawatta}}]{2019MNRAS.488.4271G}
{Gehlot}, B.~K., {Mertens}, F.~G., {Koopmans}, L.~V.~E., {$et~al$.} 2019,
  \mnras, 488, 4271

\bibitem[{{Geil} \& {Wyithe}(2008)}]{Geil+2008}
{Geil}, P.~M., \& {Wyithe}, J. S.~B. 2008, \mnras, 386, 1683

\bibitem[{{Ghara} \& {Choudhury}(2020)}]{2020MNRAS.496..739G}
{Ghara}, R., \& {Choudhury}, T.~R. 2020, \mnras, 496, 739

\bibitem[{{Ghara} {$et~al$.}(2015{\natexlab{a}}){Ghara}, {Choudhury}, \&
  {Datta}}]{Ghara+2015a}
{Ghara}, R., {Choudhury}, T.~R., \& {Datta}, K.~K. 2015{\natexlab{a}}, \mnras,
  447, 1806

\bibitem[{{Ghara} {$et~al$.}(2017){Ghara}, {Choudhury}, {Datta}, \&
  {Choudhuri}}]{ghara16}
{Ghara}, R., {Choudhury}, T.~R., {Datta}, K.~K., \& {Choudhuri}, S. 2017,
  \mnras, 464, 2234

\bibitem[{{Ghara} {$et~al$.}(2015{\natexlab{b}}){Ghara}, {Datta}, \&
  {Choudhury}}]{Ghara+2015}
{Ghara}, R., {Datta}, K.~K., \& {Choudhury}, T.~R. 2015{\natexlab{b}}, \mnras,
  453, 3143

\bibitem[{{Ghara} {$et~al$.}(2021{\natexlab{a}}){Ghara}, {Giri}, {Ciardi},
  {Mellema}, \& {Zaroubi}}]{2021MNRAS.503.4551G}
{Ghara}, R., {Giri}, S.~K., {Ciardi}, B., {Mellema}, G., \& {Zaroubi}, S.
  2021{\natexlab{a}}, \mnras, 503, 4551

\bibitem[{{Ghara} {$et~al$.}(2018){Ghara}, {Mellema}, {Giri}, {Choudhury},
  {Datta}, \& {Majumdar}}]{Ghara+2018}
{Ghara}, R., {Mellema}, G., {Giri}, S.~K., {$et~al$.} 2018, \mnras, 476, 1741

\bibitem[{{Ghara} {$et~al$.}(2021{\natexlab{b}}){Ghara}, {Mellema}, \&
  {Zaroubi}}]{2021arXiv210813593G}
{Ghara}, R., {Mellema}, G., \& {Zaroubi}, S. 2021{\natexlab{b}}, arXiv
  e-prints, arXiv:2108.13593

\bibitem[{{Ghara} {$et~al$.}(2020){Ghara}, {Giri}, {Mellema}, {Ciardi},
  {Zaroubi}, {Iliev}, {Koopmans}, {Chapman}, {Gazagnes}, {Gehlot}, {Ghosh},
  {Jeli{\'c}}, {Mertens}, {Mondal}, {Schaye}, {Silva}, {Asad}, {Kooistra},
  {Mevius}, {Offringa}, {Pandey}, \& {Yatawatta}}]{2020MNRAS.493.4728G}
{Ghara}, R., {Giri}, S.~K., {Mellema}, G., {$et~al$.} 2020, \mnras, 493, 4728

\bibitem[{Ghosh {$et~al$.}(2011{\natexlab{a}})Ghosh, Bharadwaj, Ali, \&
  Chengalur}]{Ghosh+2011a}
Ghosh, A., Bharadwaj, S., Ali, S.~S., \& Chengalur, J.~N. 2011{\natexlab{a}},
  \mnras, 411, 2426

\bibitem[{Ghosh {$et~al$.}(2011{\natexlab{b}})Ghosh, Bharadwaj, Ali, \&
  Chengalur}]{Ghosh+2011b}
---. 2011{\natexlab{b}}, \mnras, 418, 2584

\bibitem[{{Ghosh} {$et~al$.}(2012){Ghosh}, {Prasad}, {Bharadwaj}, {Ali}, \&
  {Chengalur}}]{Ghosh+2012}
{Ghosh}, A., {Prasad}, J., {Bharadwaj}, S., {Ali}, S.~S., \& {Chengalur}, J.~N.
  2012, \mnras, 426, 3295

\bibitem[{{Gillet} {$et~al$.}(2019){Gillet}, {Mesinger}, {Greig}, {Liu}, \&
  {Ucci}}]{2019MNRAS.484..282G}
{Gillet}, N., {Mesinger}, A., {Greig}, B., {Liu}, A., \& {Ucci}, G. 2019,
  \mnras, 484, 282

\bibitem[{{Giri} \& {Mellema}(2021)}]{2021MNRAS.505.1863G}
{Giri}, S.~K., \& {Mellema}, G. 2021, \mnras, 505, 1863

\bibitem[{{Giri} {$et~al$.}(2018){Giri}, {Mellema}, \&
  {Ghara}}]{2018MNRAS.479.5596G}
{Giri}, S.~K., {Mellema}, G., \& {Ghara}, R. 2018, \mnras, 479, 5596

\bibitem[{{Gnedin}(2000{\natexlab{a}})}]{Gnedin+2000a}
{Gnedin}, N.~Y. 2000{\natexlab{a}}, \apj, 535, 530

\bibitem[{{Gnedin}(2000{\natexlab{b}})}]{Gnedin+2000b}
---. 2000{\natexlab{b}}, \apj, 542, 535

\bibitem[{{Gnedin} \& {Ostriker}(1997)}]{Gnedin+1997}
{Gnedin}, N.~Y., \& {Ostriker}, J.~P. 1997, \apj, 486, 581

\bibitem[{{Gong} {$et~al$.}(2012){Gong}, {Cooray}, {Silva}, {Santos}, {Bock},
  {Bradford}, \& {Zemcov}}]{Gong+2012}
{Gong}, Y., {Cooray}, A., {Silva}, M., {$et~al$.} 2012, \apj, 745, 49

\bibitem[{Gong {$et~al$.}(2011)Gong, Cooray, Silva, Santos, \&
  Lubin}]{Gong_2011}
Gong, Y., Cooray, A., Silva, M.~B., Santos, M.~G., \& Lubin, P. 2011, \apj,
  728, L46

\bibitem[{{Greig} \& {Mesinger}(2015)}]{Greig+2015}
{Greig}, B., \& {Mesinger}, A. 2015, \mnras, 449, 4246

\bibitem[{{Greig} \& {Mesinger}(2017)}]{Greig+2017}
---. 2017, \mnras, 472, 2651

\bibitem[{{Greig} \& {Mesinger}(2018)}]{Greig+2018}
---. 2018, \mnras, 477, 3217

\bibitem[{{Greig} {$et~al$.}(2021){Greig}, {Trott}, {Barry}, {Mutch}, {Pindor},
  {Webster}, \& {Wyithe}}]{2020arXiv200802639G}
{Greig}, B., {Trott}, C.~M., {Barry}, N., {$et~al$.} 2021, \mnras, 500, 5322

\bibitem[{{Gruppioni} {$et~al$.}(2020){Gruppioni}, {B{\'e}thermin}, {Loiacono},
  {Le F{\`e}vre}, {Capak}, {Cassata}, {Faisst}, {Schaerer}, {Silverman}, {Yan},
  {Bardelli}, {Boquien}, {Carraro}, {Cimatti}, {Dessauges-Zavadsky}, {Ginolfi},
  {Fujimoto}, {Hathi}, {Jones}, {Khusanova}, {Koekemoer}, {Lagache}, {Lemaux},
  {Oesch}, {Pozzi}, {Riechers}, {Rodighiero}, {Romano}, {Talia}, {Vallini},
  {Vergani}, {Zamorani}, \& {Zucca}}]{Gruppioni+2020}
{Gruppioni}, C., {B{\'e}thermin}, M., {Loiacono}, F., {$et~al$.} 2020,
  Astronomy \& Astrophysics, 643, A8

\bibitem[{{Hammer} {$et~al$.}(2021){Hammer}, {Morris}, {Cuby}, {Kaper},
  {Steinmetz}, {Afonso}, {Barbuy}, {Bergin}, {Finogenov}, {Gallego}, {Kassin},
  {Miller}, {{\"O}stlin}, {Pentericci}, {Schaerer}, {Ziegler}, {Chemla},
  {Dalton}, {De Frondat}, {Evans}, {Le Mignant}, {Puech}, {Rodrigues},
  {Sanchez-Janssen}, {Taburet}, {Tasca}, {Yang}, {Zanchetta}, {Dohlen},
  {Dubbeldam}, {El Hadi}, {Janssen}, {Kelz}, {Larrieu}, {Lewis}, {MacIntosh},
  {Morris}, {Navarro}, \& {Seifert}}]{Hammer+2021}
{Hammer}, F., {Morris}, S., {Cuby}, J.~G., {$et~al$.} 2021, The Messenger, 182,
  33

\bibitem[{{Harker} {$et~al$.}(2009){Harker}, {Zaroubi}, {Bernardi},
  {Brentjens}, {de Bruyn}, {Ciardi}, {Jeli{\'c}}, {Koopmans}, {Labropoulos},
  {Mellema}, {Offringa}, {Pandey}, {Schaye}, {Thomas}, \&
  {Yatawatta}}]{2009MNRAS.397.1138H}
{Harker}, G., {Zaroubi}, S., {Bernardi}, G., {$et~al$.} 2009, \mnras, 397, 1138

\bibitem[{Hassan {$et~al$.}(2016)Hassan, Davé, Finlator, \&
  Santos}]{Hassan+2016}
Hassan, S., Davé, R., Finlator, K., \& Santos, M.~G. 2016, \mnras, 457, 1550

\bibitem[{{Hassan} {$et~al$.}(2019){Hassan}, {Liu}, {Kohn}, \& {La
  Plante}}]{Hassan+2019}
{Hassan}, S., {Liu}, A., {Kohn}, S., \& {La Plante}, P. 2019, \mnras, 483, 2524

\bibitem[{{Heneka} \& {Cooray}(2021)}]{heneka21}
{Heneka}, C., \& {Cooray}, A. 2021, \mnras, 506, 1573

\bibitem[{{Hothi} {$et~al$.}(2021){Hothi}, {Chapman}, {Pritchard}, {Mertens},
  {Koopmans}, {Ciardi}, {Gehlot}, {Ghara}, {Ghosh}, {Giri}, {Iliev},
  {Jeli{\'c}}, \& {Zaroubi}}]{2021MNRAS.500.2264H}
{Hothi}, I., {Chapman}, E., {Pritchard}, J.~R., {$et~al$.} 2021, \mnras, 500,
  2264

\bibitem[{{Hunacek} {$et~al$.}(2018){Hunacek}, {Bock}, {Bradford}, {Butler},
  {Chang}, {Cheng}, {Cooray}, {Crites}, {Frez}, {Hailey-Dunsheath}, {Hoscheit},
  {Kim}, {Li}, {Marrone}, {Moncelsi}, {Shirokoff}, {Steinbach}, {Sun},
  {Trumper}, {Turner}, {Uzgil}, {Weber}, \& {Zemcov}}]{2018JLTP..193..893H}
{Hunacek}, J., {Bock}, J., {Bradford}, C.~M., {$et~al$.} 2018, Journal of Low
  Temperature Physics, 193, 893

\bibitem[{{Hutter} {$et~al$.}(2017){Hutter}, {Dayal}, {M{\"u}ller}, \&
  {Trott}}]{hutter17}
{Hutter}, A., {Dayal}, P., {M{\"u}ller}, V., \& {Trott}, C.~M. 2017, \apj, 836,
  176

\bibitem[{{Hutter} {$et~al$.}(2019){Hutter}, {Dayal}, {Malhotra}, {Rhoads},
  {Choudhury}, {Ciardi}, {Conselice}, {Cooray}, {Cuby}, {Datta}, {Fan},
  {Finkelstein}, {Hirata}, {Iliev}, {Jansen}, {Kakiichi}, {Koekemoer}, {Maio},
  {Majumdar}, {Mellema}, {Mondal}, {Papovich}, {Rhodes}, {Sahl{\'e}n},
  {Schauer}, {Takahashi}, {Ucci}, {Windhorst}, \&
  {Zackrisson}}]{Hutter+2019_WhitePaper}
{Hutter}, A., {Dayal}, P., {Malhotra}, S., {$et~al$.} 2019, Bulletin of the
  American Astronomical Society, 51, 57

\bibitem[{{Ihle} {$et~al$.}(2019){Ihle}, {Chung}, {Stein}, {Alvarez}, {Bond},
  {Breysse}, {Cleary}, {Eriksen}, {Foss}, {Gundersen}, {Harper}, {Murray},
  {Padmanabhan}, {Viero}, {Wehus}, \& {COMAP Collaboration}}]{Ihle+2019}
{Ihle}, H.~T., {Chung}, D., {Stein}, G., {$et~al$.} 2019, \apj, 871, 75

\bibitem[{{Ihle} {$et~al$.}(2021){Ihle}, {Borowska}, {Cleary}, {Eriksen},
  {Foss}, {Harper}, {Kim}, {Lunde}, {Philip}, {Rasmussen}, {Stutzer}, {Uzgil},
  {Watts}, {Kathrine Wehus}, {Bond}, {Breysse}, {Catha}, {Church}, {Chung},
  {Dickinson}, {Dunne}, {Gaier}, {Ott Gundersen}, {Harris}, {Hobbs}, {Lamb},
  {Lawrence}, {Murray}, {Readhead}, {Padmanabhan}, {Pearson}, {Rennie}, \&
  {Woody}}]{2021arXiv211105930I}
{Ihle}, H.~T., {Borowska}, J., {Cleary}, K.~A., {$et~al$.} 2021, arXiv
  e-prints, arXiv:2111.05930

\bibitem[{{Iliev} {$et~al$.}(2014){Iliev}, {Mellema}, {Ahn}, {Shapiro}, {Mao},
  \& {Pen}}]{Iliev+2014}
{Iliev}, I.~T., {Mellema}, G., {Ahn}, K., {$et~al$.} 2014, \mnras, 439, 725

\bibitem[{{Iliev} {$et~al$.}(2006){Iliev}, {Mellema}, {Pen}, {Merz}, {Shapiro},
  \& {Alvarez}}]{Iliev+2006}
{Iliev}, I.~T., {Mellema}, G., {Pen}, U.~L., {$et~al$.} 2006, \mnras, 369, 1625

\bibitem[{{Iliev} {$et~al$.}(2007){Iliev}, {Mellema}, {Shapiro}, \&
  {Pen}}]{Iliev+2007}
{Iliev}, I.~T., {Mellema}, G., {Shapiro}, P.~R., \& {Pen}, U.-L. 2007, \mnras,
  376, 534

\bibitem[{{Iliev} {$et~al$.}(2012){Iliev}, {Mellema}, {Shapiro}, {Pen}, {Mao},
  {Koda}, \& {Ahn}}]{Iliev+2012}
{Iliev}, I.~T., {Mellema}, G., {Shapiro}, P.~R., {$et~al$.} 2012, \mnras, 423,
  2222

\bibitem[{{Jensen} {$et~al$.}(2013){Jensen}, {Laursen}, {Mellema}, {Iliev},
  {Sommer-Larsen}, \& {Shapiro}}]{jensen+2013}
{Jensen}, H., {Laursen}, P., {Mellema}, G., {$et~al$.} 2013, \mnras, 428, 1366

\bibitem[{{Ji} {$et~al$.}(2021){Ji}, {Hotinli}, \& {Kamionkowski}}]{Ji+2021}
{Ji}, L., {Hotinli}, S.~C., \& {Kamionkowski}, M. 2021, arXiv e-prints,
  arXiv:2110.01619

\bibitem[{{Kamran} {$et~al$.}(2021{\natexlab{a}}){Kamran}, {Ghara}, {Majumdar},
  {Mondal}, {Mellema}, {Bharadwaj}, {Pritchard}, \& {Iliev}}]{Kamran+2021}
{Kamran}, M., {Ghara}, R., {Majumdar}, S., {$et~al$.} 2021{\natexlab{a}},
  \mnras, 502, 3800

\bibitem[{{Kamran} {$et~al$.}(2021{\natexlab{b}}){Kamran}, {Majumdar}, {Ghara},
  {Mellema}, {Bharadwaj}, {Pritchard}, {Mondal}, \& {Iliev}}]{Kamran+2021b}
{Kamran}, M., {Majumdar}, S., {Ghara}, R., {$et~al$.} 2021{\natexlab{b}}, arXiv
  e-prints, arXiv:2108.08201

\bibitem[{{Kannan} {$et~al$.}(2021{\natexlab{a}}){Kannan}, {Garaldi}, {Smith},
  {Pakmor}, {Springel}, {Vogelsberger}, \& {Hernquist}}]{Kannan+2021a}
{Kannan}, R., {Garaldi}, E., {Smith}, A., {$et~al$.} 2021{\natexlab{a}},
  \mnras, arXiv:2110.00584

\bibitem[{{Kannan} {$et~al$.}(2021{\natexlab{b}}){Kannan}, {Smith}, {Garaldi},
  {Shen}, {Vogelsberger}, {Pakmor}, {Springel}, \& {Hernquist}}]{Kannan+2021}
{Kannan}, R., {Smith}, A., {Garaldi}, E., {$et~al$.} 2021{\natexlab{b}}, arXiv
  e-prints, arXiv:2111.02411

\bibitem[{{Kapahtia} {$et~al$.}(2019){Kapahtia}, {Chingangbam}, \&
  {Appleby}}]{Kapahtia+2019}
{Kapahtia}, A., {Chingangbam}, P., \& {Appleby}, S. 2019, \jcap, 2019, 053

\bibitem[{{Kapahtia} {$et~al$.}(2018){Kapahtia}, {Chingangbam}, {Appleby}, \&
  {Park}}]{Kapahtia+2018}
{Kapahtia}, A., {Chingangbam}, P., {Appleby}, S., \& {Park}, C. 2018, \jcap,
  2018, 011

\bibitem[{{Kapahtia} {$et~al$.}(2021){Kapahtia}, {Chingangbam}, {Ghara},
  {Appleby}, \& {Choudhury}}]{2021JCAP...05..026K}
{Kapahtia}, A., {Chingangbam}, P., {Ghara}, R., {Appleby}, S., \& {Choudhury},
  T.~R. 2021, \jcap, 2021, 026

\bibitem[{{Karoumpis} {$et~al$.}(2022){Karoumpis}, {Magnelli},
  {Romano-D{\'\i}az}, {Haslbauer}, \& {Bertoldi}}]{Karoumpis+2022}
{Karoumpis}, C., {Magnelli}, B., {Romano-D{\'\i}az}, E., {Haslbauer}, M., \&
  {Bertoldi}, F. 2022, \aap, 659, A12

\bibitem[{{Kashikawa} {$et~al$.}(2011){Kashikawa}, {Shimasaku}, {Matsuda},
  {Egami}, {Jiang}, {Nagao}, {Ouchi}, {Malkan}, {Hattori}, {Ota}, {Taniguchi},
  {Okamura}, {Ly}, {Iye}, {Furusawa}, {Shioya}, {Shibuya}, {Ishizaki}, \&
  {Toshikawa}}]{Kashikawa+2011}
{Kashikawa}, N., {Shimasaku}, K., {Matsuda}, Y., {$et~al$.} 2011, \apj, 734,
  119

\bibitem[{{Kazemi} {$et~al$.}(2011){Kazemi}, {Yatawatta}, {Zaroubi},
  {Lampropoulos}, {de Bruyn}, {Koopmans}, \& {Noordam}}]{2011MNRAS.414.1656K}
{Kazemi}, S., {Yatawatta}, S., {Zaroubi}, S., {$et~al$.} 2011, \mnras, 414,
  1656

\bibitem[{{Keating} {$et~al$.}(2016){Keating}, {Marrone}, {Bower}, {Leitch},
  {Carlstrom}, \& {DeBoer}}]{Keating+2016}
{Keating}, G.~K., {Marrone}, D.~P., {Bower}, G.~C., {$et~al$.} 2016, \apj, 830,
  34

\bibitem[{{Keating} {$et~al$.}(2015){Keating}, {Bower}, {Marrone}, {DeBoer},
  {Heiles}, {Chang}, {Carlstrom}, {Greer}, {Hawkins}, {Lamb}, {Leitch},
  {Miller}, {Muchovej}, \& {Woody}}]{Keating+2015}
{Keating}, G.~K., {Bower}, G.~C., {Marrone}, D.~P., {$et~al$.} 2015, \apj, 814,
  140

\bibitem[{{Kern} {$et~al$.}(2020{\natexlab{a}}){Kern}, {Dillon}, {Parsons},
  {Carilli}, {Bernardi}, {Abdurashidova}, {Aguirre}, {Alexander}, {Ali},
  {Balfour}, {Beardsley}, {Billings}, {Bowman}, {Bradley}, {Bull}, {Burba},
  {Carey}, {Cheng}, {DeBoer}, {Dexter}, {de Lera Acedo}, {Ely}, {Ewall-Wice},
  {Fagnoni}, {Fritz}, {Furlanetto}, {Gale-Sides}, {Glendenning}, {Gorthi},
  {Greig}, {Grobbelaar}, {Halday}, {Hazelton}, {Hewitt}, {Hickish}, {Jacobs},
  {Julius}, {Kerrigan}, {Kittiwisit}, {Kohn}, {Kolopanis}, {Lanman}, {La
  Plante}, {Lekalake}, {Liu}, {MacMahon}, {Malan}, {Malgas}, {Maree},
  {Martinot}, {Matsetela}, {Mesinger}, {Molewa}, {Morales}, {Mosiane},
  {Murray}, {Neben}, {Nikolic}, {Nunhokee}, {Patra}, {Pieterse}, {Pober},
  {Razavi-Ghods}, {Ringuette}, {Robnett}, {Rosie}, {Sims}, {Smith}, {Syce},
  {Thyagarajan}, {Williams}, \& {Zheng}}]{2020ApJ...890..122K}
{Kern}, N.~S., {Dillon}, J.~S., {Parsons}, A.~R., {$et~al$.}
  2020{\natexlab{a}}, \apj, 890, 122

\bibitem[{{Kern} {$et~al$.}(2020{\natexlab{b}}){Kern}, {Parsons}, {Dillon},
  {Lanman}, {Liu}, {Bull}, {Ewall-Wice}, {Abdurashidova}, {Aguirre},
  {Alexander}, {Ali}, {Balfour}, {Beardsley}, {Bernardi}, {Bowman}, {Bradley},
  {Burba}, {Carilli}, {Cheng}, {DeBoer}, {Dexter}, {de Lera Acedo}, {Fagnoni},
  {Fritz}, {Furlanetto}, {Glendenning}, {Gorthi}, {Greig}, {Grobbelaar},
  {Halday}, {Hazelton}, {Hewitt}, {Hickish}, {Jacobs}, {Julius}, {Kerrigan},
  {Kittiwisit}, {Kohn}, {Kolopanis}, {La Plante}, {Lekalake}, {MacMahon},
  {Malan}, {Malgas}, {Maree}, {Martinot}, {Matsetela}, {Mesinger}, {Molewa},
  {Morales}, {Mosiane}, {Murray}, {Neben}, {Parsons}, {Patra}, {Pieterse},
  {Pober}, {Razavi-Ghods}, {Ringuette}, {Robnett}, {Rosie}, {Sims}, {Smith},
  {Syce}, {Thyagarajan}, {Williams}, \& {Zheng}}]{2020ApJ...888...70K}
{Kern}, N.~S., {Parsons}, A.~R., {Dillon}, J.~S., {$et~al$.}
  2020{\natexlab{b}}, \apj, 888, 70

\bibitem[{{Kerrigan} {$et~al$.}(2018){Kerrigan}, {Pober}, {Ali}, {Cheng},
  {Beardsley}, {Parsons}, {Aguirre}, {Barry}, {Bradley}, {Bernardi}, {Carilli},
  {DeBoer}, {Dillon}, {Jacobs}, {Kohn}, {Kolopanis}, {Lanman}, {Li}, {Liu}, \&
  {Sullivan}}]{Kerrigan+2018}
{Kerrigan}, J.~R., {Pober}, J.~C., {Ali}, Z.~S., {$et~al$.} 2018, \apj, 864,
  131

\bibitem[{{Khusanova} {$et~al$.}(2021){Khusanova}, {Bethermin}, {Le F{\`e}vre},
  {Capak}, {Faisst}, {Schaerer}, {Silverman}, {Cassata}, {Yan}, {Ginolfi},
  {Fudamoto}, {Loiacono}, {Amorin}, {Bardelli}, {Boquien}, {Cimatti},
  {Dessauges-Zavadsky}, {Gruppioni}, {Hathi}, {Jones}, {Koekemoer}, {Lagache},
  {Maiolino}, {Lemaux}, {Oesch}, {Pozzi}, {Riechers}, {Romano}, {Talia},
  {Toft}, {Vergani}, {Zamorani}, \& {Zucca}}]{Khusanova+2021}
{Khusanova}, Y., {Bethermin}, M., {Le F{\`e}vre}, O., {$et~al$.} 2021, \aap,
  649, A152

\bibitem[{{Komatsu} {$et~al$.}(2011){Komatsu}, {Smith}, {Dunkley}, {Bennett},
  {Gold}, {Hinshaw}, {Jarosik}, {Larson}, {Nolta}, {Page}, {Spergel},
  {Halpern}, {Hill}, {Kogut}, {Limon}, {Meyer}, {Odegard}, {Tucker}, {Weiland},
  {Wollack}, \& {Wright}}]{komatsu+2011}
{Komatsu}, E., {Smith}, K.~M., {Dunkley}, J., {$et~al$.} 2011, \apjs, 192, 18

\bibitem[{{Konno} {$et~al$.}(2014){Konno}, {Ouchi}, {Ono}, {Shimasaku},
  {Shibuya}, {Furusawa}, {Nakajima}, {Naito}, {Momose}, {Yuma}, \&
  {Iye}}]{Konno+2014}
{Konno}, A., {Ouchi}, M., {Ono}, Y., {$et~al$.} 2014, \apj, 797, 16

\bibitem[{{Koopmans} {$et~al$.}(2015){Koopmans}, {Pritchard}, {Mellema},
  {Aguirre}, {Ahn}, {Barkana}, {van Bemmel}, {Bernardi}, {Bonaldi}, {Briggs},
  {de Bruyn}, {Chang}, {Chapman}, {Chen}, {Ciardi}, {Dayal}, {Ferrara},
  {Fialkov}, {Fiore}, {Ichiki}, {Illiev}, {Inoue}, {Jelic}, {Jones}, {Lazio},
  {Maio}, {Majumdar}, {Mack}, {Mesinger}, {Morales}, {Parsons}, {Pen},
  {Santos}, {Schneider}, {Semelin}, {de Souza}, {Subrahmanyan}, {Takeuchi},
  {Vedantham}, {Wagg}, {Webster}, {Wyithe}, {Datta}, \&
  {Trott}}]{koopmans+2015}
{Koopmans}, L., {Pritchard}, J., {Mellema}, G., {$et~al$.} 2015, in Advancing
  Astrophysics with the Square Kilometre Array (AASKA14), 1

\bibitem[{{Kovetz} {$et~al$.}(2019){Kovetz}, {Breysse}, {Lidz}, {Bock},
  {Bradford}, {Chang}, {Foreman}, {Padmanabhan}, {Pullen}, {Riechers}, {Silva},
  \& {Switzer}}]{Kovetz+2019}
{Kovetz}, E., {Breysse}, P.~C., {Lidz}, A., {$et~al$.} 2019, \baas, 51, 101

\bibitem[{Leung {$et~al$.}(2020)Leung, Olsen, Somerville, Dav{\'{e}}, Greve,
  Hayward, Narayanan, \& Popping}]{Daisy_Leung_2020}
Leung, T. K.~D., Olsen, K.~P., Somerville, R.~S., {$et~al$.} 2020, \apj, 905,
  102

\bibitem[{{Li} {$et~al$.}(2016){Li}, {Wechsler}, {Devaraj}, \&
  {Church}}]{Li+2016}
{Li}, T.~Y., {Wechsler}, R.~H., {Devaraj}, K., \& {Church}, S.~E. 2016, \apj,
  817, 169

\bibitem[{{Li} {$et~al$.}(2019){Li}, {Pober}, {Barry}, {Hazelton}, {Morales},
  {Trott}, {Lanman}, {Wilensky}, {Sullivan}, {Beardsley}, {Booler}, {Bowman},
  {Byrne}, {Crosse}, {Emrich}, {Franzen}, {Hasegawa}, {Horsley},
  {Johnston-Hollitt}, {Jacobs}, {Jordan}, {Joseph}, {Kaneuji}, {Kaplan},
  {Kenney}, {Kubota}, {Line}, {Lynch}, {McKinley}, {Mitchell}, {Murray},
  {Pallot}, {Pindor}, {Rahimi}, {Riding}, {Sleap}, {Steele}, {Takahashi},
  {Tingay}, {Walker}, {Wayth}, {Webster}, {Williams}, {Wu}, {Wyithe},
  {Yoshiura}, \& {Zheng}}]{2019ApJ...887..141L}
{Li}, W., {Pober}, J.~C., {Barry}, N., {$et~al$.} 2019, \apj, 887, 141

\bibitem[{{Lidz} {$et~al$.}(2011){Lidz}, {Furlanetto}, {Oh}, {Aguirre},
  {Chang}, {Dor{\'e}}, \& {Pritchard}}]{Lidz+2011}
{Lidz}, A., {Furlanetto}, S.~R., {Oh}, S.~P., {$et~al$.} 2011, \apj, 741, 70

\bibitem[{{Lidz} \& {Taylor}(2016)}]{Lidz_and_Taylor+2016}
{Lidz}, A., \& {Taylor}, J. 2016, \apj, 825, 143

\bibitem[{{Loiacono} {$et~al$.}(2021){Loiacono}, {Decarli}, {Gruppioni},
  {Talia}, {Cimatti}, {Zamorani}, {Pozzi}, {Yan}, {Lemaux}, {Riechers}, {Le
  F{\`e}vre}, {B{\`e}thermin}, {Capak}, {Cassata}, {Faisst}, {Schaerer},
  {Silverman}, {Bardelli}, {Boquien}, {Burkutean}, {Dessauges-Zavadsky},
  {Fudamoto}, {Fujimoto}, {Ginolfi}, {Hathi}, {Jones}, {Khusanova},
  {Koekemoer}, {Lagache}, {Lubin}, {Massardi}, {Oesch}, {Romano}, {Vallini},
  {Vergani}, \& {Zucca}}]{Loiacono+2021}
{Loiacono}, F., {Decarli}, R., {Gruppioni}, C., {$et~al$.} 2021, Astronomy \&
  Astrophysics, 646, A76

\bibitem[{{Lonsdale} {$et~al$.}(2009){Lonsdale}, {Cappallo}, {Morales},
  {Briggs}, {Benkevitch}, {Bowman}, {Bunton}, {Burns}, {Corey}, {Desouza},
  {Doeleman}, {Derome}, {Deshpande}, {Gopala}, {Greenhill}, {Herne}, {Hewitt},
  {Kamini}, {Kasper}, {Kincaid}, {Kocz}, {Kowald}, {Kratzenberg}, {Kumar},
  {Lynch}, {Madhavi}, {Matejek}, {Mitchell}, {Morgan}, {Oberoi}, {Ord},
  {Pathikulangara}, {Prabu}, {Rogers}, {Roshi}, {Salah}, {Sault}, {Shankar},
  {Srivani}, {Stevens}, {Tingay}, {Vaccarella}, {Waterson}, {Wayth}, {Webster},
  {Whitney}, {Williams}, \& {Williams}}]{Lonsdale+2009}
{Lonsdale}, C.~J., {Cappallo}, R.~J., {Morales}, M.~F., {$et~al$.} 2009, IEEE
  Proceedings, 97, 1497

\bibitem[{{Madau} {$et~al$.}(1997){Madau}, {Meiksin}, \& {Rees}}]{Madau+1997}
{Madau}, P., {Meiksin}, A., \& {Rees}, M.~J. 1997, \apj, 475, 429

\bibitem[{{Maity} \& {Choudhury}(2022)}]{Maity+2022}
{Maity}, B., \& {Choudhury}, T.~R. 2022, \mnras, 511, 2239

\bibitem[{{Majumdar} {$et~al$.}(2012){Majumdar}, {Bharadwaj}, \&
  {Choudhury}}]{Majumdar+2012}
{Majumdar}, S., {Bharadwaj}, S., \& {Choudhury}, T.~R. 2012, \mnras, 426, 3178

\bibitem[{{Majumdar} {$et~al$.}(2013){Majumdar}, {Bharadwaj}, \&
  {Choudhury}}]{Majumdar+2013}
---. 2013, \mnras, 434, 1978

\bibitem[{{Majumdar} {$et~al$.}(2020){Majumdar}, {Kamran}, {Pritchard},
  {Mondal}, {Mazumdar}, {Bharadwaj}, \& {Mellema}}]{Majumdar+2020}
{Majumdar}, S., {Kamran}, M., {Pritchard}, J.~R., {$et~al$.} 2020, \mnras, 499,
  5090

\bibitem[{{Majumdar} {$et~al$.}(2014){Majumdar}, {Mellema}, {Datta}, {Jensen},
  {Choudhury}, {Bharadwaj}, \& {Friedrich}}]{Majumdar+2014}
{Majumdar}, S., {Mellema}, G., {Datta}, K.~K., {$et~al$.} 2014, \mnras, 443,
  2843

\bibitem[{{Majumdar} {$et~al$.}(2018){Majumdar}, {Pritchard}, {Mondal},
  {Watkinson}, {Bharadwaj}, \& {Mellema}}]{Majumdar+2018}
{Majumdar}, S., {Pritchard}, J.~R., {Mondal}, R., {$et~al$.} 2018, \mnras, 476,
  4007

\bibitem[{{Majumdar} {$et~al$.}(2016){Majumdar}, {Jensen}, {Mellema},
  {Chapman}, {Abdalla}, {Lee}, {Iliev}, {Dixon}, {Datta}, {Ciardi},
  {Fernandez}, {Jeli{\'c}}, {Koopmans}, \& {Zaroubi}}]{Majumdar+2016}
{Majumdar}, S., {Jensen}, H., {Mellema}, G., {$et~al$.} 2016, \mnras, 456, 2080

\bibitem[{{Mangena} {$et~al$.}(2020){Mangena}, {Hassan}, \&
  {Santos}}]{Mangena+2020}
{Mangena}, T., {Hassan}, S., \& {Santos}, M.~G. 2020, \mnras, 494, 600

\bibitem[{{Mashian} {$et~al$.}(2015){Mashian}, {Sternberg}, \&
  {Loeb}}]{Mashian+2015}
{Mashian}, N., {Sternberg}, A., \& {Loeb}, A. 2015, \jcap, 2015, 028

\bibitem[{{Matthee} {$et~al$.}(2015){Matthee}, {Sobral}, {Santos},
  {R{\"o}ttgering}, {Darvish}, \& {Mobasher}}]{Matthee+2015}
{Matthee}, J., {Sobral}, D., {Santos}, S., {$et~al$.} 2015, \mnras, 451, 400

\bibitem[{{Mellema} {$et~al$.}(2006){Mellema}, {Iliev}, {Pen}, \&
  {Shapiro}}]{Mellema+2006}
{Mellema}, G., {Iliev}, I.~T., {Pen}, U.-L., \& {Shapiro}, P.~R. 2006, \mnras,
  372, 679

\bibitem[{{Mellema} {$et~al$.}(2015){Mellema}, {Koopmans}, {Shukla}, {Datta},
  {Mesinger}, \& {Majumdar}}]{mellema+2015}
{Mellema}, G., {Koopmans}, L., {Shukla}, H., {$et~al$.} 2015, in Advancing
  Astrophysics with the Square Kilometre Array (AASKA14), 10

\bibitem[{{Mercier, C.} {$et~al$.}(2006){Mercier, C.}, {Subramanian, P.},
  {Kerdraon, A.}, {Pick, M.}, {Ananthakrishnan, S.}, \& {Janardhan,
  P.}}]{Mercier+2006}
{Mercier, C.}, {Subramanian, P.}, {Kerdraon, A.}, {$et~al$.} 2006, A\&A, 447,
  1189

\bibitem[{{Mertens} {$et~al$.}(2018){Mertens}, {Ghosh}, \&
  {Koopmans}}]{Mertens+2018}
{Mertens}, F.~G., {Ghosh}, A., \& {Koopmans}, L.~V.~E. 2018, \mnras, 478, 3640

\bibitem[{{Mertens} {$et~al$.}(2020){Mertens}, {Mevius}, {Koopmans},
  {Offringa}, {Mellema}, {Zaroubi}, {Brentjens}, {Gan}, {Gehlot}, {Pand ey},
  {Sardarabadi}, {Vedantham}, {Yatawatta}, {Asad}, {Ciardi}, {Chapman},
  {Gazagnes}, {Ghara}, {Ghosh}, {Giri}, {Iliev}, {Jeli{\'c}}, {Kooistra},
  {Mondal}, {Schaye}, \& {Silva}}]{2020MNRAS.493.1662M}
{Mertens}, F.~G., {Mevius}, M., {Koopmans}, L.~V.~E., {$et~al$.} 2020, \mnras,
  493, 1662

\bibitem[{{Mesinger} \& {Furlanetto}(2007)}]{Mesinger+2007}
{Mesinger}, A., \& {Furlanetto}, S. 2007, \apj, 669, 663

\bibitem[{{Mesinger} {$et~al$.}(2011){Mesinger}, {Furlanetto}, \&
  {Cen}}]{Mesinger+2011}
{Mesinger}, A., {Furlanetto}, S., \& {Cen}, R. 2011, \mnras, 411, 955

\bibitem[{{Mevius} {$et~al$.}(2022){Mevius}, {Mertens}, {Koopmans}, {Offringa},
  {Yatawatta}, {Brentjens}, {Chapman}, {Ciardi}, {Gan}, {Gehlot}, {Ghara},
  {Ghosh}, {Giri}, {Iliev}, {Mellema}, {Pandey}, \&
  {Zaroubi}}]{2022MNRAS.509.3693M}
{Mevius}, M., {Mertens}, F., {Koopmans}, L.~V.~E., {$et~al$.} 2022, \mnras,
  509, 3693

\bibitem[{Mitchell {$et~al$.}(2008)Mitchell, Greenhill, Wayth, Sault, Lonsdale,
  Cappallo, Morales, \& Ord}]{RTC4703504}
Mitchell, D.~A., Greenhill, L.~J., Wayth, R.~B., {$et~al$.} 2008, IEEE Journal
  of Selected Topics in Signal Processing, 2, 707

\bibitem[{Mitra {$et~al$.}(2011)Mitra, Choudhury, \& Ferrara}]{Mitra+2011}
Mitra, S., Choudhury, T.~R., \& Ferrara, A. 2011, \mnras, 413, 1569

\bibitem[{{Mitra} {$et~al$.}(2015){Mitra}, {Choudhury}, \&
  {Ferrara}}]{mitra+2015}
{Mitra}, S., {Choudhury}, T.~R., \& {Ferrara}, A. 2015, \mnras, 454, L76

\bibitem[{{Mitra} {$et~al$.}(2013){Mitra}, {Ferrara}, \&
  {Choudhury}}]{mitra+2013}
{Mitra}, S., {Ferrara}, A., \& {Choudhury}, T.~R. 2013, \mnras, 428, L1

\bibitem[{{Mondal} {$et~al$.}(2018){Mondal}, {Bharadwaj}, \&
  {Datta}}]{Mondal+2018}
{Mondal}, R., {Bharadwaj}, S., \& {Datta}, K.~K. 2018, \mnras, 474, 1390

\bibitem[{{Mondal} {$et~al$.}(2019){Mondal}, {Bharadwaj}, {Iliev}, {Datta},
  {Majumdar}, {Shaw}, \& {Sarkar}}]{Mondal+2019}
{Mondal}, R., {Bharadwaj}, S., {Iliev}, I.~T., {$et~al$.} 2019, \mnras, 483,
  L109

\bibitem[{{Mondal} {$et~al$.}(2016){Mondal}, {Bharadwaj}, \&
  {Majumdar}}]{Mondal+2016}
{Mondal}, R., {Bharadwaj}, S., \& {Majumdar}, S. 2016, \mnras, 456, 1936

\bibitem[{{Mondal} {$et~al$.}(2015){Mondal}, {Bharadwaj}, {Majumdar}, {Bera},
  \& {Acharyya}}]{Mondal+2015}
{Mondal}, R., {Bharadwaj}, S., {Majumdar}, S., {Bera}, A., \& {Acharyya}, A.
  2015, \mnras, 449, L41

\bibitem[{{Mondal} {$et~al$.}(2021){Mondal}, {Mellema}, {Shaw}, {Kamran}, \&
  {Majumdar}}]{Mondal+2021}
{Mondal}, R., {Mellema}, G., {Shaw}, A.~K., {Kamran}, M., \& {Majumdar}, S.
  2021, \mnras, 508, 3848

\bibitem[{{Mondal} {$et~al$.}(2020){Mondal}, {Fialkov}, {Fling}, {Iliev},
  {Barkana}, {Ciardi}, {Mellema}, {Zaroubi}, {Koopmans}, {Mertens}, {Gehlot},
  {Ghara}, {Ghosh}, {Giri}, {Offringa}, \& {Pandey}}]{2020MNRAS.498.4178M}
{Mondal}, R., {Fialkov}, A., {Fling}, C., {$et~al$.} 2020, \mnras, 498, 4178

\bibitem[{{Moradinezhad Dizgah} \& {Keating}(2019)}]{Moradinezhad_Dizgah+2019}
{Moradinezhad Dizgah}, A., \& {Keating}, G.~K. 2019, \apj, 872, 126

\bibitem[{{Moradinezhad Dizgah} {$et~al$.}(2022{\natexlab{a}}){Moradinezhad
  Dizgah}, {Keating}, {Karkare}, {Crites}, \&
  {Choudhury}}]{Moradinezhad_Dizgah+2022a}
{Moradinezhad Dizgah}, A., {Keating}, G.~K., {Karkare}, K.~S., {Crites}, A., \&
  {Choudhury}, S.~R. 2022{\natexlab{a}}, \apj, 926, 137

\bibitem[{{Moradinezhad Dizgah} {$et~al$.}(2022{\natexlab{b}}){Moradinezhad
  Dizgah}, {Nikakhtar}, {Keating}, \& {Castorina}}]{Moradinezhad_Dizgah+2022b}
{Moradinezhad Dizgah}, A., {Nikakhtar}, F., {Keating}, G.~K., \& {Castorina},
  E. 2022{\natexlab{b}}, \jcap, 2022, 026

\bibitem[{{Mu{\~n}oz} \& {Loeb}(2018)}]{2018Natur.557..684M}
{Mu{\~n}oz}, J.~B., \& {Loeb}, A. 2018, \nat, 557, 684

\bibitem[{{Murmu} {$et~al$.}(2021{\natexlab{a}}){Murmu}, {Majumdar}, \&
  {Datta}}]{Murmu+2021}
{Murmu}, C.~S., {Majumdar}, S., \& {Datta}, K.~K. 2021{\natexlab{a}}, \mnras,
  507, 2500

\bibitem[{{Murmu} {$et~al$.}(2021{\natexlab{b}}){Murmu}, {Olsen}, {Greve},
  {Majumdar}, {Datta}, {Scott}, {Leung}, {Dave}, {Popping}, {Ortega Ochoa},
  {Vizgan}, \& {Narayanan}}]{Murmu+2021b}
{Murmu}, C.~S., {Olsen}, K.~P., {Greve}, T.~R., {$et~al$.} 2021{\natexlab{b}},
  arXiv e-prints, arXiv:2110.10687

\bibitem[{{Nambissan T.} {$et~al$.}(2021){Nambissan T.}, {Subrahmanyan},
  {Somashekar}, {Udaya Shankar}, {Singh}, {Raghunathan}, {Girish}, {Srivani},
  \& {Sathyanarayana Rao}}]{2021arXiv210401756N}
{Nambissan T.}, J., {Subrahmanyan}, R., {Somashekar}, R., {$et~al$.} 2021,
  arXiv e-prints, arXiv:2104.01756

\bibitem[{{Olsen} {$et~al$.}(2017){Olsen}, {Greve}, {Narayanan}, {Thompson},
  {Dav{\'e}}, {Niebla Rios}, \& {Stawinski}}]{Olsen+2017}
{Olsen}, K., {Greve}, T.~R., {Narayanan}, D., {$et~al$.} 2017, \apj, 846, 105

\bibitem[{{Olsen} {$et~al$.}(2016){Olsen}, {Greve}, {Brinch}, {Sommer-Larsen},
  {Rasmussen}, {Toft}, \& {Zirm}}]{Olsen+2016}
{Olsen}, K.~P., {Greve}, T.~R., {Brinch}, C., {$et~al$.} 2016, \mnras, 457,
  3306

\bibitem[{Olsen {$et~al$.}(2015)Olsen, Greve, Narayanan, Thompson, Toft, \&
  Brinch}]{Olsen+2015}
Olsen, K.~P., Greve, T.~R., Narayanan, D., {$et~al$.} 2015, \apj, 814, 76

\bibitem[{{Ota} {$et~al$.}(2017){Ota}, {Iye}, {Kashikawa}, {Konno}, {Nakata},
  {Totani}, {Kobayashi}, {Fudamoto}, {Seko}, {Toshikawa}, {Ichikawa},
  {Shibuya}, \& {Onoue}}]{Ota+2017}
{Ota}, K., {Iye}, M., {Kashikawa}, N., {$et~al$.} 2017, \apj, 844, 85

\bibitem[{{Ouchi} {$et~al$.}(2010){Ouchi}, {Shimasaku}, {Furusawa}, {Saito},
  {Yoshida}, {Akiyama}, {Ono}, {Yamada}, {Ota}, {Kashikawa}, {Iye}, {Kodama},
  {Okamura}, {Simpson}, \& {Yoshida}}]{ouchi+2010}
{Ouchi}, M., {Shimasaku}, K., {Furusawa}, H., {$et~al$.} 2010, \apj, 723, 869

\bibitem[{{Paciga} {$et~al$.}(2011){Paciga}, {Chang}, {Gupta}, {Nityanada},
  {Odegova}, {Pen}, {Peterson}, {Roy}, \&
  {Sigurdson}}]{Paciga2011MNRAS.413.1174P}
{Paciga}, G., {Chang}, T.-C., {Gupta}, Y., {$et~al$.} 2011, \mnras, 413, 1174

\bibitem[{{Paciga} {$et~al$.}(2013){Paciga}, {Albert}, {Bandura}, {Chang},
  {Gupta}, {Hirata}, {Odegova}, {Pen}, {Peterson}, {Roy}, {Shaw}, {Sigurdson},
  \& {Voytek}}]{Paciga+2013}
{Paciga}, G., {Albert}, J.~G., {Bandura}, K., {$et~al$.} 2013, \mnras, 433, 639

\bibitem[{{Padmanabhan}(2018)}]{Padmanabhan+2018}
{Padmanabhan}, H. 2018, \mnras, 475, 1477

\bibitem[{{Padmanabhan}(2019)}]{Padmanabhan+2019a}
---. 2019, \mnras, 488, 3014

\bibitem[{{Paranjape} \& {Choudhury}(2014)}]{Paranjape+2014}
{Paranjape}, A., \& {Choudhury}, T.~R. 2014, \mnras, 442, 1470

\bibitem[{{Paranjape} {$et~al$.}(2016){Paranjape}, {Choudhury}, \&
  {Padmanabhan}}]{Paranjape+2016}
{Paranjape}, A., {Choudhury}, T.~R., \& {Padmanabhan}, H. 2016, \mnras, 460,
  1801

\bibitem[{{Paranjape} \& {Sheth}(2012)}]{Paranjape+2012}
{Paranjape}, A., \& {Sheth}, R.~K. 2012, \mnras, 426, 2789

\bibitem[{{Pathak} {$et~al$.}(2022){Pathak}, {Bag}, {Majumdar}, {Mondal},
  {Kamran}, \& {Sarkar}}]{Pathak+2022}
{Pathak}, A., {Bag}, S., {Majumdar}, S., {$et~al$.} 2022, arXiv e-prints,
  arXiv:2202.03701

\bibitem[{{Patil} {$et~al$.}(2017){Patil}, {Yatawatta}, {Koopmans}, {de Bruyn},
  {Brentjens}, {Zaroubi}, {Asad}, {Hatef}, {Jeli{\'c}}, {Mevius}, {Offringa},
  {Pandey}, {Vedantham}, {Abdalla}, {Brouw}, {Chapman}, {Ciardi}, {Gehlot},
  {Ghosh}, {Harker}, {Iliev}, {Kakiichi}, {Majumdar}, {Mellema}, {Silva},
  {Schaye}, {Vrbanec}, \& {Wijnholds}}]{2017ApJ...838...65P}
{Patil}, A.~H., {Yatawatta}, S., {Koopmans}, L.~V.~E., {$et~al$.} 2017, \apj,
  838, 65

\bibitem[{{Patra} {$et~al$.}(2013){Patra}, {Subrahmanyan}, {Raghunathan}, \&
  {Udaya Shankar}}]{2013ExA....36..319P}
{Patra}, N., {Subrahmanyan}, R., {Raghunathan}, A., \& {Udaya Shankar}, N.
  2013, Experimental Astronomy, 36, 319

\bibitem[{{Patra} {$et~al$.}(2015){Patra}, {Subrahmanyan}, {Sethi}, {Udaya
  Shankar}, \& {Raghunathan}}]{2015ApJ...801..138P}
{Patra}, N., {Subrahmanyan}, R., {Sethi}, S., {Udaya Shankar}, N., \&
  {Raghunathan}, A. 2015, \apj, 801, 138

\bibitem[{{Planck Collaboration} {$et~al$.}(2020){Planck Collaboration},
  {Aghanim}, {Akrami}, {Ashdown}, {Aumont}, {Baccigalupi}, {Ballardini},
  {Banday}, {Barreiro}, {Bartolo}, {Basak}, {Battye}, {Benabed}, {Bernard},
  {Bersanelli}, {Bielewicz}, {Bock}, {Bond}, {Borrill}, {Bouchet}, {Boulanger},
  {Bucher}, {Burigana}, {Butler}, {Calabrese}, {Cardoso}, {Carron},
  {Challinor}, {Chiang}, {Chluba}, {Colombo}, {Combet}, {Contreras}, {Crill},
  {Cuttaia}, {de Bernardis}, {de Zotti}, {Delabrouille}, {Delouis}, {Di
  Valentino}, {Diego}, {Dor{\'e}}, {Douspis}, {Ducout}, {Dupac}, {Dusini},
  {Efstathiou}, {Elsner}, {En{\ss}lin}, {Eriksen}, {Fantaye}, {Farhang},
  {Fergusson}, {Fernandez-Cobos}, {Finelli}, {Forastieri}, {Frailis},
  {Fraisse}, {Franceschi}, {Frolov}, {Galeotta}, {Galli}, {Ganga},
  {G{\'e}nova-Santos}, {Gerbino}, {Ghosh}, {Gonz{\'a}lez-Nuevo}, {G{\'o}rski},
  {Gratton}, {Gruppuso}, {Gudmundsson}, {Hamann}, {Handley}, {Hansen},
  {Herranz}, {Hildebrandt}, {Hivon}, {Huang}, {Jaffe}, {Jones}, {Karakci},
  {Keih{\"a}nen}, {Keskitalo}, {Kiiveri}, {Kim}, {Kisner}, {Knox},
  {Krachmalnicoff}, {Kunz}, {Kurki-Suonio}, {Lagache}, {Lamarre}, {Lasenby},
  {Lattanzi}, {Lawrence}, {Le Jeune}, {Lemos}, {Lesgourgues}, {Levrier},
  {Lewis}, {Liguori}, {Lilje}, {Lilley}, {Lindholm}, {L{\'o}pez-Caniego},
  {Lubin}, {Ma}, {Mac{\'\i}as-P{\'e}rez}, {Maggio}, {Maino}, {Mandolesi},
  {Mangilli}, {Marcos-Caballero}, {Maris}, {Martin}, {Martinelli},
  {Mart{\'\i}nez-Gonz{\'a}lez}, {Matarrese}, {Mauri}, {McEwen}, {Meinhold},
  {Melchiorri}, {Mennella}, {Migliaccio}, {Millea}, {Mitra},
  {Miville-Desch{\^e}nes}, {Molinari}, {Montier}, {Morgante}, {Moss}, {Natoli},
  {N{\o}rgaard-Nielsen}, {Pagano}, {Paoletti}, {Partridge}, {Patanchon},
  {Peiris}, {Perrotta}, {Pettorino}, {Piacentini}, {Polastri}, {Polenta},
  {Puget}, {Rachen}, {Reinecke}, {Remazeilles}, {Renzi}, {Rocha}, {Rosset},
  {Roudier}, {Rubi{\~n}o-Mart{\'\i}n}, {Ruiz-Granados}, {Salvati}, {Sandri},
  {Savelainen}, {Scott}, {Shellard}, {Sirignano}, {Sirri}, {Spencer},
  {Sunyaev}, {Suur-Uski}, {Tauber}, {Tavagnacco}, {Tenti}, {Toffolatti},
  {Tomasi}, {Trombetti}, {Valenziano}, {Valiviita}, {Van Tent}, {Vibert},
  {Vielva}, {Villa}, {Vittorio}, {Wandelt}, {Wehus}, {White}, {White},
  {Zacchei}, \& {Zonca}}]{planck+2020}
{Planck Collaboration}, {Aghanim}, N., {Akrami}, Y., {$et~al$.} 2020, \aap,
  641, A6

\bibitem[{{Pritchard} \& {Loeb}(2012)}]{pritchard+2012}
{Pritchard}, J.~R., \& {Loeb}, A. 2012, Reports on Progress in Physics, 75,
  086901

\bibitem[{{Pullen} {$et~al$.}(2013){Pullen}, {Chang}, {Dor{\'e}}, \&
  {Lidz}}]{Pullen+2013}
{Pullen}, A.~R., {Chang}, T.-C., {Dor{\'e}}, O., \& {Lidz}, A. 2013, \apj, 768,
  15

\bibitem[{{Rahimi} {$et~al$.}(2021){Rahimi}, {Pindor}, {Line}, {Barry},
  {Trott}, {Webster}, {Jordan}, {Wilensky}, {Yoshiura}, {Beardsley}, {Bowman},
  {Byrne}, {Chokshi}, {Hazelton}, {Hasegawa}, {Howard}, {Greig}, {Jacobs},
  {Joseph}, {Kolopanis}, {Lynch}, {McKinley}, {Mitchell}, {Murray}, {Morales},
  {Pober}, {Takahashi}, {Tingay}, {Wayth}, {Wyithe}, \&
  {Zheng}}]{2021MNRAS.508.5954R}
{Rahimi}, M., {Pindor}, B., {Line}, J.~L.~B., {$et~al$.} 2021, \mnras, 508,
  5954

\bibitem[{{Robertson} {$et~al$.}(2015){Robertson}, {Ellis}, {Furlanetto}, \&
  {Dunlop}}]{robertson+2015}
{Robertson}, B.~E., {Ellis}, R.~S., {Furlanetto}, S.~R., \& {Dunlop}, J.~S.
  2015, \apjl, 802, L19

\bibitem[{{Sadoun} {$et~al$.}(2017){Sadoun}, {Zheng}, \&
  {Miralda-Escud{\'e}}}]{Sadoun+2017}
{Sadoun}, R., {Zheng}, Z., \& {Miralda-Escud{\'e}}, J. 2017, \apj, 839, 44

\bibitem[{{Santos} {$et~al$.}(2005){Santos}, {Cooray}, \& {Knox}}]{Santos+2005}
{Santos}, M.~G., {Cooray}, A., \& {Knox}, L. 2005, \apj, 625, 575

\bibitem[{Santos {$et~al$.}(2010)Santos, Ferramacho, Silva, Amblard, \&
  Cooray}]{Santos+2010}
Santos, M.~G., Ferramacho, L., Silva, M.~B., Amblard, A., \& Cooray, A. 2010,
  \mnras, 406, 2421

\bibitem[{{Saxena} {$et~al$.}(2020){Saxena}, {Majumdar}, {Kamran}, \&
  {Viel}}]{Saxena+2020}
{Saxena}, A., {Majumdar}, S., {Kamran}, M., \& {Viel}, M. 2020, \mnras, 497,
  2941

\bibitem[{{Schmit} \& {Pritchard}(2018)}]{Schmit+2018}
{Schmit}, C.~J., \& {Pritchard}, J.~R. 2018, \mnras, 475, 1213

\bibitem[{Serra {$et~al$.}(2016)Serra, Dor{\'{e}}, \& Lagache}]{Serra+2016}
Serra, P., Dor{\'{e}}, O., \& Lagache, G. 2016, \apj, 833, 153

\bibitem[{{Shaw} {$et~al$.}(2019){Shaw}, {Bharadwaj}, \& {Mondal}}]{Shaw+2019}
{Shaw}, A.~K., {Bharadwaj}, S., \& {Mondal}, R. 2019, \mnras, 487, 4951

\bibitem[{{Shibuya} {$et~al$.}(2019){Shibuya}, {Ouchi}, {Harikane}, \&
  {Nakajima}}]{Shibuya+2019}
{Shibuya}, T., {Ouchi}, M., {Harikane}, Y., \& {Nakajima}, K. 2019, \apj, 871,
  164

\bibitem[{{Sikder} {$et~al$.}(2022){Sikder}, {Barkana}, {Reis}, \&
  {Fialkov}}]{Sikder+2022}
{Sikder}, S., {Barkana}, R., {Reis}, I., \& {Fialkov}, A. 2022, arXiv e-prints,
  arXiv:2201.08205

\bibitem[{Silva {$et~al$.}(2015)Silva, Santos, Cooray, \& Gong}]{Silva+2015}
Silva, M., Santos, M.~G., Cooray, A., \& Gong, Y. 2015, The Astrophysical
  Journal, 806, 209

\bibitem[{{Silva} {$et~al$.}(2013){Silva}, {Santos}, {Gong}, {Cooray}, \&
  {Bock}}]{silva13}
{Silva}, M.~B., {Santos}, M.~G., {Gong}, Y., {Cooray}, A., \& {Bock}, J. 2013,
  \apj, 763, 132

\bibitem[{{Silva} {$et~al$.}(2021){Silva}, {Baumschlager}, {Cleary}, {Breysse},
  {Chung}, {Ihle}, {Padmanabhan}, {Keating}, {Kim}, \& {Philip}}]{Silva+2021}
{Silva}, M.~B., {Baumschlager}, B., {Cleary}, K.~A., {$et~al$.} 2021, arXiv
  e-prints, arXiv:2111.05354

\bibitem[{{Singh} {$et~al$.}(2018){Singh}, {Subrahmanyan}, {Udaya Shankar},
  {Sathyanarayana Rao}, {Fialkov}, {Cohen}, {Barkana}, {Girish}, {Raghunathan},
  {Somashekar}, \& {Srivani}}]{Singh+2018}
{Singh}, S., {Subrahmanyan}, R., {Udaya Shankar}, N., {$et~al$.} 2018, \apj,
  858, 54

\bibitem[{{Singh} {$et~al$.}(2021){Singh}, {Nambissan T.}, {Subrahmanyan},
  {Udaya Shankar}, {Girish}, {Raghunathan}, {Somashekar}, {Srivani}, \&
  {Sathyanarayana Rao}}]{2021arXiv211206778S}
{Singh}, S., {Nambissan T.}, J., {Subrahmanyan}, R., {$et~al$.} 2021, arXiv
  e-prints, arXiv:2112.06778

\bibitem[{{Sobacchi} {$et~al$.}(2016){Sobacchi}, {Mesinger}, \&
  {Greig}}]{sobacchi16}
{Sobacchi}, E., {Mesinger}, A., \& {Greig}, B. 2016, \mnras, 459, 2741

\bibitem[{Steinhardt {$et~al$.}(2021)Steinhardt, Jespersen, \&
  Linzer}]{Steinhardt+2021}
Steinhardt, C.~L., Jespersen, C.~K., \& Linzer, N.~B. 2021, \apj, 923, 8

\bibitem[{{Sullivan} {$et~al$.}(2012){Sullivan}, {Morales}, {Hazelton},
  {Arcus}, {Barnes}, {Bernardi}, {Briggs}, {Bowman}, {Bunton}, {Cappallo},
  {Corey}, {Deshpande}, {deSouza}, {Emrich}, {Gaensler}, {Goeke}, {Greenhill},
  {Herne}, {Hewitt}, {Johnston-Hollitt}, {Kaplan}, {Kasper}, {Kincaid},
  {Koenig}, {Kratzenberg}, {Lonsdale}, {Lynch}, {McWhirter}, {Mitchell},
  {Morgan}, {Oberoi}, {Ord}, {Pathikulangara}, {Prabu}, {Remillard}, {Rogers},
  {Roshi}, {Salah}, {Sault}, {Udaya Shankar}, {Srivani}, {Stevens},
  {Subrahmanyan}, {Tingay}, {Wayth}, {Waterson}, {Webster}, {Whitney},
  {Williams}, {Williams}, \& {Wyithe}}]{2012ApJ...759...17S}
{Sullivan}, I.~S., {Morales}, M.~F., {Hazelton}, B.~J., {$et~al$.} 2012, \apj,
  759, 17

\bibitem[{{Sun} {$et~al$.}(2019){Sun}, {Hensley}, {Chang}, {Dor{\'e}}, \&
  {Serra}}]{Sun+2019}
{Sun}, G., {Hensley}, B.~S., {Chang}, T.-C., {Dor{\'e}}, O., \& {Serra}, P.
  2019, \apj, 887, 142

\bibitem[{Sun {$et~al$.}(2018)Sun, Moncelsi, Viero, Silva, Bock, Bradford,
  Chang, Cheng, Cooray, Crites, Hailey-Dunsheath, Uzgil, Hunacek, \&
  Zemcov}]{Sun+2018}
Sun, G., Moncelsi, L., Viero, M.~P., {$et~al$.} 2018, \apj, 856, 107

\bibitem[{{Sun} {$et~al$.}(2021){Sun}, {Chang}, {Uzgil}, {Bock}, {Bradford},
  {Butler}, {Caze-Cortes}, {Cheng}, {Cooray}, {Crites}, {Hailey-Dunsheath},
  {Emerson}, {Frez}, {Hoscheit}, {Hunacek}, {Keenan}, {Li}, {Madonia},
  {Marrone}, {Moncelsi}, {Shiu}, {Trumper}, {Turner}, {Weber}, {Wei}, \&
  {Zemcov}}]{Sun+2021}
{Sun}, G., {Chang}, T.~C., {Uzgil}, B.~D., {$et~al$.} 2021, \apj, 915, 33

\bibitem[{{Thomas} {$et~al$.}(2009){Thomas}, {Zaroubi}, {Ciardi}, {Pawlik},
  {Labropoulos}, {Jeli{\'c}}, {Bernardi}, {Brentjens}, {de Bruyn}, {Harker},
  {Koopmans}, {Mellema}, {Pandey}, {Schaye}, \& {Yatawatta}}]{Thomas+2009}
{Thomas}, R.~M., {Zaroubi}, S., {Ciardi}, B., {$et~al$.} 2009, \mnras, 393, 32

\bibitem[{Tingay {$et~al$.}(2012)Tingay, Goeke, Hewitt, Morgan, Remillard,
  Williams, Bowman, Emrich, Ord, Booler, {$et~al$.}}]{tingay+2012}
Tingay, S., Goeke, R., Hewitt, J., {$et~al$.} 2012, arXiv preprint
  arXiv:1212.1327

\bibitem[{{Tiwari} {$et~al$.}(2021){Tiwari}, {Shaw}, {Majumdar}, {Kamran}, \&
  {Choudhury}}]{Tiwari+2021}
{Tiwari}, H., {Shaw}, A.~K., {Majumdar}, S., {Kamran}, M., \& {Choudhury}, M.
  2021, arXiv e-prints, arXiv:2108.07279

\bibitem[{{Trott} {$et~al$.}(2016){Trott}, {Pindor}, {Procopio}, {Wayth},
  {Mitchell}, {McKinley}, {Tingay}, {Barry}, {Beardsley}, {Bernardi}, {Bowman},
  {Briggs}, {Cappallo}, {Carroll}, {de Oliveira-Costa}, {Dillon}, {Ewall-Wice},
  {Feng}, {Greenhill}, {Hazelton}, {Hewitt}, {Hurley-Walker},
  {Johnston-Hollitt}, {Jacobs}, {Kaplan}, {Kim}, {Lenc}, {Line}, {Loeb},
  {Lonsdale}, {Morales}, {Morgan}, {Neben}, {Thyagarajan}, {Oberoi},
  {Offringa}, {Ord}, {Paul}, {Pober}, {Prabu}, {Riding}, {Udaya Shankar},
  {Sethi}, {Srivani}, {Subrahmanyan}, {Sullivan}, {Tegmark}, {Webster},
  {Williams}, {Williams}, {Wu}, \& {Wyithe}}]{2016ApJ...818..139T}
{Trott}, C.~M., {Pindor}, B., {Procopio}, P., {$et~al$.} 2016, \apj, 818, 139

\bibitem[{{Trott} {$et~al$.}(2020){Trott}, {Jordan}, {Midgley}, {Barry},
  {Greig}, {Pindor}, {Cook}, {Sleap}, {Tingay}, {Ung}, {Hancock}, {Williams},
  {Bowman}, {Byrne}, {Chokshi}, {Hazelton}, {Hasegawa}, {Jacobs}, {Joseph},
  {Li}, {Line}, {Lynch}, {McKinley}, {Mitchell}, {Morales}, {Ouchi}, {Pober},
  {Rahimi}, {Takahashi}, {Wayth}, {Webster}, {Wilensky}, {Wyithe}, {Yoshiura},
  {Zhang}, \& {Zheng}}]{2020MNRAS.493.4711T}
{Trott}, C.~M., {Jordan}, C.~H., {Midgley}, S., {$et~al$.} 2020, \mnras, 493,
  4711

\bibitem[{{van Haarlem, M. P.} {$et~al$.}(2013){van Haarlem, M. P.}, {Wise, M.
  W.}, {Gunst, A. W.}, {Heald, G.}, {McKean, J. P.}, {Hessels, J. W. T.}, {de
  Bruyn, A. G.}, {Nijboer, R.}, {Swinbank, J.}, {Fallows, R.}, {Brentjens, M.},
  {Nelles, A.}, {Beck, R.}, {Falcke, H.}, {Fender, R.}, {H\"orandel, J.},
  {Koopmans, L. V. E.}, {Mann, G.}, {Miley, G.}, {R\"ottgering, H.}, {Stappers,
  B. W.}, {Wijers, R. A. M. J.}, {Zaroubi, S.}, {van den Akker, M.}, {Alexov,
  A.}, {Anderson, J.}, {Anderson, K.}, {van Ardenne, A.}, {Arts, M.}, {Asgekar,
  A.}, {Avruch, I. M.}, {Batejat, F.}, {B\"ahren, L.}, {Bell, M. E.}, {Bell, M.
  R.}, {van Bemmel, I.}, {Bennema, P.}, {Bentum, M. J.}, {Bernardi, G.}, {Best,
  P.}, {B\^{\i}rzan, L.}, {Bonafede, A.}, {Boonstra, A.-J.}, {Braun, R.},
  {Bregman, J.}, {Breitling, F.}, {van de Brink, R. H.}, {Broderick, J.},
  {Broekema, P. C.}, {Brouw, W. N.}, {Br\"uggen, M.}, {Butcher, H. R.}, {van
  Cappellen, W.}, {Ciardi, B.}, {Coenen, T.}, {Conway, J.}, {Coolen, A.},
  {Corstanje, A.}, {Damstra, S.}, {Davies, O.}, {Deller, A. T.}, {Dettmar,
  R.-J.}, {van Diepen, G.}, {Dijkstra, K.}, {Donker, P.}, {Doorduin, A.},
  {Dromer, J.}, {Drost, M.}, {van Duin, A.}, {Eisl\"offel, J.}, {van Enst, J.},
  {Ferrari, C.}, {Frieswijk, W.}, {Gankema, H.}, {Garrett, M. A.}, {de
  Gasperin, F.}, {Gerbers, M.}, {de Geus, E.}, {Grie\ss{}meier, J.-M.}, {Grit,
  T.}, {Gruppen, P.}, {Hamaker, J. P.}, {Hassall, T.}, {Hoeft, M.}, {Holties,
  H. A.}, {Horneffer, A.}, {van der Horst, A.}, {van Houwelingen, A.},
  {Huijgen, A.}, {Iacobelli, M.}, {Intema, H.}, {Jackson, N.}, {Jelic, V.}, {de
  Jong, A.}, {Juette, E.}, {Kant, D.}, {Karastergiou, A.}, {Koers, A.},
  {Kollen, H.}, {Kondratiev, V. I.}, {Kooistra, E.}, {Koopman, Y.}, {Koster,
  A.}, {Kuniyoshi, M.}, {Kramer, M.}, {Kuper, G.}, {Lambropoulos, P.}, {Law,
  C.}, {van Leeuwen, J.}, {Lemaitre, J.}, {Loose, M.}, {Maat, P.}, {Macario,
  G.}, {Markoff, S.}, {Masters, J.}, {McFadden, R. A.}, {McKay-Bukowski, D.},
  {Meijering, H.}, {Meulman, H.}, {Mevius, M.}, {Middelberg, E.}, {Millenaar,
  R.}, {Miller-Jones, J. C. A.}, {Mohan, R. N.}, {Mol, J. D.}, {Morawietz, J.},
  {Morganti, R.}, {Mulcahy, D. D.}, {Mulder, E.}, {Munk, H.}, {Nieuwenhuis,
  L.}, {van Nieuwpoort, R.}, {Noordam, J. E.}, {Norden, M.}, {Noutsos, A.},
  {Offringa, A. R.}, {Olofsson, H.}, {Omar, A.}, {Orr\'u, E.}, {Overeem, R.},
  {Paas, H.}, {Pandey-Pommier, M.}, {Pandey, V. N.}, {Pizzo, R.}, {Polatidis,
  A.}, {Rafferty, D.}, {Rawlings, S.}, {Reich, W.}, {de Reijer, J.-P.},
  {Reitsma, J.}, {Renting, G. A.}, {Riemers, P.}, {Rol, E.}, {Romein, J. W.},
  {Roosjen, J.}, {Ruiter, M.}, {Scaife, A.}, {van der Schaaf, K.}, {Scheers,
  B.}, {Schellart, P.}, {Schoenmakers, A.}, {Schoonderbeek, G.}, {Serylak, M.},
  {Shulevski, A.}, {Sluman, J.}, {Smirnov, O.}, {Sobey, C.}, {Spreeuw, H.},
  {Steinmetz, M.}, {Sterks, C. G. M.}, {Stiepel, H.-J.}, {Stuurwold, K.},
  {Tagger, M.}, {Tang, Y.}, {Tasse, C.}, {Thomas, I.}, {Thoudam, S.}, {Toribio,
  M. C.}, {van der Tol, B.}, {Usov, O.}, {van Veelen, M.}, {van der Veen,
  A.-J.}, {ter Veen, S.}, {Verbiest, J. P. W.}, {Vermeulen, R.}, {Vermaas, N.},
  {Vocks, C.}, {Vogt, C.}, {de Vos, M.}, {van der Wal, E.}, {van Weeren, R.},
  {Weggemans, H.}, {Weltevrede, P.}, {White, S.}, {Wijnholds, S. J.},
  {Wilhelmsson, T.}, {Wucknitz, O.}, {Yatawatta, S.}, {Zarka, P.}, {Zensus,
  A.}, \& {van Zwieten, J.}}]{Haarlem+2013}
{van Haarlem, M. P.}, {Wise, M. W.}, {Gunst, A. W.}, {$et~al$.} 2013, A\&A,
  556, A2

\bibitem[{{Villanueva-Domingo} \&
  {Villaescusa-Navarro}(2021)}]{Villanueva-Domingo+2021}
{Villanueva-Domingo}, P., \& {Villaescusa-Navarro}, F. 2021, \apj, 907, 44

\bibitem[{{Vrbanec} {$et~al$.}(2016){Vrbanec}, {Ciardi}, {Jeli{\'c}}, {Jensen},
  {Zaroubi}, {Fernandez}, {Ghosh}, {Iliev}, {Kakiichi}, {Koopmans}, \&
  {Mellema}}]{vrbanec16}
{Vrbanec}, D., {Ciardi}, B., {Jeli{\'c}}, V., {$et~al$.} 2016, \mnras, 457, 666

\bibitem[{{Watkinson} {$et~al$.}(2021){Watkinson}, {Trott}, \&
  {Hothi}}]{Watkinson+2021}
{Watkinson}, C.~A., {Trott}, C.~M., \& {Hothi}, I. 2021, \mnras, 501, 367

\bibitem[{Weinberger {$et~al$.}(2018)Weinberger, Kulkarni, Haehnelt, Choudhury,
  \& Puchwein}]{weinberger18}
Weinberger, L.~H., Kulkarni, G., Haehnelt, M.~G., Choudhury, T.~R., \&
  Puchwein, E. 2018, \mnras, 479, 2564

\bibitem[{{Wouthuysen}(1952)}]{Wouthuysen+1952}
{Wouthuysen}, S.~A. 1952, The Astronomical Journal, 57, 31

\bibitem[{{Wyithe} \& {Loeb}(2009)}]{Wyithe+2009}
{Wyithe}, J. S.~B., \& {Loeb}, A. 2009, \mnras, 397, 1926

\bibitem[{{Yan} {$et~al$.}(2020){Yan}, {Sajina}, {Loiacono}, {Lagache},
  {B{\'e}thermin}, {Faisst}, {Ginolfi}, {F{\`e}vre}, {Gruppioni}, {Capak},
  {Cassata}, {Schaerer}, {Silverman}, {Bardelli}, {Dessauges-Zavadsky},
  {Cimatti}, {Hathi}, {Lemaux}, {Ibar}, {Jones}, {Koekemoer}, {Oesch}, {Talia},
  {Pozzi}, {Riechers}, {Tasca}, {Toft}, {Vallini}, {Vergani}, {Zamorani}, \&
  {Zucca}}]{Yan+2020}
{Yan}, L., {Sajina}, A., {Loiacono}, F., {$et~al$.} 2020, \apj, 905, 147

\bibitem[{{Yang} {$et~al$.}(2022){Yang}, {Popping}, {Somerville}, {Pullen},
  {Breysse}, \& {Maniyar}}]{Yang+2022}
{Yang}, S., {Popping}, G., {Somerville}, R.~S., {$et~al$.} 2022, \apj, 929, 140

\bibitem[{{Yang} {$et~al$.}(2021){Yang}, {Somerville}, {Pullen}, {Popping},
  {Breysse}, \& {Maniyar}}]{Yang+2021}
{Yang}, S., {Somerville}, R.~S., {Pullen}, A.~R., {$et~al$.} 2021, \apj, 911,
  132

\bibitem[{{Yatawatta}(2015)}]{2015MNRAS.449.4506Y}
{Yatawatta}, S. 2015, \mnras, 449, 4506

\bibitem[{{Yue} {$et~al$.}(2015){Yue}, {Ferrara}, {Pallottini}, {Gallerani}, \&
  {Vallini}}]{Yue+2015}
{Yue}, B., {Ferrara}, A., {Pallottini}, A., {Gallerani}, S., \& {Vallini}, L.
  2015, \mnras, 450, 3829

\bibitem[{{Zackrisson} {$et~al$.}(2020){Zackrisson}, {Majumdar}, {Mondal},
  {Binggeli}, {Sahl{\'e}n}, {Choudhury}, {Ciardi}, {Datta}, {Datta}, {Dayal},
  {Ferrara}, {Giri}, {Maio}, {Malhotra}, {Mellema}, {Mesinger}, {Rhoads},
  {Rydberg}, \& {Shimizu}}]{Zackrisson+2020}
{Zackrisson}, E., {Majumdar}, S., {Mondal}, R., {$et~al$.} 2020, \mnras, 493,
  855

\bibitem[{{Zahn} {$et~al$.}(2007){Zahn}, {Lidz}, {McQuinn}, {Dutta},
  {Hernquist}, {Zaldarriaga}, \& {Furlanetto}}]{Zahn+2007}
{Zahn}, O., {Lidz}, A., {McQuinn}, M., {$et~al$.} 2007, \apj, 654, 12

\bibitem[{{Zheng} {$et~al$.}(2017){Zheng}, {Wang}, {Rhoads}, {Infante},
  {Malhotra}, {Hu}, {Walker}, {Jiang}, {Jiang}, {Hibon}, {Gonzalez}, {Kong},
  {Zheng}, {Galaz}, \& {Barrientos}}]{Zheng+2017}
{Zheng}, Z.-Y., {Wang}, J., {Rhoads}, J., {$et~al$.} 2017, \apjl, 842, L22

\bibitem[{{Zhou} \& {La Plante}(2021)}]{Zhou+2021}
{Zhou}, Y., \& {La Plante}, P. 2021, arXiv e-prints, arXiv:2112.03443

\end{thebibliography}


\end{document}